%% file: main.tex
\documentclass[fleqn]{cas-sc}

\usepackage[square,sort,comma,numbers]{natbib}
\usepackage{tabularx}
\usepackage{lineno}

\def\tsc#1{\csdef{#1}{\textsc{\lowercase{#1}}\xspace}}
\tsc{WGM}
\tsc{QE}
\tsc{EP}
\tsc{PMS}
\tsc{BEC}
\tsc{DE}

\input{symbols.tex}

\begin{document}
\let\WriteBookmarks\relax
\def\floatpagepagefraction{1}
\def\textpagefraction{.001}

\shorttitle{The Imaging Time-of-Propagation Detector at Belle~II}
\shortauthors{H. Atmacan et~al.}


\title [mode = title]{The Imaging Time-of-Propagation Detector at Belle~II}                      

\author[1]{H.~Atmacan}
[orcid=0000-0003-2435-501X]
\author[1]{M.~Belhorn} 
[orcid=0000-0003-4489-7668]
\fnmark[1]
\author[1]{Y.~Guan}
[orcid=0000-0002-5541-2278]
\author[1]{L.~Li}
[orcid=0000-0002-7366-1307]
\fnmark[2]
\author[1]{B.~Pal}
[orcid=0000-0003-0737-1528]
\author[1,4]{S.~Sandilya}
[orcid=0000-0002-4199-4369]
\author[1]{A.~J.~Schwartz}
[type=editor, orcid=0000-0002-7310-1983]
\cormark[1]
\author[1]{B.~Wang}
[orcid=0000-0001-6136-6952]
\author[1]{S.~Watanuki}
[orcid=0000-0002-5241-6628]
%
\author[2]{M.~Andrew}
\author[2,3]{M.~Barrett}
[orcid=0000-0002-2095-603X]
\author[2]{M.~Bessner}
[orcid=0000-0003-1776-0439]
\author[2]{T.~E.~Browder}
[orcid=0000-0001-7357-9007]
\author[2]{J.~Bynes}
\author[2]{J.~Cercillieux}
\fnmark[3]
\author[2]{S.~Dubey}
[orcid=0000-0002-1345-0970]
\author[2]{O.~Hartbrich}
[orcid=0000-0001-7741-4381]
\fnmark[1]
\author[2]{C.~Ketter}
[orcid=0000-0002-5161-9722]
\author[2]{B.~Kirby}
\fnmark[4]
\author[2]{S.~Kohani}
[orcid=0000-0003-3869-6552]
\author[2]{D.~Kotchetkov}
\author[2]{L.~Macchiarulo}
\fnmark[4]
\author[2]{B.~Ma\u{c}ek}
\author[2]{K.~Nishimura}
[orcid=0000-0001-8818-8922]
\fnmark[5]
\author[2]{H.~Purwar}
[orcid=0000-0002-3876-7069]
\author[2]{Z.~Sang}
\author[2]{V.~Shebalin}
[orcid=0000-0003-1012-0957]
\author[2]{S.~Tripathi}
\author[2]{G.~Varner}
[orcid=0000-0002-0302-8151]
\fnmark[6]
\author[2]{K.~Yoshihara}
[orcid=0000-0002-3656-2326]
%
\author[3]{T.~Kohriki}
\author[3]{X.~Wang}
\fnmark[7]
%
\author[4]{S.~P.~Maharana}
[orcid=0000-0002-1746-4683]
\author[4]{V.~S.~Vismaya}
[orcid=0000-0002-1606-5349]
%
\author[5]{G.~Visser}
[orcid=0000-0003-2495-758X]
%
\author[6]{A.~Gaz}
[orcid=0000-0001-6754-3315]
\author[6]{S.~Lacaprara}
[orcid=0000-0002-0551-7696]
\author[6]{R.~Stroili}
[orcid=0000-0002-3453-142X]
\author[6]{E.~Torassa}
[orcid=0000-0003-2321-0599]
%
\author[7]{R.~Mussa}
[orcid=0000-0002-0294-9071]
\author[7]{U.~Tamponi}
[orcid=0000-0001-6651-0706]
%
\author[8]{P.~Kri\v{z}an}
[orcid=0000-0002-4967-7675]
\author[8]{S.~Korpar}
[orcid=0000-0003-0971-0968]
\author[8]{T.~Nanut}
[orcid=0000-0002-5728-9867]
\author[8]{A.~Novosel}
[orcid=0000-0002-7308-8950]
\author[8]{M.~Petri\u{c}}
[orcid=0000-0002-1331-2578]
\author[8]{L.~Rizzuto}
[orcid=0000-0001-6621-6646]
\author[8]{M.~Stari\u{c}}
[orcid=0000-0001-8751-5944]
%
\author[9]{L.~M.~Cremaldi}
[orcid=0000-0001-5550-7827]
\author[9]{D.~A.~Sanders}
[orcid=0000-0002-4902-966X]
%
\author[10]{Y.~Adachi}
\author[10]{Y.~Arita}
\author[10]{D.~Furumura}
\author[10]{S.~Hara}
\author[10]{T.~Hayakawa}
\author[10]{S.~Hirose}
\author[11,10]{Y.~Horii}
\author[11,10]{T.~Iijima}
[orcid=0000-0002-4271-711X]
\author[10,11]{K.~Inami}
[orcid=0000-0003-2765-7072]
\author[10]{Y.~Ito}
\author[10]{Y.~Kato}
[orcid=0000-0001-6314-4288]
\author[10]{N.~Kiribe}
\author[10]{R.~Koga}
\author[10]{K.~Kojima}
[orcid=0000-0002-3638-0266]
\author[10]{A.~Maeda}
\author[10]{Y.~Maeda}
\author[10]{K.~Maehara}
\author[10]{R.~Maeshima}
\author[10,3]{K.~Matsuoka}
[orcid=0000-0003-1706-9365]
\author[10]{G.~Muroyama}
\author[10]{T.~Nakano}
\author[10]{R.~Okubo}
[orcid=0009-0009-0912-0678]
\author[10]{R.~Okuto}
\author[10]{R.~Omori}
\author[10]{S.~Senga}
\author[11]{K.~Suzuki}
\author[10]{H.~Takeichi}
\author[10]{N.~Tsuzuki}
[orcid=0000-0003-1141-1908]
\author[10]{T.~Yonekura}
%
\author[12]{R.~Conrad}
\author[12]{J.~Fast}
[orcid=0000-0003-2258-701X]
\fnmark[8]
\author[12]{B.~Fulsom}
[orcid=0000-0002-5862-9739]
\author[12]{J.~Strube}
[orcid=0000-0001-7470-9301]
\author[12]{L.~Wood}
[orcid=0000-0002-5235-8181]
%
\author[13]{I.~Danko}
\author[13]{T.~Gu}
[orcid=0000-0002-1470-6536]
\author[13]{N.~Herring}
\author[13]{N.~K.~Nisar}
[orcid=0000-0001-9562-1253]
\fnmark[9]
\author[13]{S.~Roffe}
\author[13]{V.~Savinov}
[orcid=0000-0002-9184-2830]
\author[13]{E.~Wang}
[orcid=0000-0001-6391-5118]
%
\author[14]{V.~Bhardwaj}
[orcid=0000-0001-8857-8621]
\fnmark[10]
\author[14]{A.~Loos}
\author[14]{M.~Purohit}
[orcid=0000-0002-8381-8689]
\author[14]{C.~Rosenfeld}
[orcid=0000-0003-3857-1223]

\affiliation[1]{organization={University of Cincinnati}, 
                city={Cincinnati}, 
                state={Ohio}, 
                postcode={45221},
                country={USA}}

\affiliation[2]{organization={University of Hawaii}, 
               city={Honolulu}, 
               state={Hawaii}, 
               postcode={96822},
               country={USA}}

\affiliation[3]{organization={High Energy Accelerator Research Organization (KEK)}, 
                city={Tsukuba}, 
                postcode={305-0801},
                country={Japan}}

\affiliation[4]{organization={Indian Institute of Technology},
                city={ Hyderabad}, 
                state={ Telangana},
                postcode={502285}, 
                country={India}}

\affiliation[5]{organization={Indiana University}, 
                city={Bloomington}, 
                state={Indiana}, 
                postcode={47408},
                country={USA}}

\affiliation[6]{organization={Istituto Nazionale di Fisica Nucleare, Sezione di Padova}, 
                city={Padova}, 
               postcode={I-35131},
                country={Italy}}

\affiliation[7]{organization={Istituto Nazionale di Fisica Nucleare, Sezione di Torino}, 
                city={Torino}, 
               postcode={I-10125},
                country={Italy}}

\affiliation[8]{organization={Josef Stefan Institute, University of Ljubljana}, 
                city={Ljubljana}, 
               postcode={1000},
                country={Slovenia}}

\affiliation[9]{organization={University of Mississippi}, 
                city={Oxford}, 
                state={Mississippi}, 
               postcode={38677},
                country={USA}}

\affiliation[10]{organization={Nagoya University}, 
                city={Nagoya}, 
               postcode={464-8602},
                country={Japan}}

\affiliation[11]{organization={Kobayashi-Maskawa Institute for the Origin of Particles and the Universe, Nagoya University}, 
                city={Nagoya}, 
               postcode={464-8602},
                country={Japan}}

\affiliation[12]{organization={Pacific Northwest National Laboratory}, 
                 city={Richland}, 
                 state={Washington}, 
               postcode={99352},
                 country={USA}}

\affiliation[13]{organization={University of Pittsburgh}, 
                 city={Pittsburgh}, 
                 state={Pennsylvania}, 
               postcode={15260},
                 country={USA}}

\affiliation[14]{organization={University of South Carolina}, 
                 city={Columbia}, 
                 state={South Carolina}, 
               postcode={29208},
                 country={USA}}

\cortext[cor1]{Corresponding author: 
Physics Department, University of Cincinnati, P.O. Box 210011, Cincinnati, Ohio 45221 (alan.j.schwartz@uc.edu).}

\fntext[fn1]{now at Oak Ridge National Laboratory, Oak Ridge, Tennessee, USA.}
\fntext[fn2]{now at Hunan Normal University, Changsha, China.}
\fntext[fn3]{now at Nevis Laboratory, Columbia University, New York, New York, USA.}
\fntext[fn4]{now at Nalu Scientific, Honolulu, Hawaii, USA.}
\fntext[fn5]{now at Paradromics, Austin, Texas, USA.}
\fntext[fn6]{deceased}
\fntext[fn7]{now at Fudan University, Shanghai, China.}
\fntext[fn8]{now at Thomas Jefferson National Accelerator Facility, Newport News, Virginia, USA.}
\fntext[fn9]{now at Brookhaven National Laboratory, Upton, New York, USA.}
\fntext[fn10]{now at Indian Institute of Science Education and Research, Mohali, Punjab, India.}

\begin{abstract}
We report on the construction, operation, and performance of the  
Time-of-Propagation detector with imaging used for the Belle~II 
experiment running at the Super-KEKB $e^+e^-$ collider. This detector 
is located in the central barrel region and uses \cherenkov\ light
to provide particle identification among hadrons. The \cherenkov\ 
light is radiated in highly polished bars of synthetic fused 
silica (quartz) and transported to the ends of the bars via total 
internal reflection. One bar end is instrumented with finely segmented
micro-channel-plate photomultiplier tubes to record the light,
while the other end has a mirror attached to reflect the photons 
back to the instrumented end.
Both the propagation times and hit positions of the \cherenkov\ photons
are measured; these depend on the \cherenkov\ angle and together 
provide good discrimination among charged pions, kaons, and protons 
with momenta up to around 4~GeV/$c$. To date, the detector has been 
used to record and analyze almost 600~fb$^{-1}$ of Belle~II data.
\end{abstract}


\maketitle

\begin{quote}
This paper is dedicated in memory of Gary Varner, who played a fundamental role in 
the design, fabrication, and operation of the electronics for the TOP detector.
\end{quote}

\newpage
\tableofcontents
\newpage


\input{intro.tex}
\input{optics.tex}

\input{barbox.tex}

\input{PMTs.tex}

\input{electronics.tex}

\input{gas_system.tex}
\input{calibration.tex}

\input{recon_software.tex}

\input{operations.tex}
\input{performance.tex}
\input{conclusion.tex}
\input{acknowledgements.tex}

\bibliographystyle{unsrt}
\bibliography{main}

\end{document}

%% file: symbols.tex
\def\invfb   {\ensuremath{\mbox{\,fb}^{-1}}\xspace}
\usepackage{relsize}

\let\bar=\overbar

\RequirePackage{xspace}

\usepackage{relsize}
\def\babar{\mbox{\slshape B\kern-0.1em{\smaller A}\kern-0.1em
    B\kern-0.1em{\smaller A\kern-0.2em R}}\xspace}

\newcommand{\tev}{\ensuremath{\mathrm{Te\kern -0.1em V}}\xspace}
\newcommand{\gev}{\ensuremath{\mathrm{Ge\kern -0.1em V}}\xspace}
\newcommand{\mev}{\ensuremath{\mathrm{Me\kern -0.1em V}}\xspace}
\newcommand{\kev}{\ensuremath{\mathrm{ke\kern -0.1em V}}\xspace}
\newcommand{\ev}{\ensuremath{\mathrm{e\kern -0.1em V}}\xspace}
\newcommand{\gevc}{\ensuremath{{\mathrm{Ge\kern -0.1em V\!/}c}}\xspace}
\newcommand{\mevc}{\ensuremath{{\mathrm{Me\kern -0.1em V\!/}c}}\xspace}
\newcommand{\gevcc}{\ensuremath{{\mathrm{Ge\kern -0.1em V\!/}c^2}}\xspace}
\newcommand{\gevgevcccc}{\ensuremath{{\mathrm{Ge\kern -0.1em V^2\!/}c^4}}\xspace}
\newcommand{\mevcc}{\ensuremath{{\mathrm{Me\kern -0.1em V\!/}c^2}}\xspace}

\newcommand{\cherenkov}{Cherenkov\xspace}

\def\simge{\mathrel{%
   \rlap{\raise 0.511ex \hbox{$>$}}{\lower 0.511ex \hbox{$\sim$}}}}
\def\simle{\mathrel{
   \rlap{\raise 0.511ex \hbox{$<$}}{\lower 0.511ex \hbox{$\sim$}}}}

%% file: intro.tex
\section{Introduction}
\label{sec:intro}

The Belle~II detector~\cite{Belle-II:2010dht} is a large, multi-purpose particle 
detector that operates at the SuperKEKB $e^+e^-$ collider~\cite{Akai:2018mbz}
located at the KEK accelerator laboratory in Tsukuba, Japan. 
The experiment records $e^+e^-$ collisions at center-of-mass energies at or near
the $\Upsilon(4S)$ resonance, and studies weak decays of $B$ and $D$ mesons and $\tau$ leptons.
The detector includes silicon pixels and strips to measure decay vertices, a large 
central tracking chamber in a magnetic field to reconstruct particle tracks and measure  
particle momenta, an electromagentic calorimeter to measure energies of electrons and photons, 
and an array of resistive plate chambers and scintillator modules to detect 
muons and neutral $K^0_L$ mesons. Positioned radially between the central tracker 
and the calorimeter is a ``time-of-propagation'' (TOP) detector with imaging, which 
measures the propagation time and hit positions of \cherenkov\ photons in order to 
identify pions, kaons, protons, and deuterons. This paper describes the construction, 
operation, and performance of the TOP detector.

The design of the imaging TOP is based on the principle 
of total internal reflection (TIR) of \cherenkov\ 
light~\cite{Akatsu:1999hi,Ohshima:2000ya,Enari:2002jz,Nishimura:2009dp,Arita:2012dxa,inami2014top}.
This principle was employed by the DIRC detector~\cite{ADAM2005281} 
of the {\it BABAR\/} experiment~\cite{Babar_experiment}.
Like {\it BABAR}'s DIRC, the Belle~II TOP uses optically polished 
bars of synthetic fused silica (quartz) as the radiator material.\footnote{The {\it BABAR\/} 
quartz bars were narrower -- only 3.5~cm wide -- to reduce the probability that two tracks
in an event traverse the same bar.}
When charge tracks 
traverse these bars, they radiate \cherenkov\ photons that propagate via TIR to the ends. 
At one end is attached a mirror to reflect photons, such that all TIR photons 
ultimately arrive at the other end. At that end the photons refract out of the bar 
and onto an array of photomultiplier tubes (PMTs), where their hit positions and
arrival times are recorded. Design work on the imaging TOP began in 2009, 
construction began in 2012, and installation was completed in the spring 
of~2016~\cite{Wang:2017ajq,Wang:2017rjt}. Since that time, the detector has 
been operating continuously and has thus far recorded almost 600~fb$^{-1}$ of 
$e^+e^-$ collision data~\cite{Tamponi:2018czd,BESSNER:2020,BESSNER:2024}. Its 
ability to discriminate among pions, kaons, and protons has been used in almost 
all Belle~II physics analyses to date.

This paper is organized as follows. 
In Section~\ref{sec:optics} we describe the optical components of the TOP, their testing, and assembly;
in Section~\ref{sec:barbox} we describe the light-tight enclosure within which the optical components reside, 
and the support structure;
in Section~\ref{sec:pmts} we describe the photomultiplier tubes used to record \cherenkov\ photons;
in Section~\ref{sec:electronics} we describe the readout electronics;
in Section~\ref{sec:gas} we describe the gas system;
in Section~\ref{sec:calibration} we describe the procedures for calibrating the detector;
in Section~\ref{sec:reconstruction} we describe the reconstruction algorithm;
in Section~\ref{sec:operations} we discuss the operation of the detector; and
in Section~\ref{sec:performance} we discuss its performance.

%% file: optics.tex
\section{Optical components}
\label{sec:optics}

\begin{figure}[tb]
\begin{center}
\hbox{
\includegraphics[width=0.38\textwidth,angle=-90.]{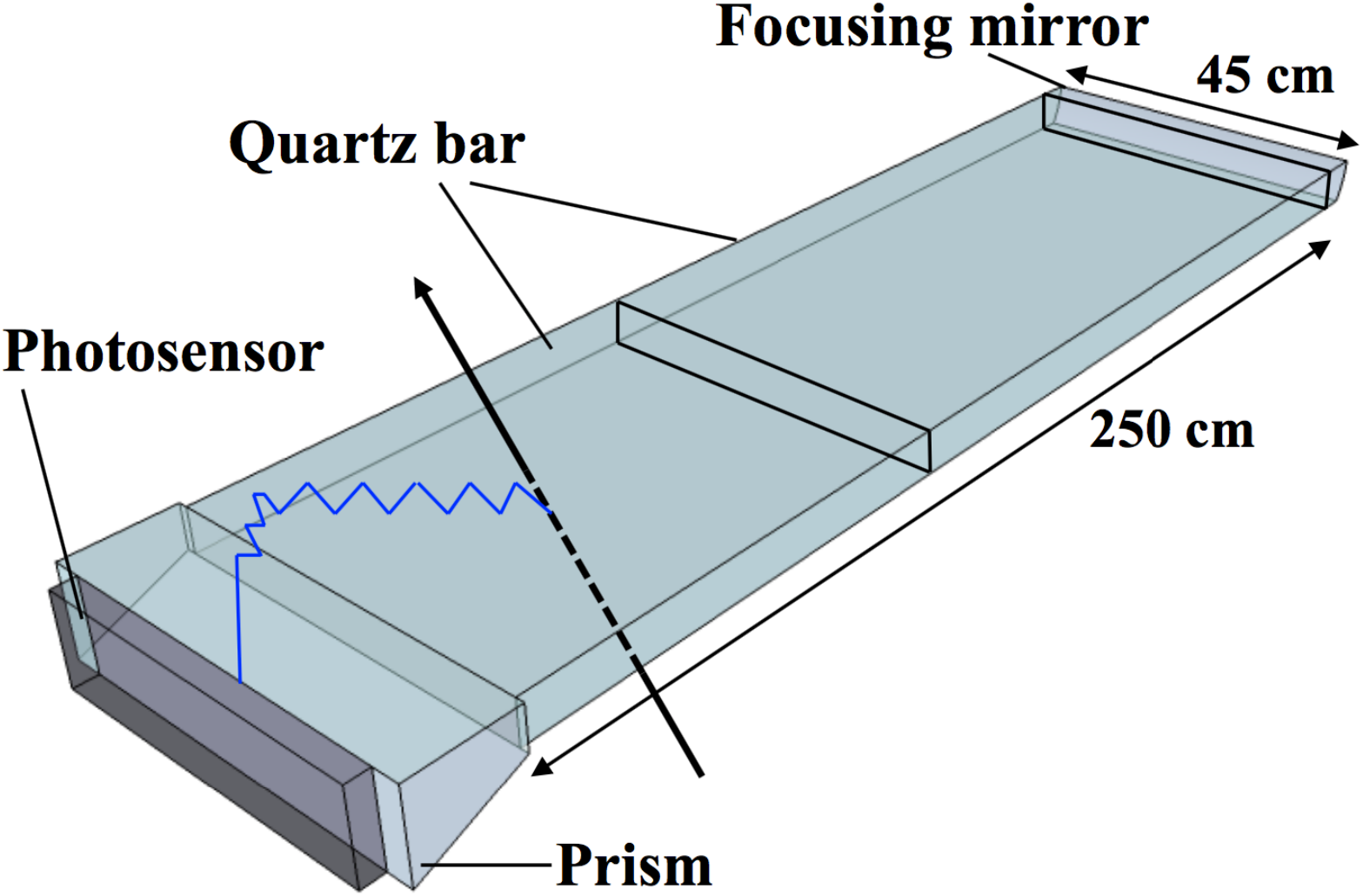}
\hskip-1.8in
\vbox{
\includegraphics[width=0.35\textwidth,angle=-90.]{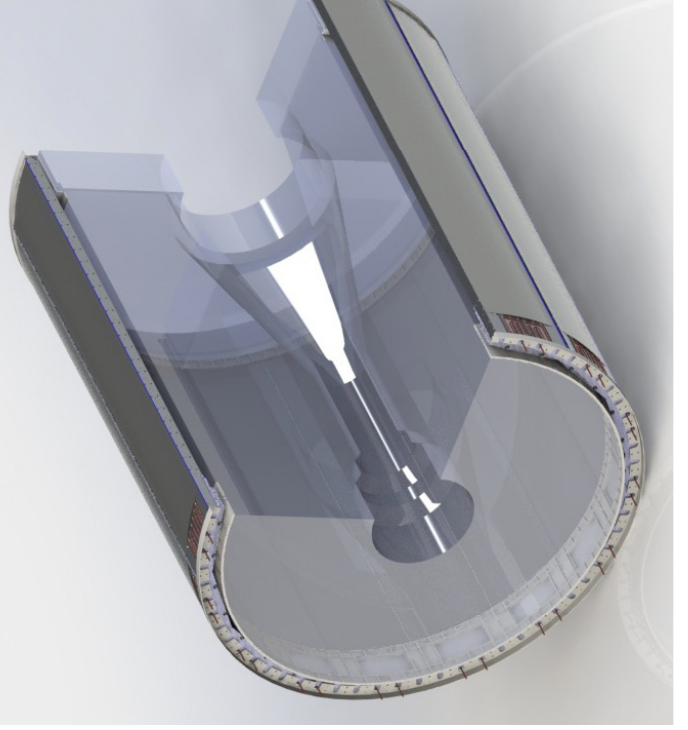}
\vskip-2.4in
}
}
\end{center}
\caption{Components of a TOP optics module~(left), and cutaway view 
of the imaging TOP detector and central tracking chamber showing the 
cylindrical arrangement of detector modules~(right).}
\label{fig:itop}
\end{figure}

A TOP optics module consists of two
bars of length 125~cm, width 45~cm, and thickness 2.0~cm. The bars are
glued together on their narrow faces, as shown in Fig.~\ref{fig:itop}~(left).
A reflective mirror with a spherical surface is mounted at one end, 
and a prism is mounted at the other end. Attached to this
prism is an array of micro-channel-plate photomuliplier tubes (MCP-PMTs).
The optical components are fabricated from Corning 7980 synthetic fused 
silica (quartz), which has high purity and minimal striae; together these
components are referred to as an optics module. The optics module 
is sealed within a light-tight and gas-tight ``quartz bar box'' (QBB),
and this entire assembly, along with attached readout electronics, is 
referred to as a detector module. 
Inside the QBB, the optical components are supported by small buttons 
fabricated from polyetheretherketone (PEEK) in order to minimize 
contact with the quartz surfaces. There are 16 detector modules in total, 
arranged azimuthally around the beampipe as shown in Fig.~\ref{fig:itop}~(right).

\begin{figure}[tb]
	\centering
	\includegraphics[width=0.98\textwidth]{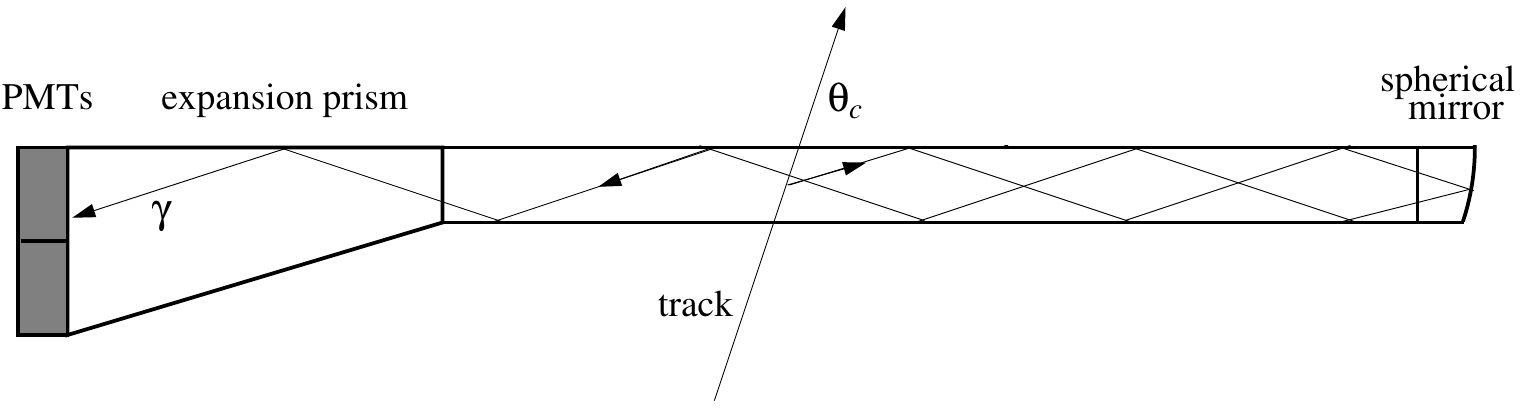}
	\caption{The working principle of a TOP module: \cherenkov\ light is transported via total 
          internal reflection to the PMTs. Light radiated in the forward direction reflects off the 
          spherical mirror.}
	\label{fig:itop-work}
\end{figure}

The working principle of the detector is illustrated in
Fig.~\ref{fig:itop-work}. \cherenkov\ photons are radiated when a
charged particle traverses the quartz bar, with the \cherenkov\ angle 
$\theta^{}_c$ depending on the velocity of the particle. The \cherenkov\ 
photons undergo total internal reflection (TIR) from the bar surfaces,
usually traveling first to the spherical mirror, from which they
reflect. They subsequently travel via TIR down to the prism,
through which they refract out of the bar onto an array of MCP-PMTs. 
These PMTs are multi-anode and record both the position and time of 
the photon hit. The timing resolution of the MCP-PMTs and the front-end
electronics is approximatly 50~ps; this resolution is needed to
resolve the difference in time-of-propagation between photons 
radiated from a $\pi^{\pm}$ track and those radiated from a $K^{\pm}$ track. 
The photon hit position and time, given the incident track's momentum, 
direction, and point of incidence on the bar, determines the particle's identity.

One advantage of a spherical, rather than flat, mirror is that it focuses parallel light 
rays to a point; thus, \cherenkov\ photons radiated at different points along a track 
trajectory but at the same azimuthal angle are imaged to the same point (PMT pixel). 
This reduces smearing of the \cherenkov\ angle due to the bar thickness.
Another advantage is that a spherical mirror allows one to correct for
chromatic disperson as follows. Photons radiated at different wavelengths~$\lambda$
can have notably different propagation times due to the dispersion $dn^{}_g/d\lambda$
of the group index $n^{}_g$, which determines the velocity of propagation.
However, such photons will also have slightly different \cherenkov\ angles due to 
the dispersion of the phase index $n$. Because of these angular differences, such photons, 
upon reflection from the spherical mirror, will be focused to different points, 
i.e., onto different PMT pixels. This spacial separation allows one to correct for the 
difference in propagation times, subsequently improving the resolution in~$\theta^{}_c$.

After inverse ray tracing, the hit position of a \cherenkov\ photon determines
the \cherenkov\ angle (and emission point) of the photon. Thus, it is crucial that the 
numerous total internal reflections preserve the angular information on the photon's 
initial trajectory. This requires that the surfaces of the quartz bars satisfy strict 
requirements on flatness, parallelism, and roughness. 
To achieve these specifications, the bar surfaces were polished to very high optical quality.

\subsection{Construction overview}

The assembly of a full detector module consists of numerous steps:  
\begin{enumerate}
\setlength\itemsep{0em}
\item acceptance testing (``quality assurance'' or QA) of quartz bars, 
prisms, and mirrors;
\item attaching to the prism with epoxy a thin frame machined from PEEK material. 
This frame fixes the position of the optics within the QBB, preventing
movement, and also makes a light-tight and gas-tight seal around the prism.
\item alignment and gluing together a mirror, two bars, and a prism to make 
an optics module;
\item assembly of a QBB, with careful alignment of the siderails and positioning 
of $2\times 19=38$ support buttons; 
\item placement of the optics module into the QBB, and sealing the QBB to make 
it light-tight and gas-tight;
\item assembling ``PMT modules,'' with each PMT module comprising four MCP-PMTs; 
\item attaching eight PMT modules to the outer face of the prism using optical ``cookies''; and
\item mounting the front-end readout electronics (``boardstacks'') to the PMT modules. 
\end{enumerate}
The construction of detector modules began at the beginning of 2014 and was
completed in April, 2016. In total, 16 detector modules were built to be installed
in Belle~II, and one additional module was built as a prototype and for testing purposes. 
Each detector module was individually tested with laser pulses and cosmic rays.
Module installation within Belle~II was completed in May, 2016. 
The specific optical components used for each detector module, along with the 
module's slot assignment (azimuthal position around the central tracking chamber), 
are listed in Table~\ref{tab:modules}. The table also lists results from 
QA measurements of mirrors and prisms, as discussed below.

\begin{table}
\begin{center}
\resizebox{0.99\textwidth}{!}{
\begin{tabular}{c|ccccc|ccc}
\hline \hline
Slot & Module & Prism & Bar (front) & Bar (rear) & Mirror & Prism angle & Mirror radius & Mirror \\
\#   & \#     &       &             &            &        & ($^\circ$) & (mm) & reflect. (\%) \\
\hline
--  & 1 &	449006 & OOW SN-03 &	OOW SN-01 &	SN-002 & 18.089 & 6500.15 &	87.81 \\
1  & 17 &	451252 & 452339 &	452338 &	SN-020 & 18.075 & 6521.72 &	87.15 \\
2  & 16 &	450946 & 450689 &	451368 &	SN-019 & 18.077 & 6507.82 &	86.40 \\
3  & 11 &	450939 & 450687 &	450690 &	SN-013 & 18.082 & 6507.46 &	85.87 \\
4  & 12 &	450942 & 451635 &	451849 &	SN-014 & 18.080 & 6513.36 &	85.07 \\
5  & 15 &	450945 & 451370 &	452291 &	SN-018 & 18.076 & 6514.26 &	85.59 \\
6  & 2 &	449009 & OOW SN-02 &	OOW SN-04 &	SN-004 & 18.073 & 6499.54 &	87.76 \\
7  & 14 &	450944 & 450694 &	451369 &	SN-016 & 18.077 & 6502.49 &	85.04 \\
8  & 7 &	450935 & 450681 &	450481 &	SN-010 & 18.055 & 6507.20 &	85.13 \\
9  & 5 &	450932 & Zygo SN-01 &	450478 &	SN-007 & 18.069 & 6503.16 &	81.94 \\
10 & 13 &	450943 & 450691 &	450693 &	SN-015 & 18.080 & 6508.32 &	84.86 \\
11 & 8 &	450941 & 450683 &	450680 &	SN-011 & 18.058 & 6505.43 &	85.13 \\
12 & 6 &	450937 & 450479 &	450480 &	SN-008 & 18.080 & 6505.25 &	82.33 \\
13 & 9 &	450938 & 450688 &	450685 &	SN-017 & 18.071 & 6506.98 &	85.53 \\
14 & 4 &	449010 & Zygo SN-02 &	Zygo SN-04 &	SN-006 & 18.086 & 6504.60 &	85.85 \\
15 & 3 &	449008 & Zygo SN-03 &	Zygo SN-05 &	SN-003 & 18.091 & 6503.11 &	88.43 \\
16 & 10 &	450936 & 450684 &	450686 &	SN-012 & 18.069 & 6502.32 &	85.21 \\
\hline \hline
\end{tabular}
}
\end{center}
\caption{Optical components for modules, and QA measurements for mirrors and prisms.
The first module (\#1) served as a prototype and was not installed in the Belle~II detector.}
\label{tab:modules}
\end{table}

\subsection{Quality Assurance (QA) testing of optics}

The optics of the imaging TOP detector were fabricated to very high
mechanical and optical tolerances to achieve the requisite performance in discriminating
charged kaons from pions. As the \cherenkov\ photons typically reflect 50--100
times inside the bar before being recorded by a PMT, to preserve information 
on the \cherenkov\ angle, the bar surfaces must be parallel and 
highly polished. The requirement for surface flatness was $<6.3~\mu$m 
run-out over the 125~cm length. The requirement for surface roughness
was $<5$~\AA~r.m.s. Of the 34 bars fabricated, the first four were produced 
by Okamoto Optics Works (OOW) in Japan, while the remaining 30 were produced by 
Zygo Corporation in the US. After production, the bars underwent metrology 
testing at OOW and Zygo to qualify them for meeting specifications.
The bars were then shipped to the KEK laboratory, where they 
underwent additional QA testing as described below. These tests
measured the bulk transmittance (attenuation length) and 
the coefficient of total internal reflection for each bar. 

The prisms were also fabricated by Zygo. 
The mirrors were fabricated by IIT Exelis Corporation in the US.
All prisms and mirrors were delivered to the University of Cincinnati,
where they underwent QA testing as described below. For each prism, 
the bulk transmittance and the angle of the tilted face were
measured. For each mirror, the radius of curvature and the 
reflectivity of the spherical surface were measured.

After prism testing, a thin frame fabricated from PEEK material was
glued to the outer edge of each prism such that one edge of the 
PEEK frame was flush with the large face of the prism.
This frame sits in a recess machined into the front section 
of the QBB (referred to as the ``prism enclosure'') and fixes the position of
the prism (and thus the optics module) within the QBB. It also provides 
a light-tight and gas-tight seal for the prism enclosure. After this gluing, 
mirrors and prisms were transported in Pelican cases~\cite{Pelican_case} 
to the KEK laboratory for module assembly.

\subsubsection{Quartz bar measurements}

All bars were tested in a clean room at KEK. Bars were first inspected for
chips and scratches; any found were recorded and compared to those listed in the
vendors' metrology report. Subsequently, each bar was mounted on an optical table and 
a 405~nm laser injected into one end at normal incidence such that it emerged at the far 
end without any reflections and impinged on a photodiode. The photodiode current
was recorded by a small computer. Measuring this current with and without the bar in place yielded 
a measurement of the bulk transmittance ($T$) via
\begin{eqnarray}
T & = & \frac{I^{}_1/(1-R^{}_1)}{I^{}_0\cdot (1-R^{}_0)}\,,
\label{eqn:trans}
\end{eqnarray}
where $I^{}_1$ and $I^{}_0$ are the beam intensities recorded with and without, 
respectively, the bar in the beam. The reflectances $R^{}_0$ and $R^{}_1$ 
correspond to reflections at the quartz-air boundaries where the beam respectively 
enters and exits the bar and were calculated via the Fresnel equations. 
The transmittance was normalized to a standard length of 1.00~m via the formula
\begin{eqnarray}
T' & = & T^{\ell'/\ell} \,,
\label{eqn:trans_corr}
\end{eqnarray}
where $\ell$ is the bar length (1.25~m) and $\ell' = 1.00$~m.
The requirement for acceptance was $T'>0.9850$, which corresponds to 
an attentuation length of $\Lambda = -1/\log(T') >66.2$~m.

\begin{figure}[hbt]
  \centering
  \includegraphics[width=0.195\textwidth,angle=-90]{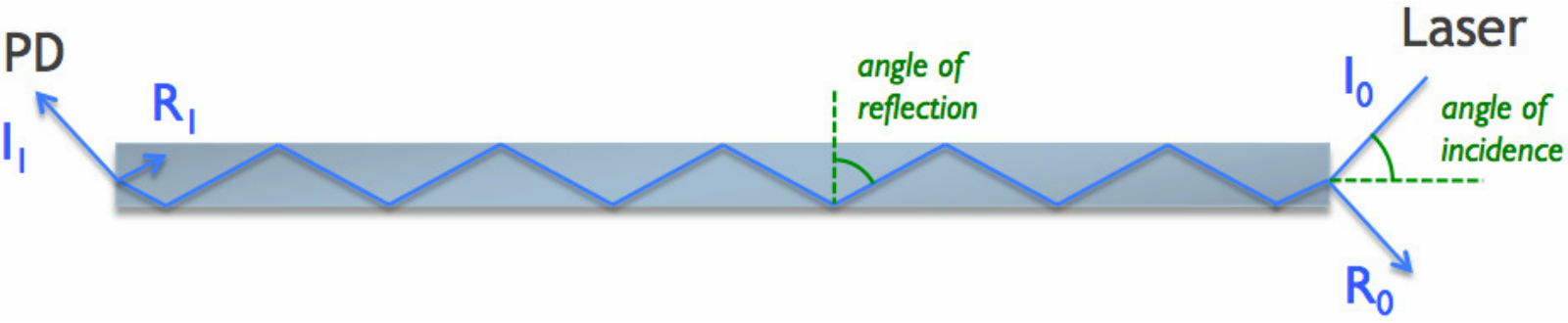}
  \caption{Set-up used to measure the coefficient of total internal reflection of a bar.}
  \label{fig:ctir_setup}
\end{figure}

\begin{figure}[hbt]
  \centering
  \includegraphics[width=0.55\textwidth,angle=-90]{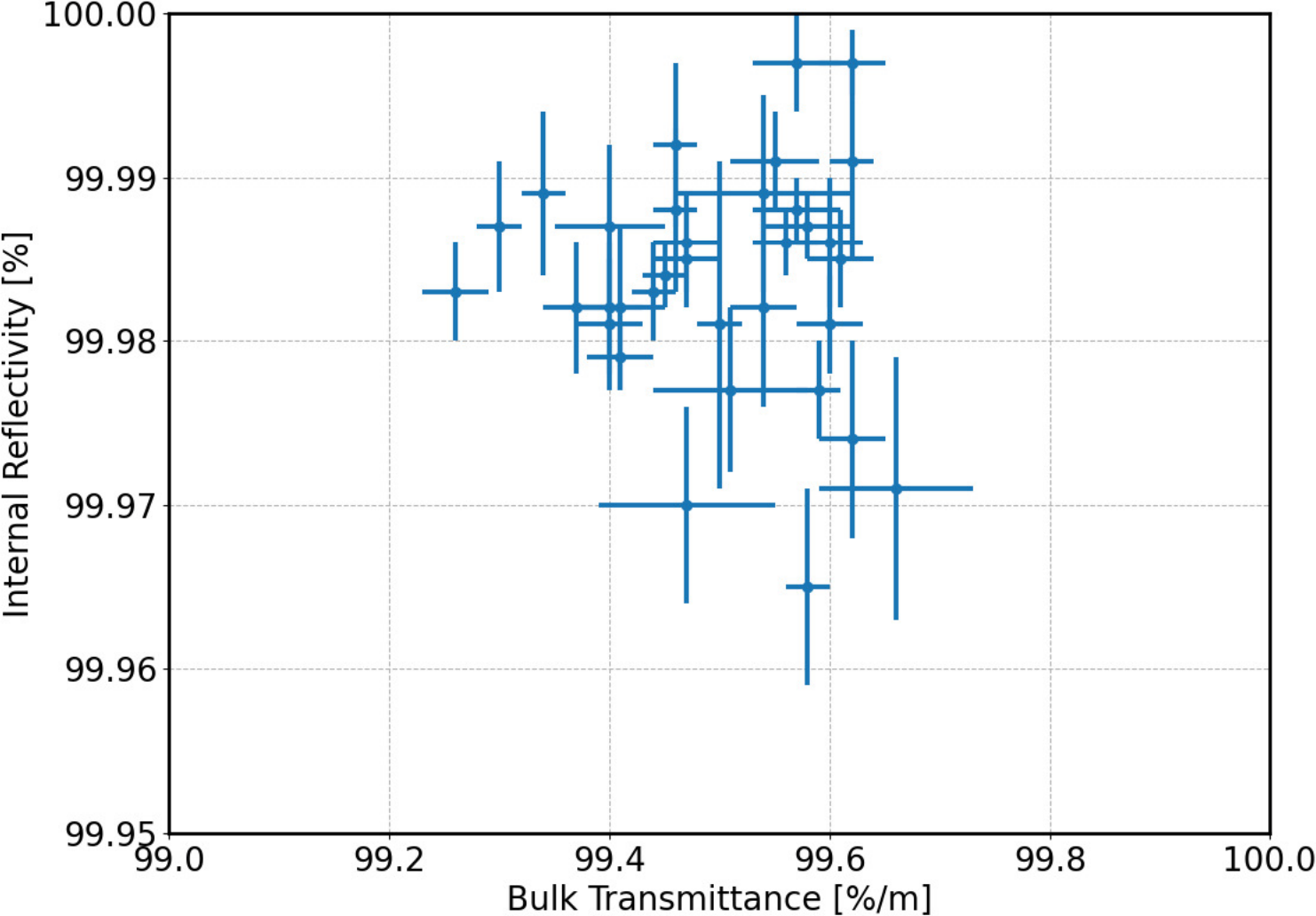}
  \vskip0.10in
  \caption{Summary of QA measurements of the quartz bars, for $\lambda = 405$~nm.}
  \label{fig:bar_qa}
\end{figure}

To measure the coefficient of total internal reflection, the same 405~nm laser and
set-up was used but the laser was incident on the bar face at an angle such that 
the refracted beam underwent TIR within the bar to reach the far end, as shown in 
Fig.~\ref{fig:ctir_setup}. The coefficient of total internal reflection ($\alpha$) 
was calculated via
\begin{eqnarray}
\frac{I^{}_1}{1 - R^{}_1} & = & I^{}_0 (1 - R^{}_0)\,\alpha^N 
\exp{\left[-\frac{L}{\Lambda}\sqrt{1+\left(\frac{Nh}{L}\right)^2}\right]}\,,
\end{eqnarray}
where $\Lambda$ is the previously measured attenuation length, $L\!=\!1.25$~m is 
the bar length, $h\!=\!2.0$~cm is the bar height, and $N$ is the number of 
reflections within the bar, typically 27--28. 
The criterion for acceptance was $\alpha > 0.9990$. The measurements 
of $T'$ and $\alpha$ are listed in Table~\ref{tab:bar_qa} and plotted 
in Fig.~\ref{fig:bar_qa}. All bars fulfilled the acceptance criteria.
The attenuation lengths are listed for a wavelength of 442~nm
by assuming Rayleigh scattering and scaling by $1/\lambda^4$; this
facilitates comparison with measurements of the quartz bars of the DIRC 
detector built for the {\it BABAR\/} experiment~\cite{Cohen-Tanugi:2003kyc}. 

\begin{table}
\begin{center}
\begin{tabular}{cc|rccc}
\hline \hline
Vendor	& Serial Number	& Module & Transmittance   & Atten. length	& Internal reflection \\
	&               &        &  405 nm (\%/m)  & 	442 nm (m)	& (\%/bounce) \\
\hline
OOW	&  SN-01  &  1  &  $99.54\pm 0.08$	&  307.69	&  $99.989\pm  0.006$ \\  
OOW	&  SN-02  &  2	&  $99.62\pm  0.03$	&  372.61	&  $99.997\pm  0.002$ \\  
OOW	&  SN-03  &  1	&  $99.62\pm  0.02$	&  372.61	&  $99.991\pm  0.006$ \\  
OOW	&  SN-04  &  2	&  $99.57\pm  0.04$	&  329.20	&  $99.997\pm  0.003$ \\  
Zygo	&  SN-1	  &  5	&  $99.66\pm  0.07$	&  416.53	&  $99.971\pm  0.008$ \\  
Zygo	&  SN-2	  &  4	&  $99.47\pm  0.08$	&  266.96	&  $99.970\pm  0.006$ \\  
Zygo	&  SN-3	  &  3	&  $99.51\pm  0.07$	&  288.81	&  $99.977\pm  0.005$ \\  
Zygo	&  SN-4	  &  4	&  $99.58\pm  0.04$	&  337.06	&  $99.987\pm  0.002$ \\  
Zygo	&  SN-5	  &  3	&  $99.57\pm  0.04$	&  329.20	&  $99.988\pm  0.002$ \\  
Zygo	&  450478 &  5	&  $99.55\pm  0.04$	&  314.54	&  $99.991\pm  0.002$ \\  
Zygo	&  450481 &  7	&  $99.58\pm  0.02$	&  337.06	&  $99.965\pm  0.006$ \\  
Zygo	&  450479 &  6	&  $99.54\pm  0.03$	&  307.69	&  $99.984\pm  0.004$ \\  
Zygo	&  450480 &  6	&  $99.61\pm  0.03$	&  363.04	&  $99.988\pm  0.003$ \\  
Zygo	&  450680 &  8	&  $99.62\pm  0.03$	&  372.61	&  $99.974\pm  0.006$ \\  
Zygo	&  450681 &  7	&  $99.60\pm  0.03$	&  353.95	&  $99.981\pm  0.003$ \\  
Zygo	&  450683 &  8	&  $99.41\pm  0.04$	&  239.74	&  $99.982\pm  0.005$ \\  
Zygo	&  450684 &  10	&  $99.60\pm  0.03$	&  353.95	&  $99.986\pm  0.004$ \\  
Zygo	&  450685 &  9	&  $99.40\pm  0.05$	&  235.73	&  $99.987\pm  0.005$ \\  
Zygo	&  450686 &  10	&  $99.56\pm  0.03$	&  321.71	&  $99.986\pm  0.002$ \\  
Zygo	&  450688 &  9	&  $99.47\pm  0.03$	&  266.96	&  $99.985\pm  0.003$ \\  	
Zygo	&  450687 &  11	&  $99.44\pm  0.02$	&  252.62	&  $99.983\pm  0.003$ \\  
Zygo	&  450690 &  11	&  $99.47\pm  0.03$	&  266.96	&  $99.986\pm  0.003$ \\  
Zygo	&  451635 &  12 &  $99.45\pm  0.02$	&  257.22	&  $99.984\pm  0.002$ \\  
Zygo	&  451849 &  12 &  $99.46\pm  0.02$	&  262.00	&  $99.988\pm  0.005$ \\  
Zygo	&  450691 &  13 &  $99.50\pm  0.02$	&  283.02	&  $99.981\pm  0.010$ \\  
Zygo	&  450693 &  13 &  $99.59\pm  0.02$	&  345.30	&  $99.977\pm  0.003$ \\  
Zygo	&  450694 &  14 &  $99.41\pm  0.03$	&  239.74	&  $99.979\pm  0.002$ \\  
Zygo	&  451369 &  14	&  $99.37\pm  0.03$	&  224.47	&  $99.982\pm  0.004$ \\  
Zygo	&  451370 &  15	&  $99.34\pm  0.02$	&  214.23	&  $99.989\pm  0.005$ \\  
Zygo	&  452291 &  15	&  $99.40\pm  0.03$	&  235.73	&  $99.981\pm  0.004$ \\  
Zygo	&  450689 &  16	&  $99.26\pm  0.03$	&  191.00	&  $99.983\pm  0.003$ \\  
Zygo	&  451368 &  16	&  $99.30\pm  0.02$	&  201.95	&  $99.987\pm  0.004$ \\  
Zygo	&  452339 &  17	&  $99.40\pm  0.03$	&  235.73	&  $99.982\pm  0.004$ \\  
Zygo	&  452338 &  17	&  $99.46\pm  0.02$	&  262.00	&  $99.992\pm  0.005$ \\  
\hline
{\bf Average}  &   &    &               &  {\bf 292.93}	&  {\bf 99.984} \\
\hline \hline
\end{tabular}
\end{center}
\caption{Measurements of the transmittance, attenuation length, and 
coefficient of total internal reflection for the quartz bars.
All measurements were performed at a wavelength $\lambda = 405$~nm.
The results listed for $\lambda=442$~nm are obtained by scaling by 
$1/\lambda^4$ to facilitate comparison with results from 
Ref.~\cite{Cohen-Tanugi:2003kyc}. }
\label{tab:bar_qa}
\end{table}

\subsubsection{Prism measurements}

For each prism, the transmittance and the angle of the tilted face were measured. 
As \cherenkov\ photons reflect off the tilted face, precise knowledge of the tilt 
angle is important for tracing photons back to the track and determining
the \cherenkov\ angle $\theta^{}_c$ with high precision.
The specification for the tilt angle was $(18.07\pm0.04)^{\circ}$.

\begin{figure}[htb]
\centering
\includegraphics[width=1.7in,angle=90.]{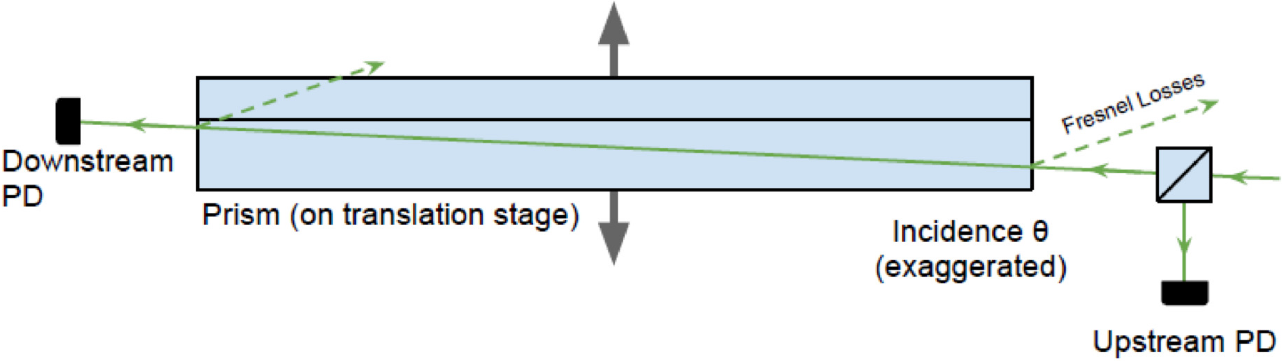}
\caption{Set-up used to measure the transmittance (bulk attenuation) of a prism.}
\label{fig:transmittance_schematic}
\end{figure}

\begin{figure}[htb]
\centering
\includegraphics[width=3.5in,angle=-90]{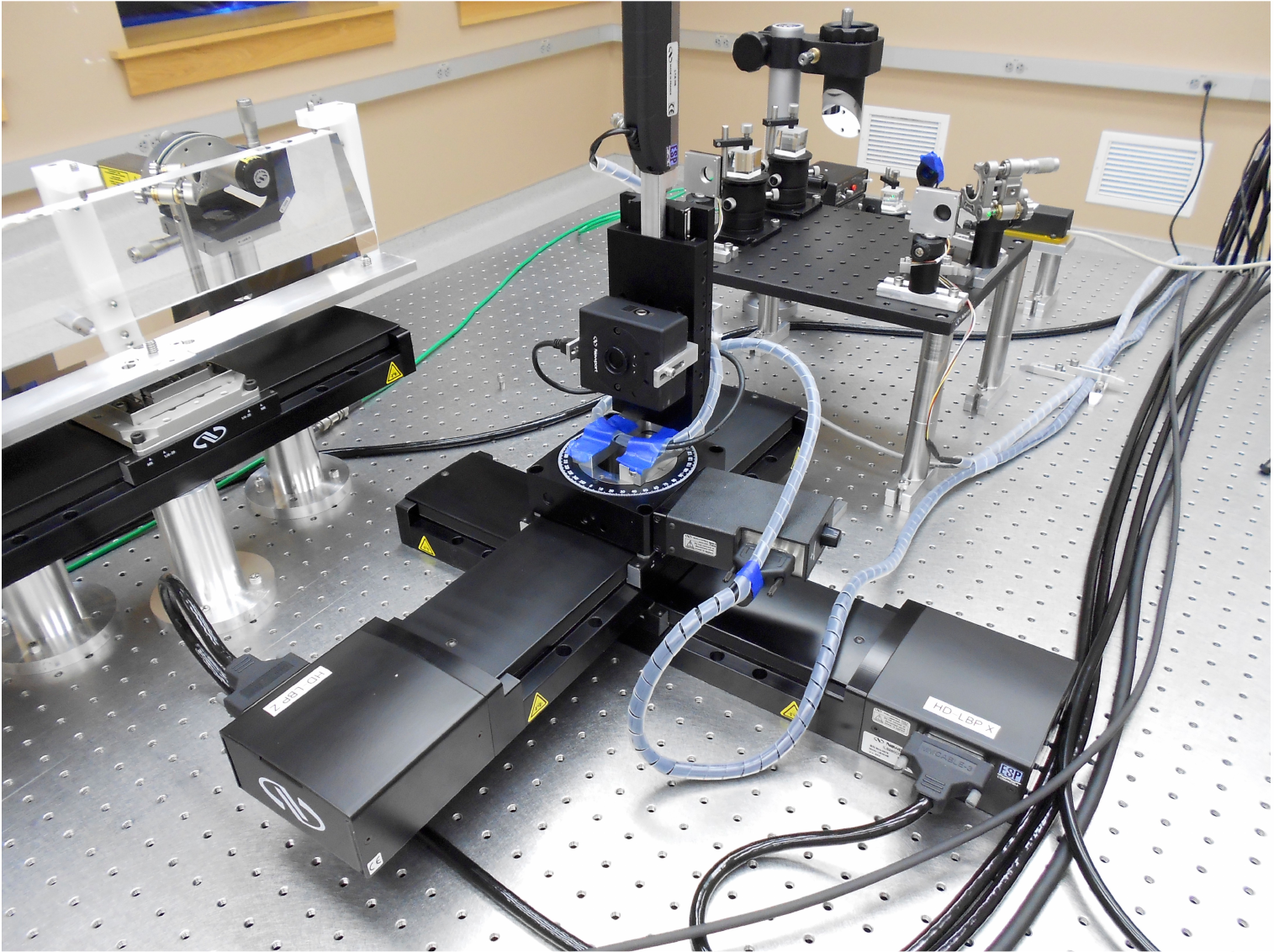}
\caption{Apparatus used for prism and mirror QA measurements.
A prism being measured is mounted on the left.}
\label{fig:UC_optics_lab}
\end{figure}

All measurements were performed in a class 100 cleanroom on a large optical table 
using a 532~nm laser, filter, polarimeter, beam splitter, mirrors, translation stages, 
and photodiodes. The set-up for measuring transmittance is shown in 
Fig.~\ref{fig:transmittance_schematic}, and the apparatus is shown in 
Fig.~\ref{fig:UC_optics_lab}.
A laser beam was incident on the narrow face of the prism, 
propagated down the 45~cm length, and refracted out onto
a photodiode. The laser intensity was monitored by a separate 
photodiode positioned upstream of the prism.
The beam was incident at a small angle ($8\pm 3$~mrad) to minimize
recording light reflected at the surface boundaries.  The transmittance 
per meter was calculated using Eqs.~(\ref{eqn:trans}) and (\ref{eqn:trans_corr}),
where $I^{}_1$ and $I^{}_0$ are the respective beam intensities recorded with 
and without the prism in the beam. The length $\ell$ in this case is 45~cm.

\begin{figure}[htb]
\centering
\includegraphics[width=2.5in,angle=90.]{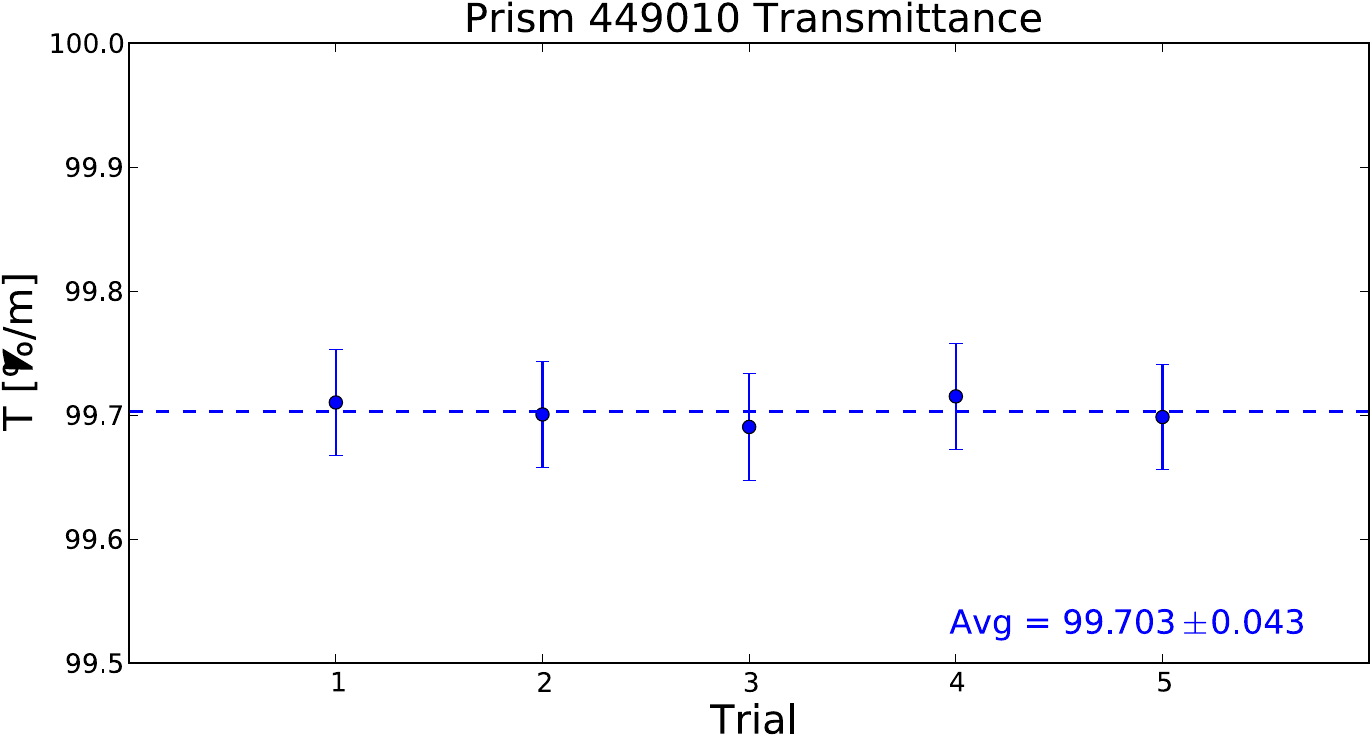}
\caption{Measurements of transmittance for prism~\#449010. 
Each point is the mean of 50 samples, and the error bar corresponds to
the r.m.s.\ of the samples. The dashed line shows the weighted average 
of the measurements.}
\label{fig:transmittance}
\end{figure}

For each prism, 50 measurements of $I^{}_0$ and $I^{}_1$ were recorded, 
and the mean values calculated to obtain~$T$. The currents 
$I^{}_0$ and $I^{}_1$ were measured using high quality photodiodes
operating in their linear range. The procedure yielded repeatable results: 
those obtained for five trials of a single prism (\#449010) are 
plotted in Fig.~\ref{fig:transmittance} and show excellent agreement.
The deviation among trials indicates the systematic uncertainty and
is much smaller than the uncertainty for individual trials, which is
taken to be the r.m.s. of the 50 measurements.

\begin{figure}[htb]
\begin{center}
\includegraphics[width=2.2in,angle=90.]{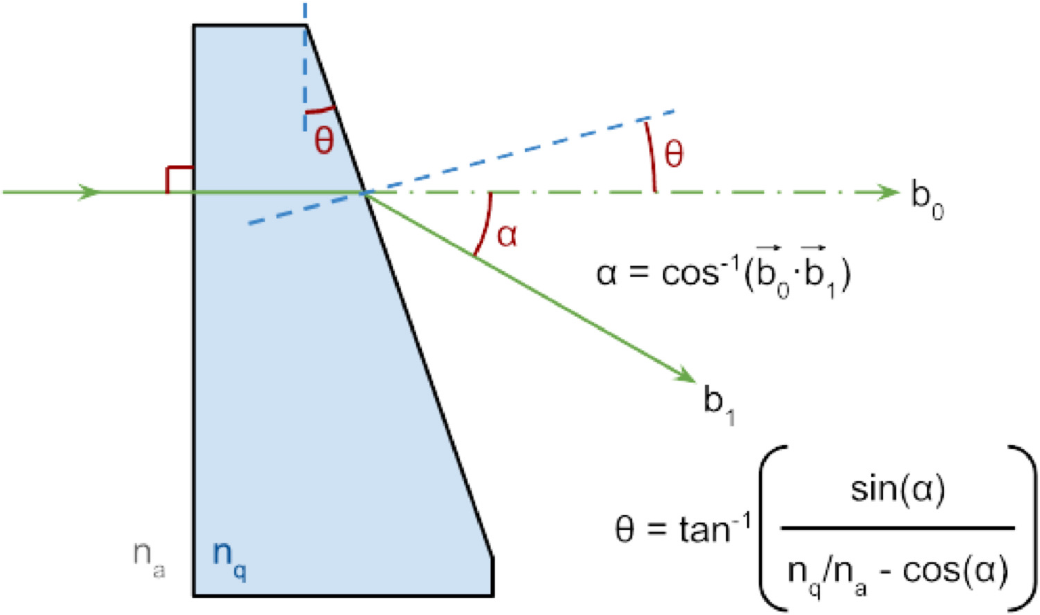}
\end{center}
\caption{Set-up used to measure the angle $\theta$ of the tilted face of a prism.}
\label{fig:angle_schematic}
\end{figure}

\begin{figure}[htb]
\begin{center}
\includegraphics[width=2.6in,angle=90]{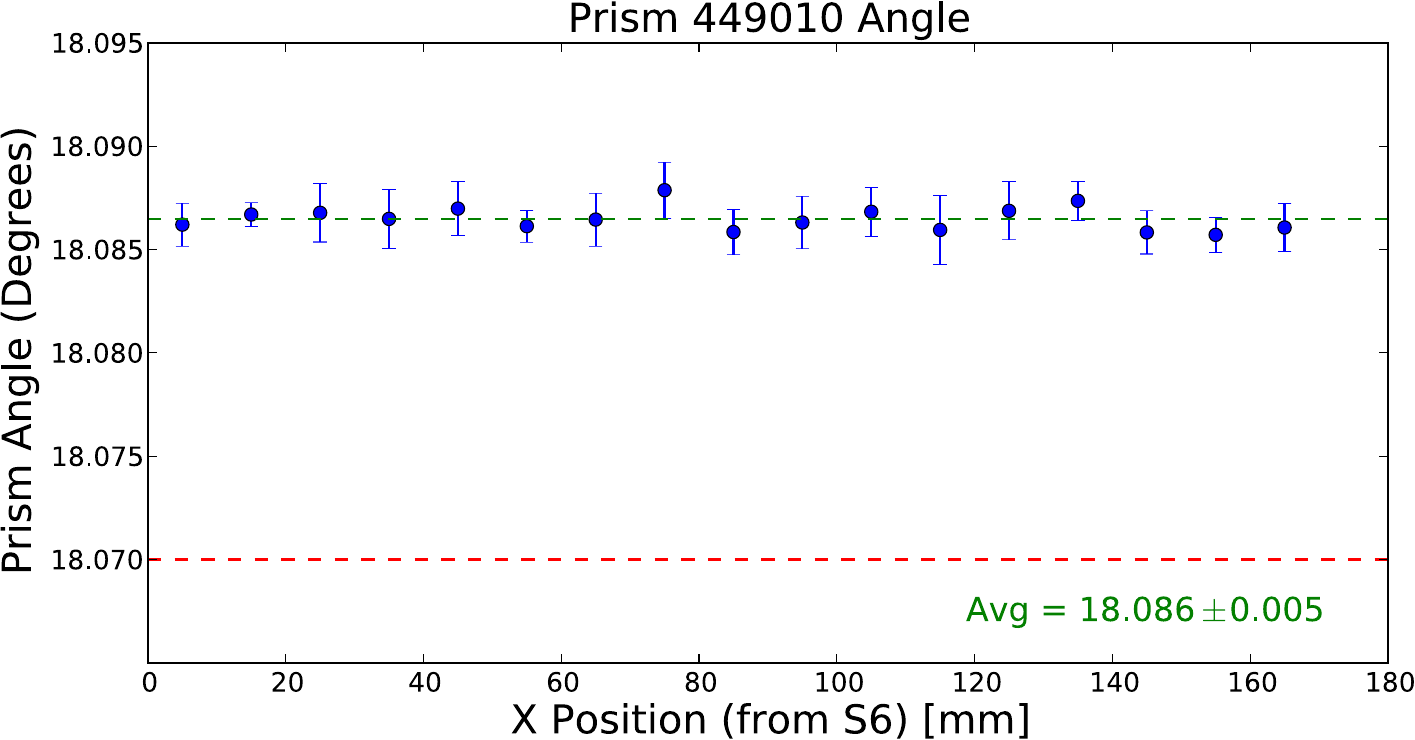}
\end{center}
\caption{Measurements of the  tilt angle of prism \#449010. Each point represents 
the mean of six separate measurements; the error bar corresponds to the sample standard 
deviation. The upper dashed line (green) is the weighted average of the 17 measurements, 
and the lower dashed line (red) is the Zygo measured value.}
\label{fig:angle}
\end{figure}

Measurement of the angle of the tilted face was performed using the same
set-up. However, for this measurement the beam entered the back 
face of the prism at normal incidence and refracted out through the tilted 
face as shown in Fig.~\ref{fig:angle_schematic}. The incident beam was 
normal to the prism face to high precision ($\pm 0.40$~mrad); this was 
achieved by adjusting the laser light reflected from the surface to pass 
through an 0.5~mm aperture located 1.3~m upstream.
The incident and refracted beam directions were determined by 
measuring the corresponding beamspots with a CCD camera
mounted on a rotation stage that itself was mounted on 
$x$-$y$ translation stages, as shown in Fig.~\ref{fig:UC_optics_lab}.
As illustrated in Fig.~\ref{fig:angle_schematic}, the angle subtended 
by the two beams ($\alpha$) is related to the 
tilt angle $\theta$ by Snell's law: 
$n^{}_{\rm q} \sin\theta = n^{}_{\rm a}\sin(\alpha+\theta)$, where
$n^{}_{\rm q}=1.4603$ and $n^{}_{\rm a}=1.000$ are the respective
refractive indices for Corning 7980 synthetic fused silica and 
air at a wavelength of~532~nm. Thus,
\begin{eqnarray}
\theta & = & \tan^{-1}\left(\frac{\sin\alpha}{n_{\rm q}/n_{\rm a}-\cos\alpha}\right) .
\end{eqnarray}
For each prism, the tilt angle was measured at 17 positions 
along the tilted surface.
The results for a typical prism (\#449010) are shown in Figure~\ref{fig:angle}.
Each point represents the average of six samplings, and the r.m.s.\ of the samplings
is plotted as the error bar. The final result for the tilt angle was taken 
as the weighted average of the 17 measurements. All results, listed 
in Table~\ref{tab:modules}, were consistent with the specification 
of~18.07$^\circ$.

\subsubsection{Spherical mirror measurements}

\begin{figure}[bth]
\centering
\includegraphics[width=0.55\textwidth,angle=90]{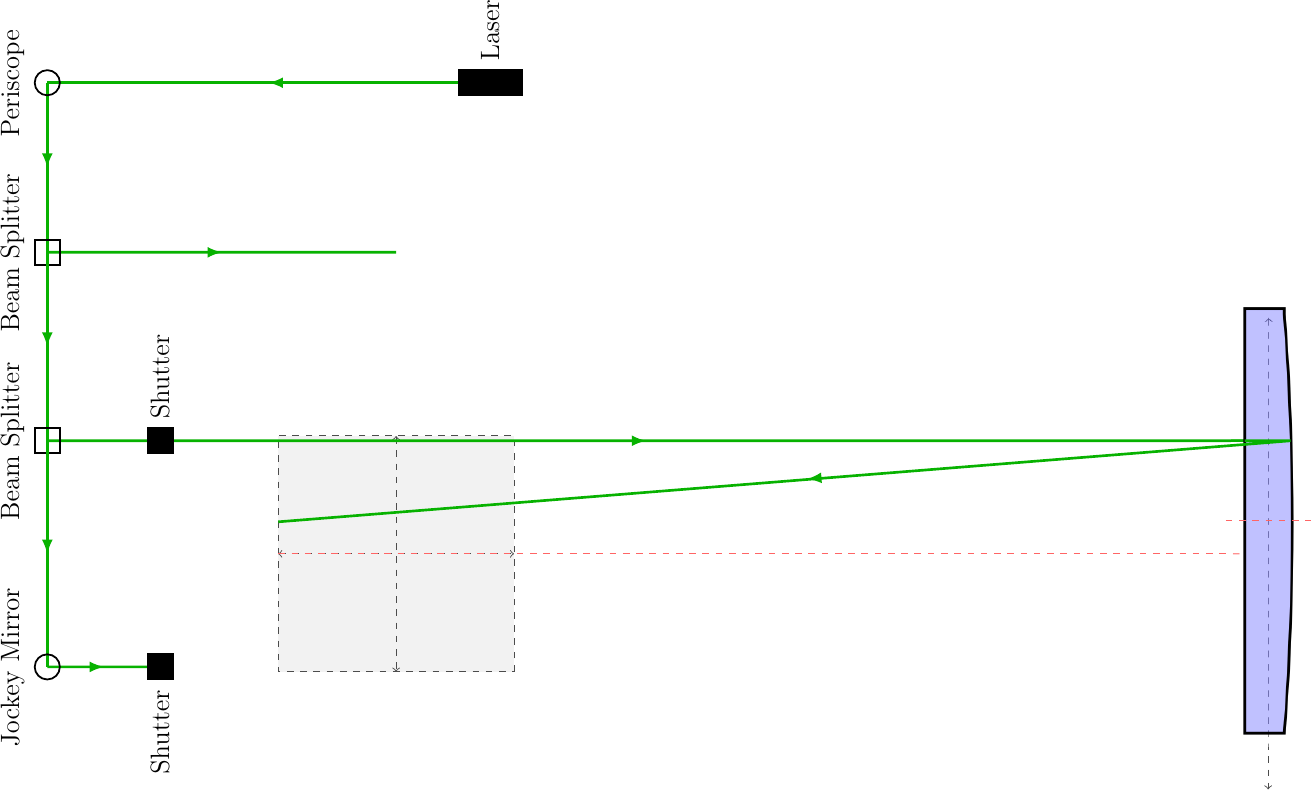}
\caption{Set-up used for measuring the reflectivity and radius of curvature of the spherical mirrors.}
\label{fig:mirror_test_setup}
\end{figure}

All mirrors were shipped to the University of Cincinnati for QA measurements,
which consisted of measuring the reflectivity and radius of curvature.
The set-up used consisted of a 532~nm laser, 
filter, polarimeter, beam splitter, mirrors, translation stages, a rotation stage,
photodiodes, and a laser beam profiler. The beam profiler was based on a 
7.6~mm $\times$ 6.2~mm 1.4~megapixel CCD. The apparatus is shown in 
Fig.~\ref{fig:mirror_test_setup}. 

Mirror reflectivity was measured using a photodiode. The photodiode current 
was recorded with the laser incident on it ($I^{}_0$) and after reflection 
from the mirror surface ($I^{}_1$). After background levels were subtracted,
and small corrections applied to account for beam losses at the air-quartz
boundaries, the reflectivity was calculated as the ratio~$I^{}_1/I^{}_0$. 
For each mirror, the laser was translated across the mirror and $I^{}_0$ 
and $I^{}_1$ were recorded at several positions. The mean value of $I^{}_1/I^{}_0$ 
was taken as the reflectivity of the mirror.

\begin{figure}[bth]
\centering
\vbox{
\vskip0.30in
\includegraphics[width=0.6\textwidth]{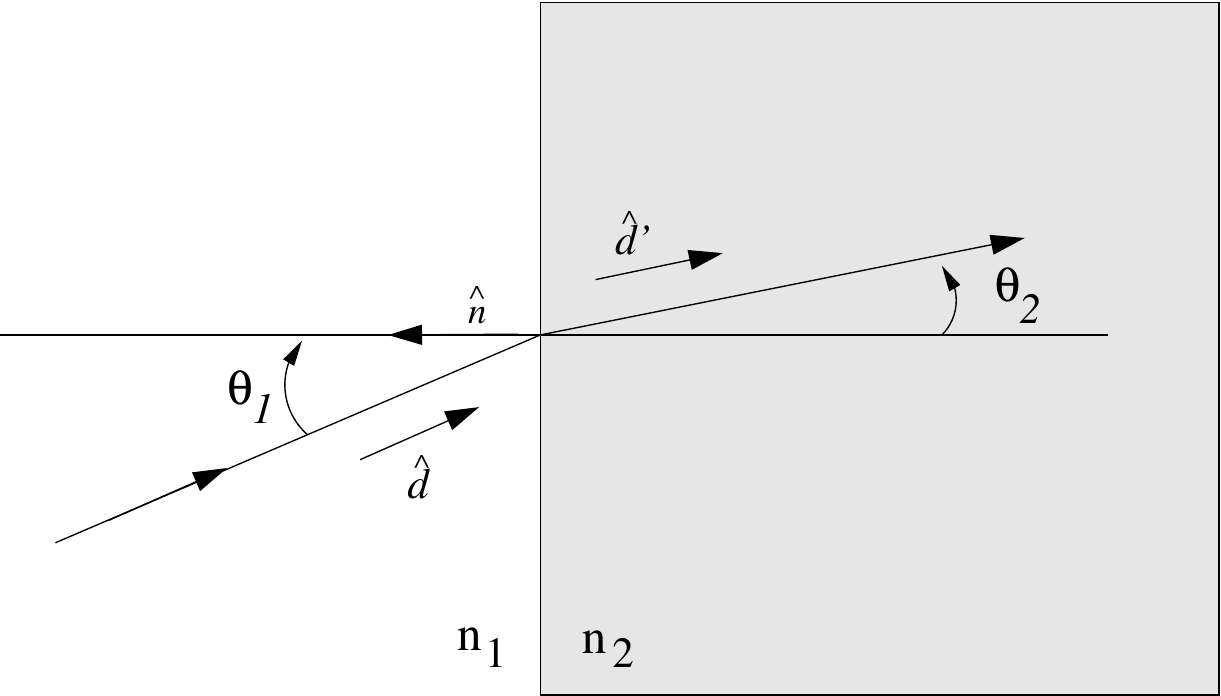}
}
\vskip0.10in
\caption{Refraction of a laser beam when entering (or exiting) the mirror substrate.
The initial direction is denoted by $\hat{d}$, and the final direction by $\hat{d}\,'$. }
\label{fig:mirror_test_refraction}
\end{figure}

The radius of curvature ($R$) was measured in three steps. In the first step, 
the laser was translated across the mirror and, at several fixed positions, the 
directions of the incoming and reflected beams were measured. These directions 
were corrected for the change due to refraction into and out of the mirror 
substrate, as shown in Fig.~\ref{fig:mirror_test_refraction}. 
Denoting the direction of light before refraction by the unit vector
$\hat{d}$, the unit direction after refraction $\hat{d}\,'$ is obtained
by Snell's law:
\begin{eqnarray}
\hat{d}\,' & = & \frac{1}{\rho}
\left[ \hat{d} - (\hat{d}\cdot\hat{n})\,\hat{n} - 
\sqrt{\rho^2 - 1 +(\hat{d}\cdot\hat{n})^2}\,\,\hat{n}\right] ,
\label{eqn:refraction}
\end{eqnarray}
where $\hat{n}$ is a unit vector normal to the surface through which the light refracts,
and $\rho = n^{}_2/n^{}_1$ is the ratio of final to initial indices of refraction. For our 
measurements, the incoming beam was adjusted to be normal to the surface of the substrate,
as shown in Fig.~\ref{fig:mirror_test_setup}. Thus, $\hat{d}_{\rm in} = -\hat{n}$, 
$\hat{d}\,'_{\rm in} = \hat{d}^{}_{\rm in}$, and Eq.~(\ref{eqn:refraction}) was used 
only to correct the direction $\hat{d}_{\rm out}$ of the reflected beam. 

In the second step, the direction vectors before and after reflection were used 
to calculate the unit vector normal to the reflecting surface at the point of reflection:
\begin{eqnarray}
\hat{n}^{}_r & = & \frac{\hat{d}\,'_{\rm out} - \hat{d}\,'_{\rm in}} 
{\bigl| \hat{d}\,'_{\rm out} - \hat{d}\,'_{\rm in}\bigr| }\,.
\label{eqn:refl_normal}
\end{eqnarray}
Neglecting spherical aberration, which is a small effect and negligible near the optical axis, 
all such vectors $\hat{n}^{}_r$ point to the center of curvature.

\begin{figure}[bth]
\centering
\includegraphics[width=0.38\textwidth,angle=90]{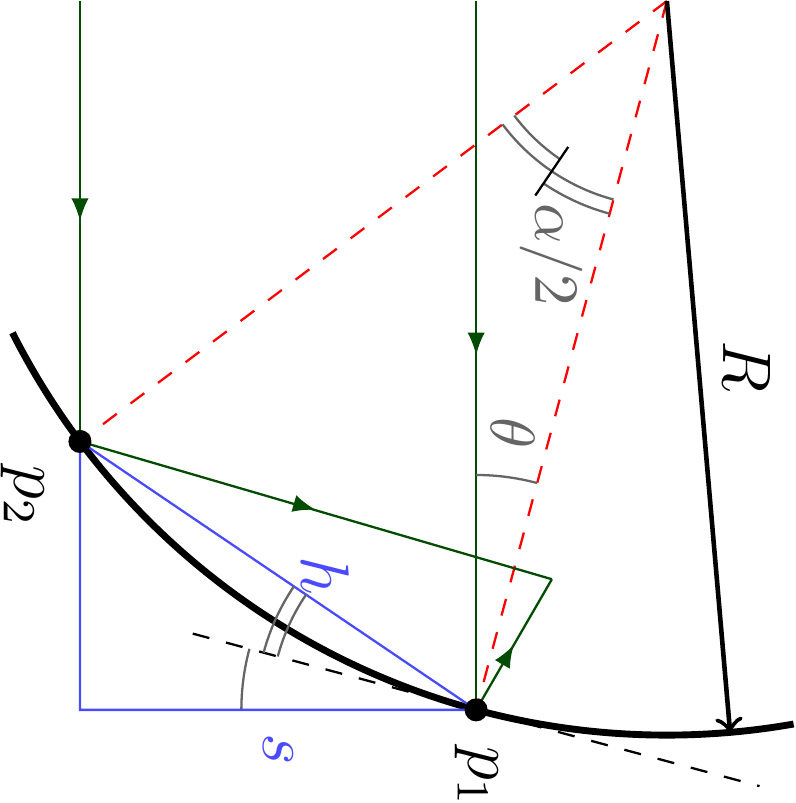}
\caption{Geometry for calculating the radius of curvature of a spherical mirror (see text).}
\label{fig:mirror_test_geometry}
\end{figure}

\begin{figure}[tb]
\begin{center}
\includegraphics[width=0.42\textwidth,angle=-90]{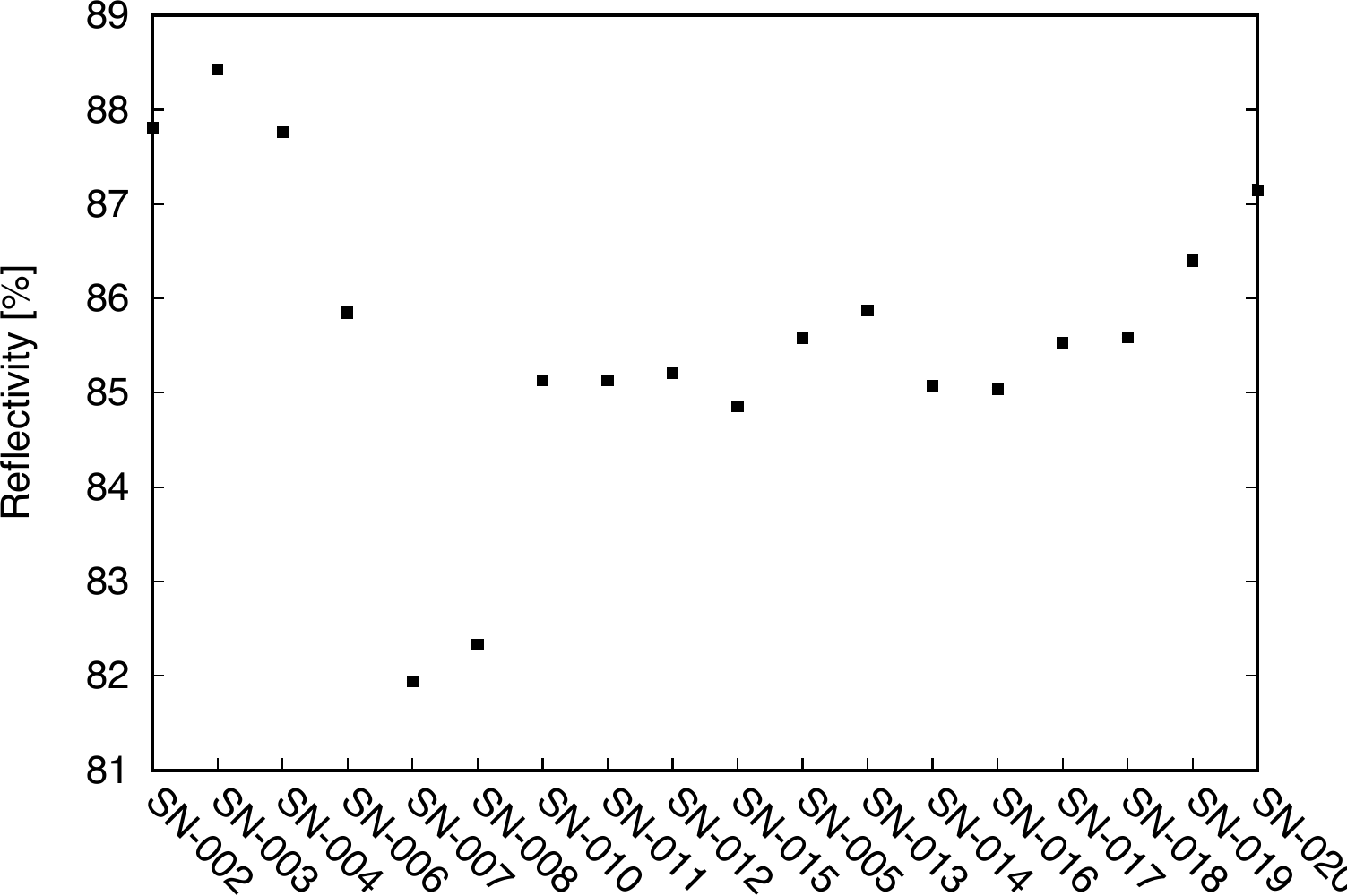}
\vskip0.20in
\includegraphics[width=0.42\textwidth,angle=-90]{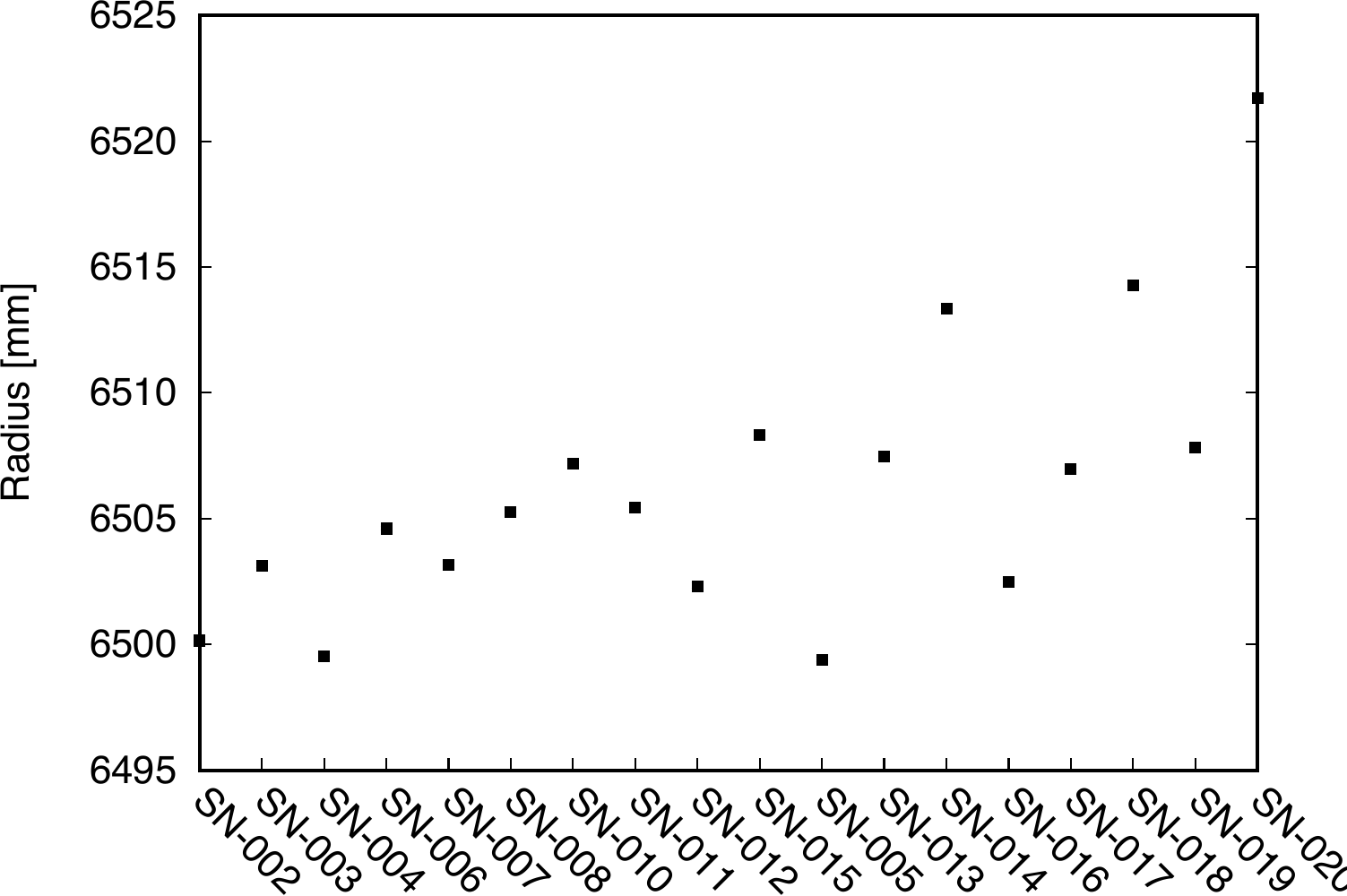}
\end{center}
\caption{Measurements of reflectivity~(top) and radius of curvature~(bottom) 
for the 18 spherical mirrors fabricated.} 
\label{fig:mirror_results}
\end{figure}

In the third step, the radius of curvature ($R$) was calculated from pairs of 
$\hat{n}^{}_r$ measurements as shown in Fig.~\ref{fig:mirror_test_geometry}.
From this figure, we obtain the relations
\begin{eqnarray}
h^2 & = & 2R^2 - 2R^2\cos\alpha \ \Rightarrow\ h = 2R\sin\frac{\alpha}{2} \\
h & = & \frac{s}{\cos (\theta + \alpha/2)} \,,
\end{eqnarray}
where $\alpha$ is the angle subtended by the two reflected beams, 
$\theta$ is the angle of incidence of the innermost beam 
(angle between $\hat{d}\,'_1$ and $\hat{n}^{}_{r1}$),
and $s$ is the separation between the two incoming beams. 
Eliminating $h$ yields
\begin{eqnarray}
R & = & \frac{s}{2\sin(\alpha/2)\cos(\theta + \alpha/2)}\,,
\label{eqn:R}
\end{eqnarray}
where $s$ is known and angles $\alpha$ and $\theta$ are calculated via
\begin{eqnarray}
\alpha & = & \cos^{-1} (\hat{n}^{}_{r1}\cdot \hat{n}^{}_{r2}) \label{eqn:alpha} \\
\theta & = & \cos^{-1} (-\hat{d}^{}_1\cdot \hat{n}^{}_{r1})\,. \label{eqn:theta}
\end{eqnarray}
Inserting these values into Eq.~(\ref{eqn:R}) yielded the radius of curvature~$R$.
The measurement was performed at several positions along the length of the mirror.
At each position, the directions of the incoming and reflected beams 
were measured, and the reflection vector $\hat{n}^{}_r$ calculated 
using Eq.~(\ref{eqn:refl_normal}). 
The resulting set of $\hat{n}^{}_r$ vectors were combined pair-wise to
calculate $\alpha$, $\theta$, and $R$ using Eqs.~(\ref{eqn:alpha}), 
(\ref{eqn:theta}), and (\ref{eqn:R}). There were typically 6--8  
separate measurements of $R$; the mean value was taken as the 
final radius of curvature. The r.m.s.~(spread) of the values 
gave a measure of the mirror's spherical aberration.
All results for mirror reflectivity and the radius of curvature are listed 
in Table~\ref{tab:modules} and plotted in Fig.~\ref{fig:mirror_results}. 
The measured radii are consistent with the specification of $(6500\pm 100)$~mm.

\subsection{Gluing PEEK frame to quartz prism}
\label{sec:optics:peek_frame}

After a prism underwent QA measurements and was graded satisfactory, a thin 
frame fabricated from PEEK material was glued around the outer edge near the prism
face to which the PMTs were mounted. The edge of the frame was aligned
flush with this face. This frame sits in a channel in the QBB, thus positioning 
the prism (and attached optics) and preventing any movement of optics 
within the QBB. The frame also provides a gas-tight and light-tight seal for
the prism enclosure. The gas seal prevents purified nitrogen gas circulating
around the bars from reaching the electronics, and the light seal prevents 
spurious light in the QBB {\it not\/} originating from \cherenkov\ radiation 
within the bars from reaching the PMTs.

\begin{figure}[bth]
\hbox{
\includegraphics[width=4.4in]{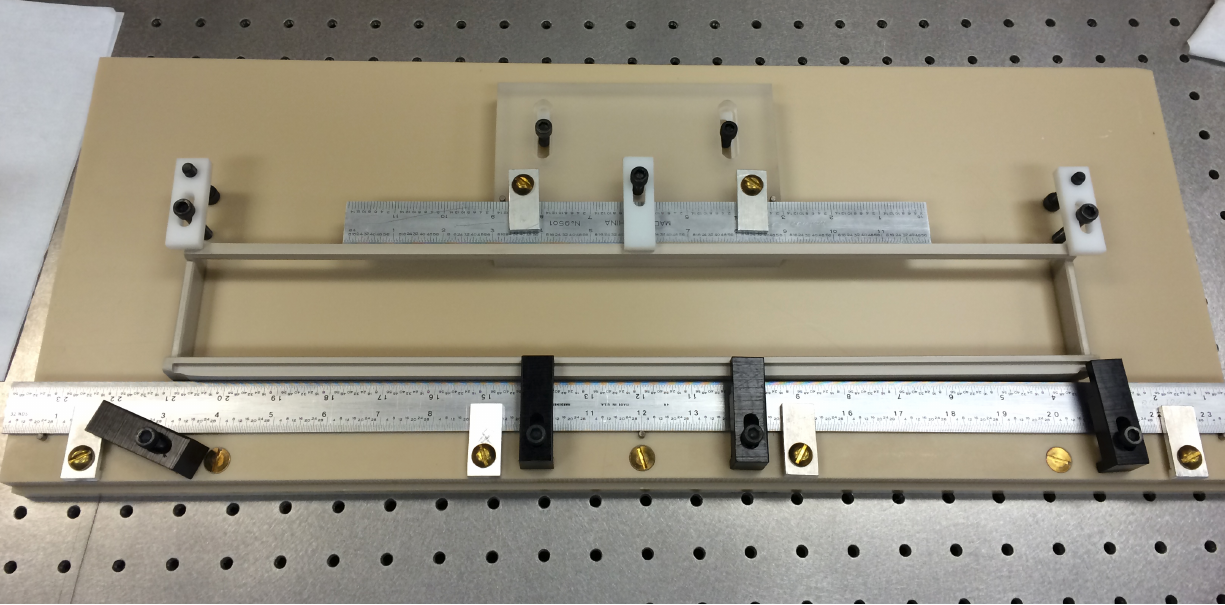}
\hspace*{0.20in}
\includegraphics[width=1.8in]{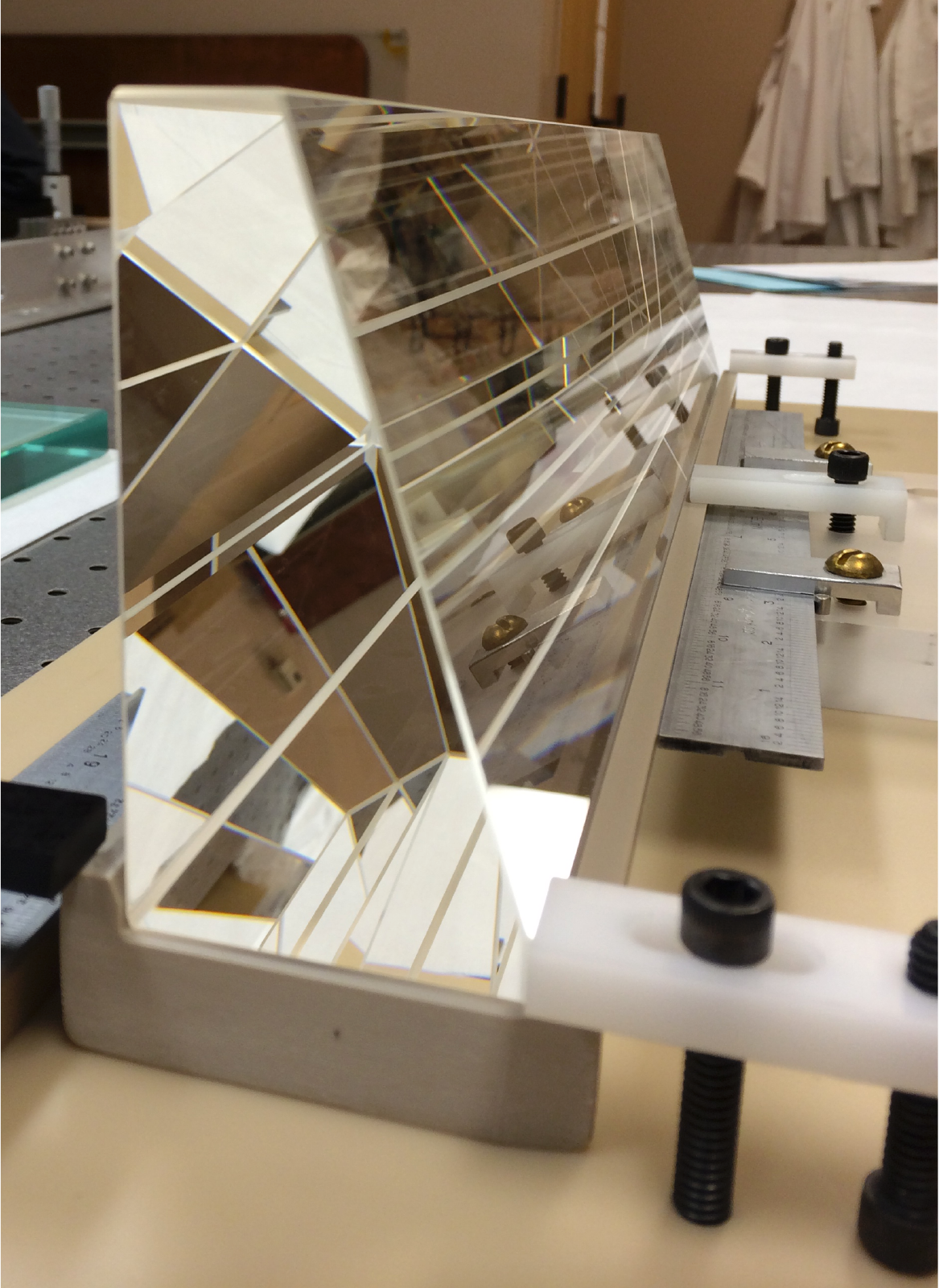}}
\caption{A PEEK frame afixed within the gluing jig~(left), and 
the same frame after a prism was glued~(right). }
\label{fig:peek-frame-gluing}
\end{figure}

The PEEK frames were fabricated at Pacific Northwest National Laboratory (PNNL) and
glued onto the prisms at the University of Cincinnati. At Cincinnati, each frame
was machined further to achieve a uniform gap of 100-200~$\mu$m between the frame 
and the prism surface it surrounds. After machining, the frame was cleaned with 
methanol and afixed within a custom-made jig. A prism was positioned within the 
frame and aligned, and the gap between frame and prism checked with shims to
ensure the requisite 100-200~$\mu$m of clearance. After alignment, the prism 
and frame were dusted with purified nitrogen and epoxied together with 
Masterbond EP21LV-LO~\cite{MasterbondEP21LV}. This epoxy has good shear 
strength and radiation hardness, 
and a small coefficient of thermal expansion. Its two components were mixed, 
centrifuged to remove air bubbles, and then dispensed with a syringe into the 
gap between prism and frame. The glue joint was inspected to check that no air 
bubbles were trapped within the epoxy.
A period of 48--72 hours was required for the epoxy to fully cure.
A typical PEEK frame afixed within the gluing jig is shown in 
Fig.~\ref{fig:peek-frame-gluing} on the left, and the same frame with 
a prism glued is shown in Fig.~\ref{fig:peek-frame-gluing} on the right.

\subsection{Alignment of optical components and gluing}

After all QA was completed and prisms and mirrors transported to KEK, 
the optical components were mounted on an optical table, aligned, and 
glued together to make an optics module. Each module consisted of one prism, 
two bars, and one mirror, as shown in Fig.~\ref{fig:itop}~(left). 
There were three separate joints to be glued: between the first bar 
and prism, between the two bars, and between the second bar and mirror. 
For each joint, the gluing procedure was similar.

\begin{figure}[tb]
\begin{center}
\includegraphics[width=0.75\textwidth]{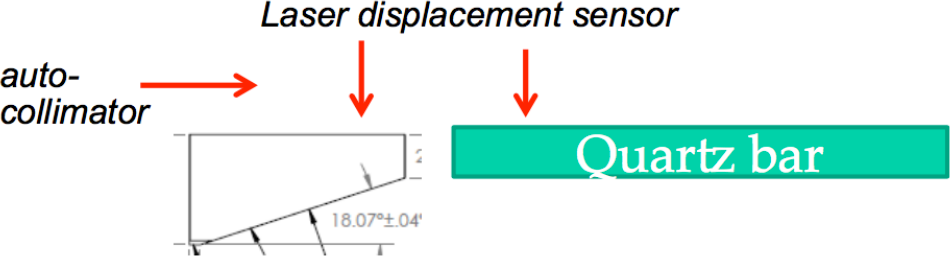}
\end{center}
\caption{Procedure to align a prism and a bar before gluing.}
\label{fig:align}
\end{figure}

The components were mounted on custom-made platforms that allowed precise
translations in the $x$, $y$, and $z$ directions, and whose tilt angles around 
these axes could be adjusted. All adjustments were made using built-in 
micrometers. The alignment procedure for a glue joint is illustrated in 
Figure~\ref{fig:align}. First, the top surfaces were aligned to the same 
height to within $\pm 20~\mu$m using a laser displacement sensor (LDS). 
The side surfaces were then aligned to within $\pm 30~\mu$m using the LDS.
Following this, the tilt angles of the top surfaces around the horizontal 
$x$ and $y$ axes were adjusted to match to within $\pm 40$~arcseconds;
this was achieved using an electronic autocollimator and a mirror placed 
successively on each top surface. Finally, the angles of the side surfaces 
were adjusted to match to within $\pm 20$~arcseconds using the autocollimator. 
As adjusting the angles often disturbed the previous surface alignments
attained with the LDS, the entire procedure was typically repeated 2--3 
times in order to have all surfaces satisfactorily aligned within 
specifications. 

\begin{figure}[htb]
\begin{center}
\hspace*{-0.40in}
\includegraphics[width=4.2in,angle=90.]{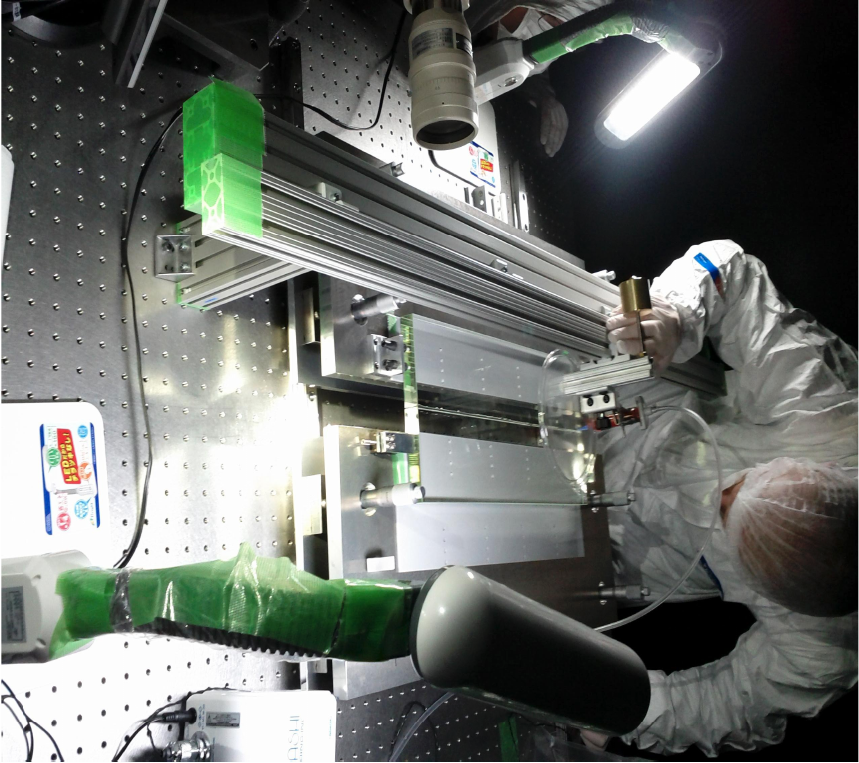}
\end{center}
\caption{For gluing, the syringe dispensing glue is mounted on a cart that
rolls on a rail aligned parallel to the glue joint. }
\label{fig:boqun_gluing}
\end{figure}

\begin{figure}[htb]
\begin{center}
\includegraphics[width=3.5in,angle=-90.]{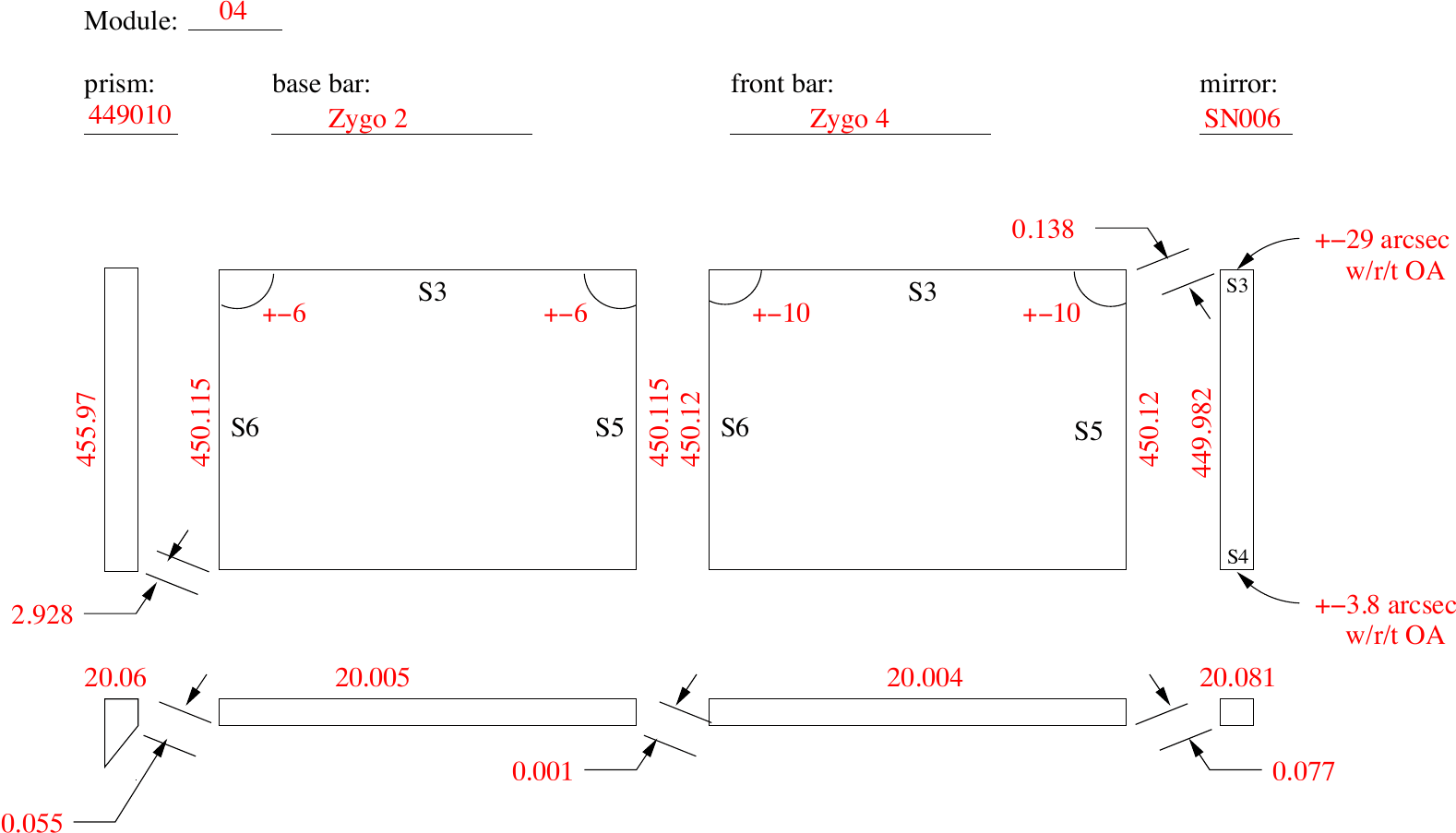}
\end{center}
\caption{Alignment measurements taken after epoxy curing, using a laser 
displacement sensor and autocollimator. The units are mm and arcseconds.}
\label{fig:alignment_module4}
\end{figure}

After alignment, the two components to be glued were translated towards
each other until there was a narrow gap of 50--100~$\mu$m between them. 
Teflon tape was applied to the bottom and sides of this gap, such 
that epoxy dispensed into the gap would not leak out. After taping, 
the two-component epoxy was mixed, centrifuged to remove air bubbles, and 
dispensed into the gap using a Nordson EFD fluid dispenser and syringe~\cite{EFD}.
The syringe was mounted on a movable cart that rolled on a rail aligned 
parallel to the glue joint, as shown in Fig.~\ref{fig:boqun_gluing}. The 
epoxy used was EPOTEK 301-2~\cite{Epotek301-2}, which has high transmission in the 
relevant range of wavelengths (340--600~nm), high strength, and good radiation 
resistance. This epoxy required 3--4 days to fully cure. After curing, the tape
was removed and any excess epoxy cleaned off using purified acetone. 
As the alignment sometimes changed slightly during the curing process, 
all surface heights and angles were re-measured after curing and documented. 
A typical ``alignment report'' is shown in Fig.~\ref{fig:alignment_module4}.

\subsection{Moving an optics module into a QBB}

\begin{figure}[tb]
\begin{center}
\includegraphics[width=0.50\textwidth,angle=90.]{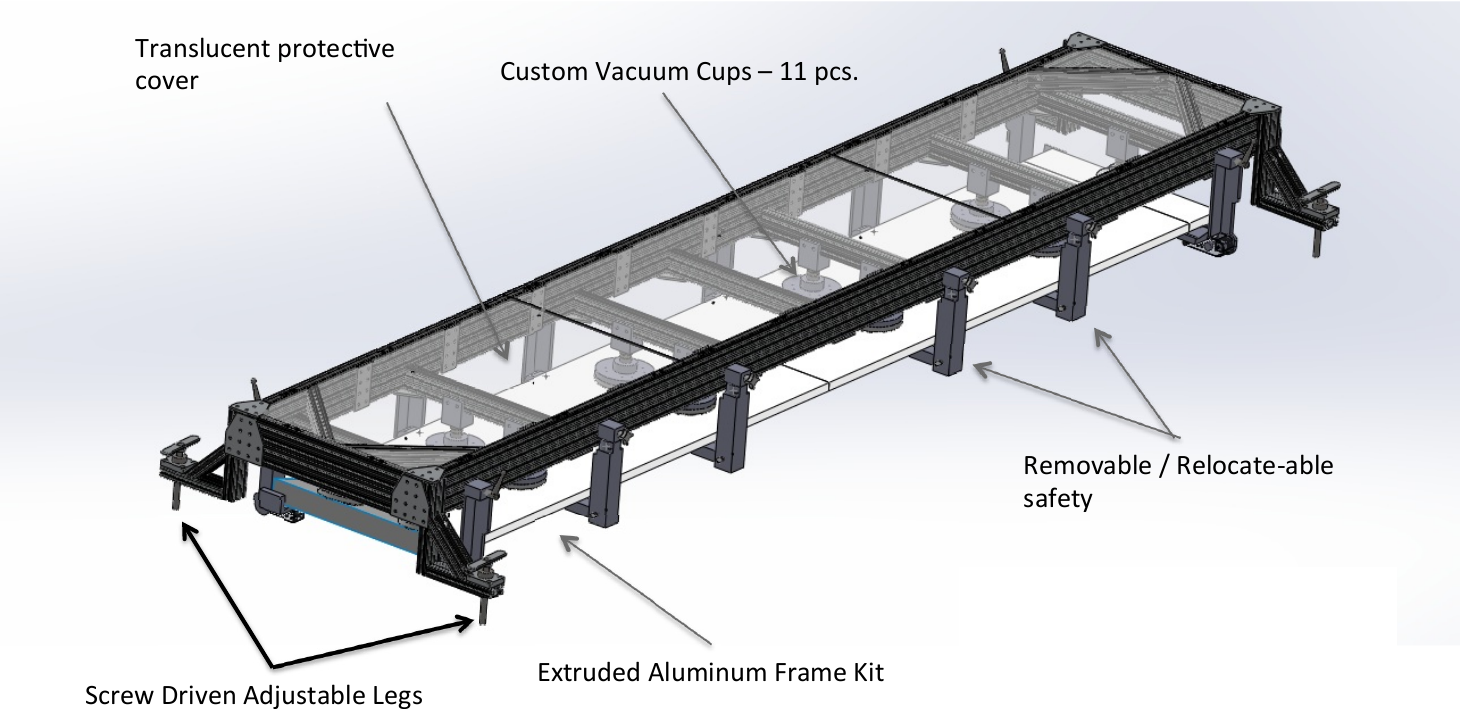}
\end{center}
\caption{Drawing of the lifting jig used to lift an optics module and 
place it into a QBB. The jig was suspended from a pulley system that 
rolled on wheels (not shown).}
\label{fig:lifting_jig}
\end{figure}

After two bars, a prism, and a mirror were glued together, the resulting
optics module was placed inside a QBB. This required lifting and moving the 
2.7~m-long module from the optical table (upon which it was glued) to
a QBB positioned on an adjacent granite table. To perform this movement, 
a specially designed lifting jig was used. A drawing of the lifting jig is 
shown in Fig.~\ref{fig:lifting_jig}.
The jig used 11 suction cups operated under vacuum, and a pulley system
that rolled on wheels, to lift the module, translate it approximately 3~m, 
and lower it into a QBB. The suction cups were positioned such that the optics' 
glue joints were subjected to minimum stresses. 
Protective tape was applied to the polished quartz surface beneath each suction 
cup to prevent scratches to the surface. After the module was lowered into the QBB, 
the vacuum was released and the lifting jig moved away. The tape was then removed 
and the quartz surface cleaned with purified acetone. 

%% file: barbox.tex
\section{Quartz bar box}
\label{sec:barbox}

\subsection{Design}
\label{sec:barbox_design}

\begin{figure}
\centering
\includegraphics[width=15cm]{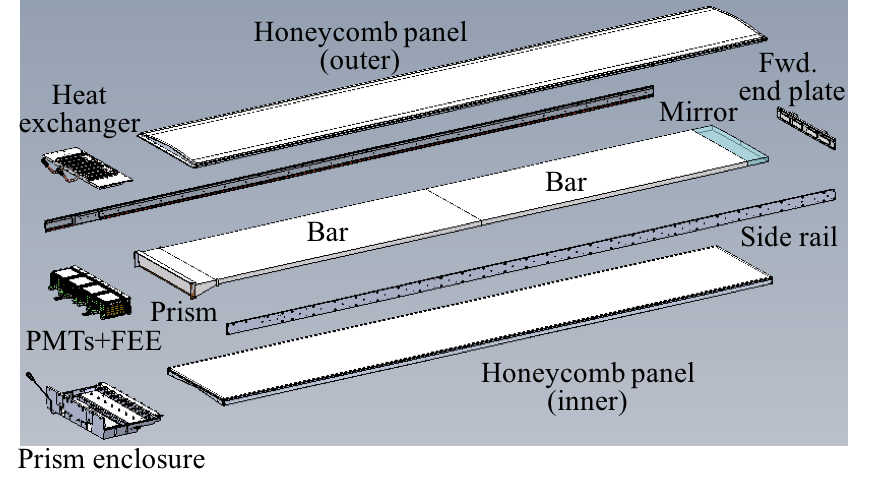}
\caption{Schematic view of the components of a quartz bar box (QBB).}
\label{fig:mechanics:components}
\end{figure}

An optics module is supported inside a quartz bar box (QBB). The components of a QBB 
are shown in Fig.~\ref{fig:mechanics:components} and consist of the following:
\begin{enumerate}
\item aluminum honeycomb panels that run the length of the bars\,+\,mirror;
\item a box-like enclosure machined from a block of aluminum that 
surrounds and supports the prism; this is referred to as the ``prism enclosure.''
\item aluminum plates, referred to as ``side rails,'' that run the length of the QBB; and
\item a forward endplate that covers the end of the spherical mirror.
\end{enumerate}
Four spring-loaded blocks fabricated from PEEK were mounted to the forward endplate such
that they press against the backside of the mirror; this prevents horizontal movement of the 
module within the QBB. The prism enclosure was glued to the inner honeycomb panel.
The enclosure has an access hatch on its inner side to allow for replacing 
front-end electronics and photomultiplier tubes (PMTs) in situ. On the outer side 
of the prism enclosure, a heat exchanger with copper tubing was mounted to cool 
the front-end electronics via water circulation.

To support the optics module, PMTs, and front-end electronics with
minimal movement, the QBB must have high rigidity. However, 
low mass is desired to minimize particle interactions. Thus, aluminum honeycomb panels 
were used for the large-area sides through which charged particles pass. The panel shape 
was curved to increase rigidity, with the curvature matching the cylindrical 
arrangement of the modules. The two side rails run the length of the QBB and
were fabricated from aluminum. They have a thickness of only 6~mm but contribute
significantly to the rigidity. As the detector is located within a strong magnetic 
field, all components (including screws, washers, etc.) are nonmagnetic.

The QBB's were positioned within the Belle II detector cylindrically around the beampipe and
supported by flanges at the forward and backward ends of the central (barrel) region. 
To prevent sagging between the flanges, each QBB was attached to adjacent QBB's by
bolting together the side rails; this resulted in a ``roman arch'' structure, 
greatly increasing the rigidity. A finite element analysis of this structure 
with a simple flat-panel design predicted a maximum sag of only 0.1~mm, and
the addition of curvature to the honeycomb panels increased the rigidity 
even further (by a factor of~2.7).

\begin{figure}
\centering
\includegraphics[width=9cm]{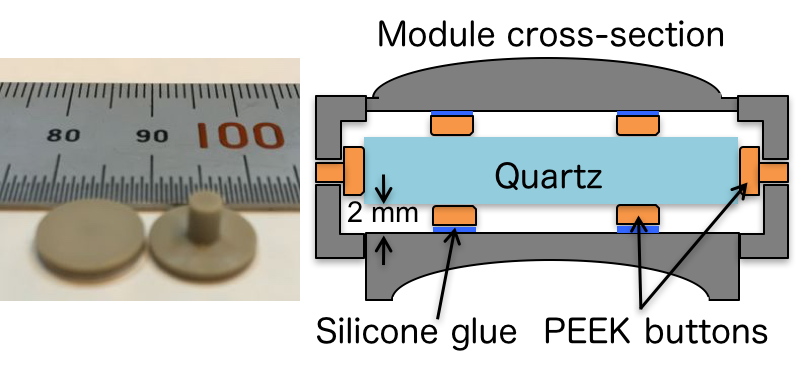}
\caption{PEEK buttons~(left), and a schematic view of buttons inside a QBB~(right).}
\label{fig:mechanics:buttons}
\end{figure}

The optics module is supported on all faces by small buttons fabricated from PEEK. The
buttons have a diameter of 10~mm and a thickness of 1.5--1.9~mm. They were glued to the
top and bottom honeycomb panels, and also to the side rails, in order to maintain a 
uniform gap of 2~mm around the optics as shown in Fig.~\ref{fig:mechanics:buttons}. 
This gap is filled with purified, circulating nitrogen gas to ensure that the 
bar surfaces remain clean and dry; this maximizes the total internal reflection 
of \cherenkov\ light radiated within the bars. There were 
14 buttons glued to each honeycomb panel in a $2\times7$ 
configuration, and five buttons glued to each side rail in a line. 
The height alignment of buttons is important for uniformly supporting 
the module without causing excessive stress or deformation. 
Thus, the heights of the panel buttons were adjusted using shims 
and gluing jigs to match the alignment of the optics after gluing.
The button heights were measured using a precision height gauge 
and adjusted as necessary. The standard deviation ($\sigma$) of 
the distribution of button heights for a panel was required to be 
less than 0.02~mm. Typically, $\sigma < 0.01$~mm was achieved.

\begin{figure}
\centering
\includegraphics[width=14cm]{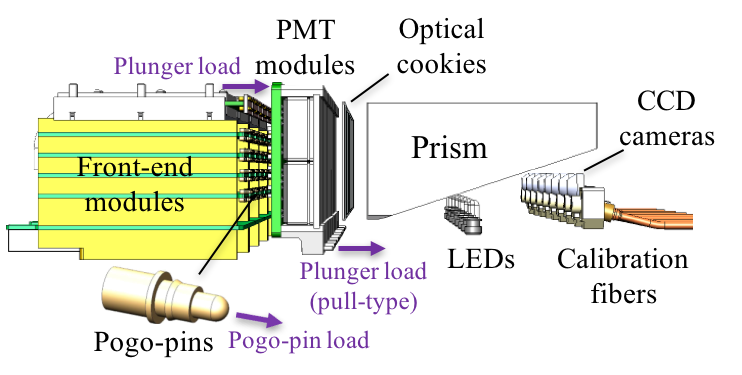}
\caption{Components mounted within the front-end region of a QBB (see text).}
\label{fig:mechanics:enclosure}
\end{figure}

The prism enclosure was glued to the inner honeycomb panel and fixed in position by 
the side rails. The prism itself is supported by PEEK inserts mounted inside the 
prism enclosure using silicone glue. The forward edges of the prism are also supported 
by PEEK inserts. The prism enclosure, side rails, and heat exchanger form the front-end 
region. Within this region, eight ``PMT modules'' and four front-end electronics modules 
were mounted. A schematic view of these components as arranged within the front-end region 
is shown in Figure~\ref{fig:mechanics:enclosure}.

Each PMT module consists of four MCP-PMTs arranged in 
a $2\times2$ configuration; see Section~\ref{subsec:pmt_module}.
The modules were mounted to the wide face of the prism using 
silicone optical ``cookies'' and optical oil. They are pressed against 
the prism with spring-loaded plungers in order to maintain good optical coupling. 
Each PMT module connects 
electrically to front-end electronics via spring-loaded pogo pins; 
this scheme allows for either the PMT module or the front-end electronics module to 
be separately replaced. On the inner wall of the prism enclosure, 
optical fiber heads~\cite{fiber_heads}, 
CCD cameras~\cite{CCD_cameras}, and LEDs were mounted. The CCD cameras and LEDs were 
used to inspect the quality of the optical coupling between the prism and PMT modules.
The optical fibers are used for calibration when the experiment is running.
This is achieved by firing single-photon laser pulses that enter the prism through 
the tilted face, internally reflect off the top surface, and strike the MCP-PMTs.

\subsection{Production}

\begin{figure}
\centering
\vbox{
\includegraphics[width=11cm]{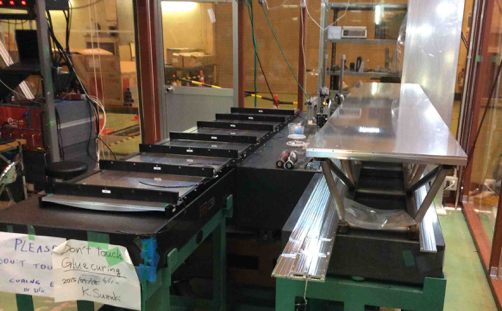}
\vskip0.20in
\includegraphics[width=11cm]{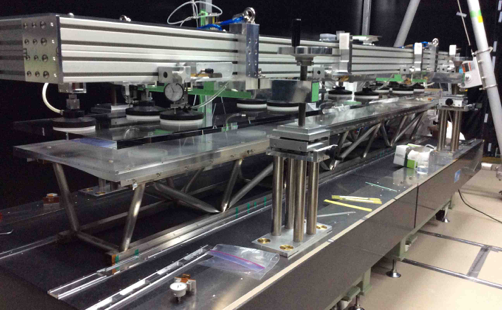}
}
\vspace*{0.20in}
\caption{Assembling a QBB with a strongback attached underneath~(top), 
and using the lifting jig to lower an optics module into a QBB~(bottom). }
\label{fig:mechanics:assembly}
\end{figure}

All QBB parts were cleaned with ethanol and acetone and assembled on a flat granite 
table to ensure that the honeycomb panels and siderails align to a precision of 
0.1~mm or less. During QBB assembly, 
an optics module was placed on the lower honeycomb panel before the side rails 
and top honeycomb panel were added. At this point, the rigidity of the lower panel 
was insufficient to support the optics within the acceptable deformation limit. 
Thus, the first step of QBB assembly was to attach a truss-frame support structure 
called a ``strongback'' to the lower honeycomb panel as shown in 
Fig.~\ref{fig:mechanics:assembly}~(left), in order to keep this panel from flexing. 
The strongback was removed only after a QBB was fully assembled, installed 
in the Belle II detector, and had its siderails attached to adjacent QBBs. 
After the strongback was attached, the PEEK buttons were glued and their heights 
checked and adjusted as necessary. 
The prism enclosure, along with fiber connectors, CCD cameras, and LEDs, was then glued 
to the inner honeycomb panel. The side rails were attached, and the structure was moved 
into the clean room where the optics module was added.

To add the optics module, the siderails were removed and, using the lifting jig 
described previously, the optics module was placed onto the PEEK buttons of the 
lower honeycomb panel [see Fig.~\ref{fig:mechanics:assembly}~(right)].
The side rails were re-attached, and the front end plate and top honeycomb panel were attached. 
All seams were sealed with a silicone sealant to ensure both a gas seal and light seal.
A gas leak check was performed using grade-1 pure Argon gas and a leak detector. The sealing work
was completed when all leaks were sealed and the output gas flow rate equaled the input rate.
After this assembly, the QBB was put on a cart and moved outside the clean room, where it 
was again checked for gas tightness. Nitrogen gas from a liquid nitrogen evaporator was flowed 
through the QBB, and the dew point of the output gas measured. If the QBB was gas tight, 
then the dew point should decrease by about $-40^\circ$~C at a flow rate of 0.20~liters/min.
Each QBB was checked that this requirement was met and, if not, additional sealing was 
performed. After all gas sealing work was completed, the PMT modules and front-end electronics 
were attached.

The finished assembly, referred to as a detector module, was then transported to the 
Belle II experiment using a custom-designed pallet with a wire suspension (to minimize vibrations) 
and a special truck with an air suspension. The distance from the assembly site (Fuji Hall) 
to the Belle II experiment site (Tsukuba Hall) was~1.5~km.

\subsection{Installation in the Belle II detector}

\begin{figure}
\centering
\includegraphics[height=4.2cm]{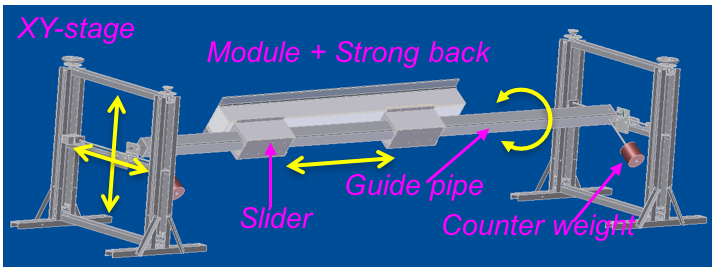}
\hspace*{0.10in}
\includegraphics[height=5.5cm]{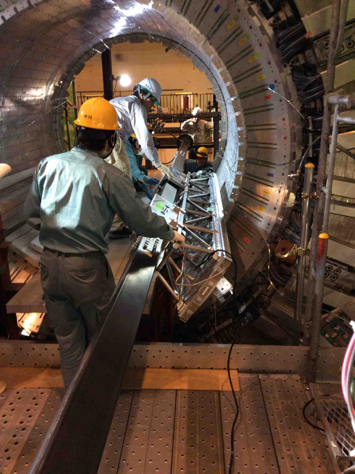}
\caption{Schematic view of the installation jig~(left), and installation of a 
detector module into Belle~II~(right).}
\label{fig:mechanics:installation}
\end{figure}

The detector module was installed using a specialized jig consisting 
of mounting plates that slid on a guide pipe, as shown in 
Fig.~\ref{fig:mechanics:installation}~(left). These plates are
referred to as ``sliders.'' The guide pipe had a square cross section, 
a length of 8~m, and spanned the central region of the Belle~II detector. 
On the guide pipe were mounted $x$-$y$ stages that translated the pipe ends in 
two orthogonal directions. The guide pipe could also rotate azimuthally.
A detector module was mounted on the sliders via the strongback.
The strongback was supported at the Bessel points, where the sag with 
two-point supports is minimized.
Each module was then slid into the Belle II barrel, positioned close to the
outer cylinder using the $x$-$y$ stages, and fixed at the two ends of 
the module to flanges on the Belle~II structure.

\begin{figure}
\centering
\includegraphics[width=10cm]{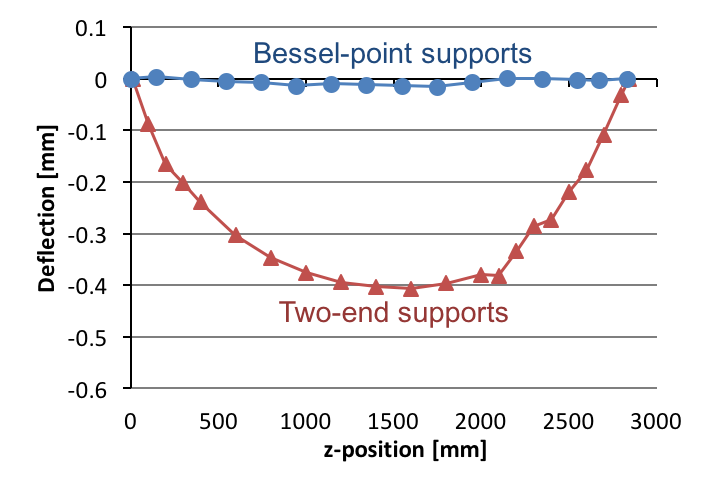}
\caption{Deflection of a prototype detector module for two cases: 
support at the two ends (red triangles), and support at the 
Bessel points (blue circles). The horizontal axis lists the 
position along the module from the forward end.}
\label{fig:mechanics:sagsupports}
\end{figure}

During installation, module deflection was monitored 
using a strain gauge fixed to a side rail and several dial gauges.
Measurements performed with a prototype detector module 
supported at the Bessel points indicated a maximum sag of 
only 0.02~mm -- see Fig.~\ref{fig:mechanics:sagsupports}.
However, after a module's ends were fixed to the flanges and 
the installation jig removed, but before the side rails were 
linked to adjacent modules, this sag was measured to be 
about 0.4~mm. This larger value is still within mechanical tolerances. 

\begin{figure}
\centering
\includegraphics[width=16cm]{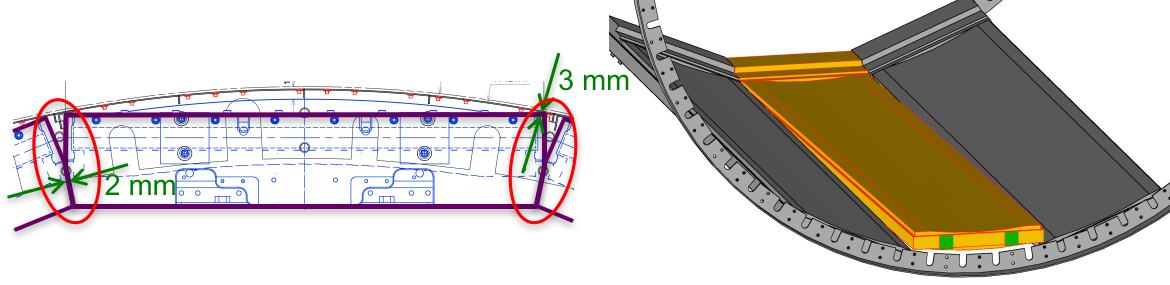}
\caption{Schematic view of the clearance around the prism enclosure
after module installation in Belle~II~(left), and a conceptual drawing
of the installation of the last module via tilting~(right).
}
\label{fig:mechanics:lastmodule}
\end{figure}

\begin{figure}
\centering
\includegraphics[width=12cm]{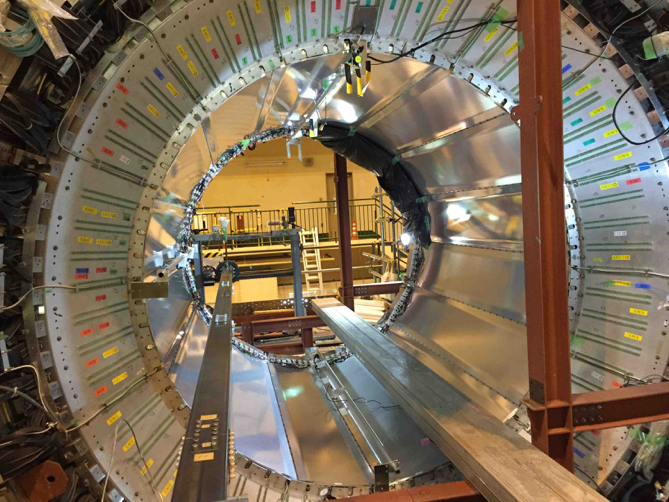}
\vskip0.20in
\caption{Installed detector modules as viewed from the forward end of 
the barrel region, after all strongbacks were removed.}
\label{fig:mechanics:installation2}
\end{figure}

Figure~\ref{fig:mechanics:lastmodule}~(left) shows the cross section of 
the prism enclosure region. The clearance between adjacent modules is 2~mm, 
and that between the edges of the prism enclosure and the barrel structure is 3~mm.
Because of the cylindrical arrangement of modules, the last module could not be moved 
into position by simply translating in the radial direction. Thus, this module 
required tilting during installation 
as shown in Fig.~\ref{fig:mechanics:lastmodule}~(right).
After installation, adjacent modules were connected to each other via 
their side rails using small bars and screws, and the strongbacks were
removed. Figure~\ref{fig:mechanics:installation2} shows all detector 
modules after installation was complete.

%% file: PMTs.tex
\section{Photomultiplier tubes}
\label{sec:pmts}

\subsection{Introduction}

To achieve the required single-photon timing resolution, 
we use micro-channel-plate photomultiplier tubes (MCP-PMTs) developed
at Nagoya University in collaboration with Hamamatsu Photonics K.K.~\cite{HamamatsuKK}.
An MCP-PMT is equipped with two micro-channel plates (MCPs) that function
as electron multipliers, utilizing pores of diameter 10~$\mu$m and
length 400~$\mu$m. The gain is typically $10^5$--$10^6$, and the pulse 
rise time is approximately 200~ps.

\subsection{Specification}

\begin{figure}[htb]
 \centering
 \includegraphics[height=4.0cm]{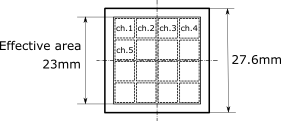}
\hskip0.40in
 \includegraphics[height=4.0cm]{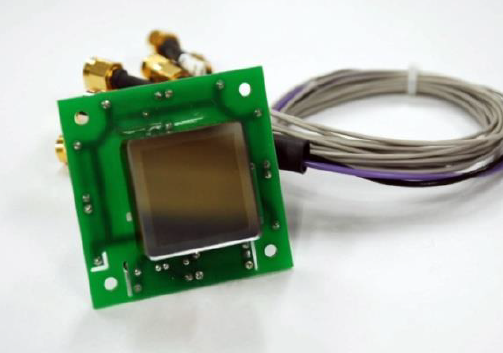}
 \caption{Schematic drawing of an MCP-PMT~(left), and one MCP-PMT 
  mounted to a test printed circuit board~(right).}
 \label{fig:MCP-PMT:drawings}
\end{figure}

The PMT has a square form factor and 16 individual anodes 
arranged in a $4\times 4$ array, as shown in Fig.~\ref{fig:MCP-PMT:drawings}.
The overall dimensions are $(27.6\!\times\!27.6)$~mm$^2$, with an active area 
of $(23\!\times\!23)$~mm$^2$. The technical specifications from Hamamatsu 
Photonics are given in Ref.~\cite{Hamamatsu:R10754-07-M16}.
The 16 modules of the TOP require 512 such MCP-PMTs.

\begin{figure}[htb]
 \centering
 \includegraphics[height=5.2cm]{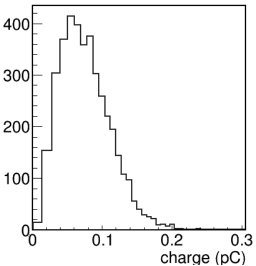}
 \includegraphics[height=5.2cm]{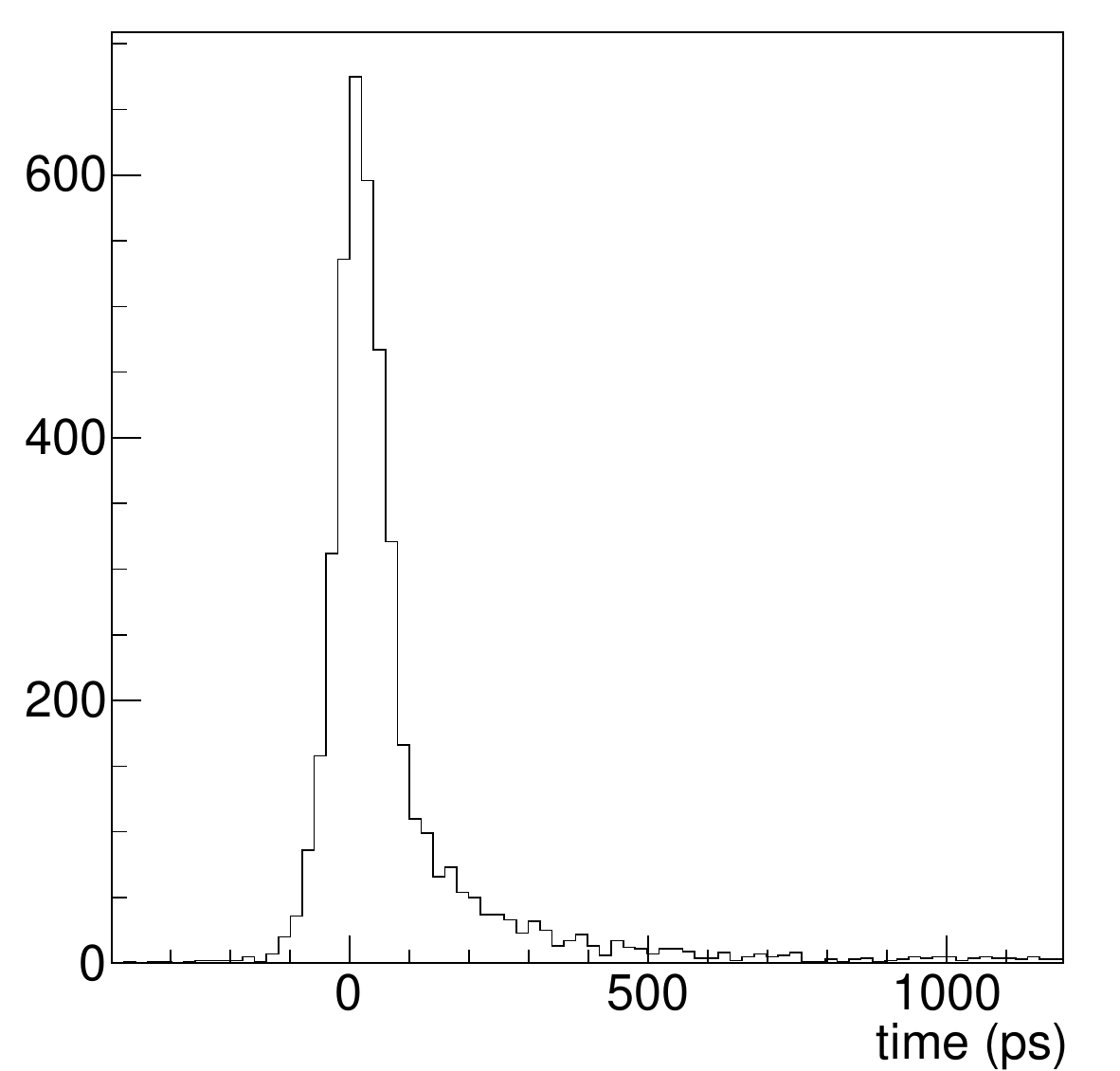}
 \includegraphics[height=5.2cm]{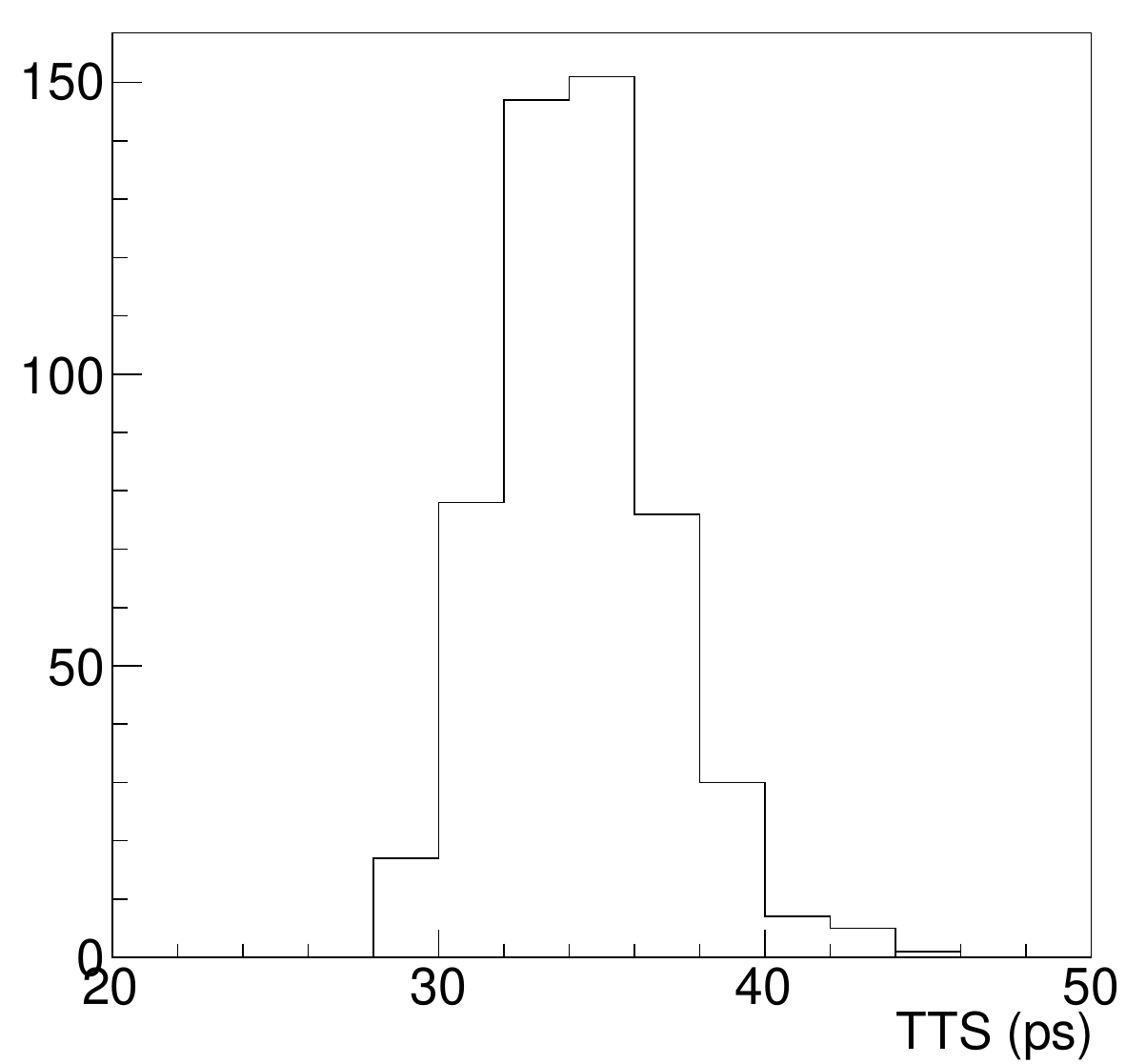}
\vskip0.10in
 \caption{Typical MCP-PMT output charge distribution~(left) and timing distribution~(middle),
and the TTS distribution~(right) for the 512 MCP-PMTs installed in Belle~II.}
 \label{fig:MCP-PMT:gainTTS}
\end{figure}

Typical distributions of output charge and timing for 
single photon detection are shown in Fig.~\ref{fig:MCP-PMT:gainTTS}.
These measurements are performed with an attenuated picosecond pulse laser.
The gain is defined as the mean of the output charge distribution divided 
by the elementary charge. 
The transit time spread (TTS) is defined as the standard 
deviation ($\sigma$) of the main Gaussian when the time 
distribution is fitted with a double Gaussian function.
The timing distribution exhibits a long tail,
which is due to electrons reflecting off the first MCP surface.
The typical TTS is 30--40~ps, as shown in Fig.~\ref{fig:MCP-PMT:gainTTS}~(right).

\begin{figure}[htb]
 \centering
 \includegraphics[height=5.8cm]{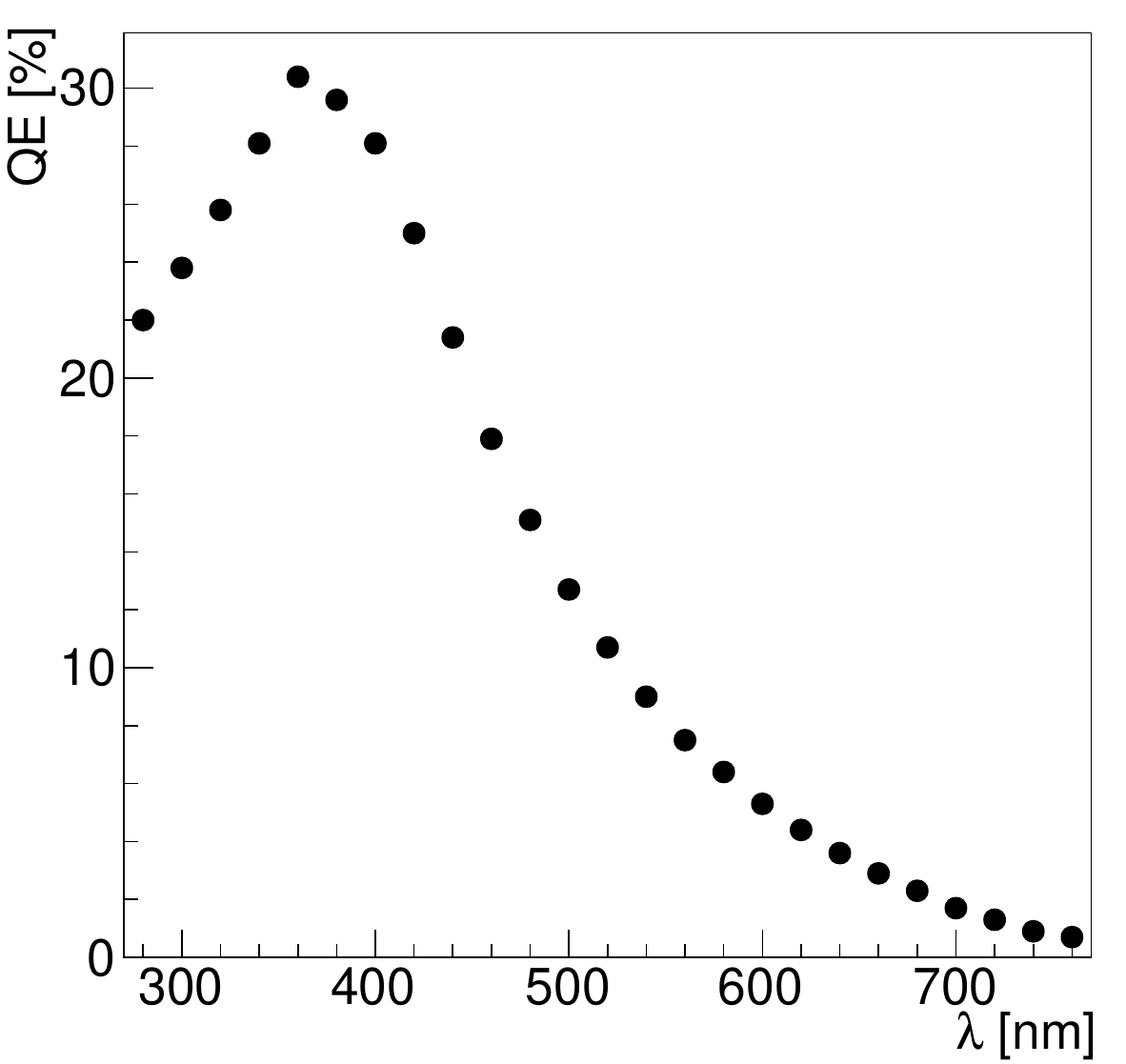}
\hskip0.20in
 \includegraphics[height=5.8cm]{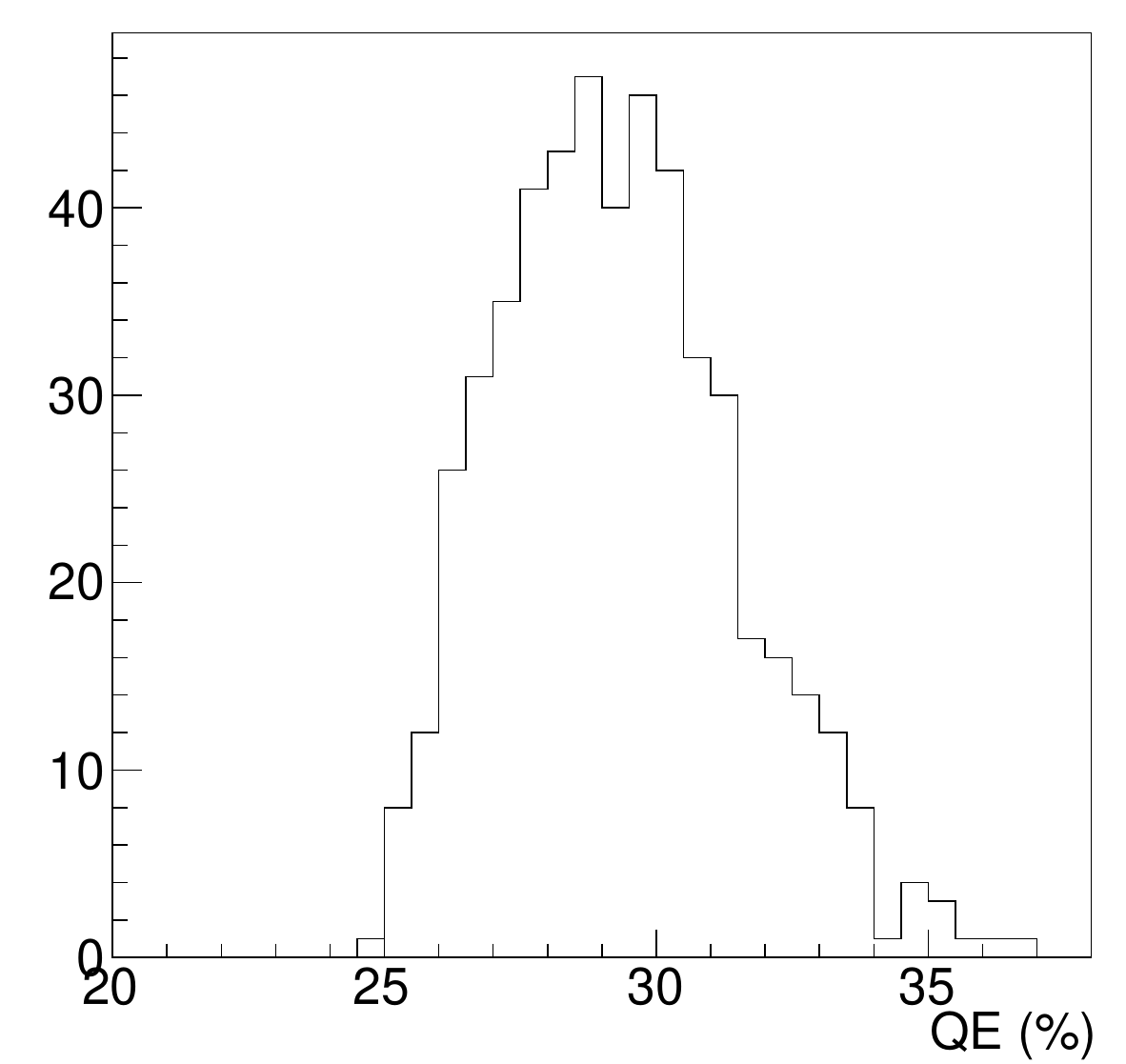}
\vskip0.10in
 \caption{MCP-PMT QE as a function of wavelength~(left);
 QE distribution at the peak wavelength $\lambda\!=\!360$~nm~(right)
 for the 512 MCP-PMTs installed in Belle~II. }
 \label{fig:MCP-PMT:QE}
\end{figure}

The photocathode consists of a multi-alkali substrate (NaKSbCs). 
A typical distribution of quantum efficiency (QE) as a function of 
wavelength is shown in Fig.~\ref{fig:MCP-PMT:QE}~(left).
To reduce ``ion feedback'' between the second MCP and the photocathode, 
which corresponds to ions knocked out from the MCP by secondary electrons 
that hit the photocathode and produce further charge multiplication,
a thin layer of aluminum is added to the surface of the second MCP. 
Aluminum is not added to the first MCP to avoid reducing collection 
efficiency~(CE). The typical CE is about~60\%, which is determined 
by the ratio of pore area to total area of an MCP.

\subsection{Lifetime improvements}
\label{sbsctn:pmt_lifetime}

The photocathode efficiency degrades with integrated charge, shortening
the effective lifetime of the MCP-PMT. The dependence of the lifetime
on integrated charge is approximately quadratic. This degradation is caused 
mainly by residual neutral gasses and ions released from the MCPs that 
hit the photocathode~\cite{Jinno:2010cj}.
In Belle~II, many photons are produced via $e^+e^-$ 
beam scattering and collisions and convert to electrons in the quartz 
bars. These electrons produce additional (background) Cherenkov photons, 
and the resulting hits in the MCP-PMTs contribute to the integrated charge, 
thus reducing the lifetime. 
To perform satisfactorily until the end of Belle~II running, an MCP-PMT 
must be able to reach an integrated output charge of 10~C/cm$^2$ with 
an efficiency drop of 20\% or less.

During development and also production of the MCP-PMT, several
refinements were made to the internal structure and fabrication 
process~\cite{Kishimoto:2006mg, Jinno:2010cj, MATSUOKA201793}
in order to improve the lifetime of the photocathodes.
The first prototype MCP-PMT had a gap between the MCP plates and 
the PMT wall. By introducing a small ceramic plate into this gap
to block neutral gas from reaching the photocathode, 
the lifetime was improved and reached 1.1~C/cm$^2$. This design is 
referred to as a conventional-type PMT. The MCP-PMT lifetime was 
subsequently (and significantly) improved by applying a special coating,
referred to as Atomic Layer Deposition (ALD), to the MCP 
surface~\cite{Wetstein:2011zz,Siegmund2012168}. This coating 
reduces outgassing from the MCP and thus the rate of neutral gas 
and ion feedback~\cite{Conneely:2013mva}.
Test results for several types of MCP-PMTs were reported in 
Refs.~\cite{Conneely:2013mva,Lehmann:2014aqa,Lehmann:2016rbg}.
Our own test results~\cite{MATSUOKA201793} show that the mean 
lifetime of ALD-type PMTs exceeds the desired 10~C/cm$^2$.
A final improvement in lifetime was obtained by decreasing
the level of residual gas during MCP-PMT fabrication. This
increased the lifetime further to $>$13.6~C/cm$^2$~\cite{MATSUOKA201793}.
This last type of MCP-PMT is referred to as a life-extended ALD type.

\subsection{Mass production}
\label{sbsctn:pmt_massproduction}

The mass production of MCP-PMTs began in 2011 with conventional-type PMTs. 
After the first batch (260) of these were produced, production switched to
ALD-type PMTs. After 224 of these were produced, production switched to
life-extended ALD-type PMTs. By May 2016, a total of 549 life-extended 
ALD-type PMTs were produced, which is enough to instrument the entire TOP detector.
However, this production was not completed before the TOP detector needed to be
installed in April 2016, and 224 conventional-type PMTs were utilized for the 
installation. These conventional-type PMTs were used to commission the detector 
and record the first Belle~II data sets; they were almost all replaced with 
life-extended ALD-type PMTs during the long Belle~II shutdown in 2022-2023.

\begin{figure}[htb]
 \centering
 \includegraphics[height=7cm]{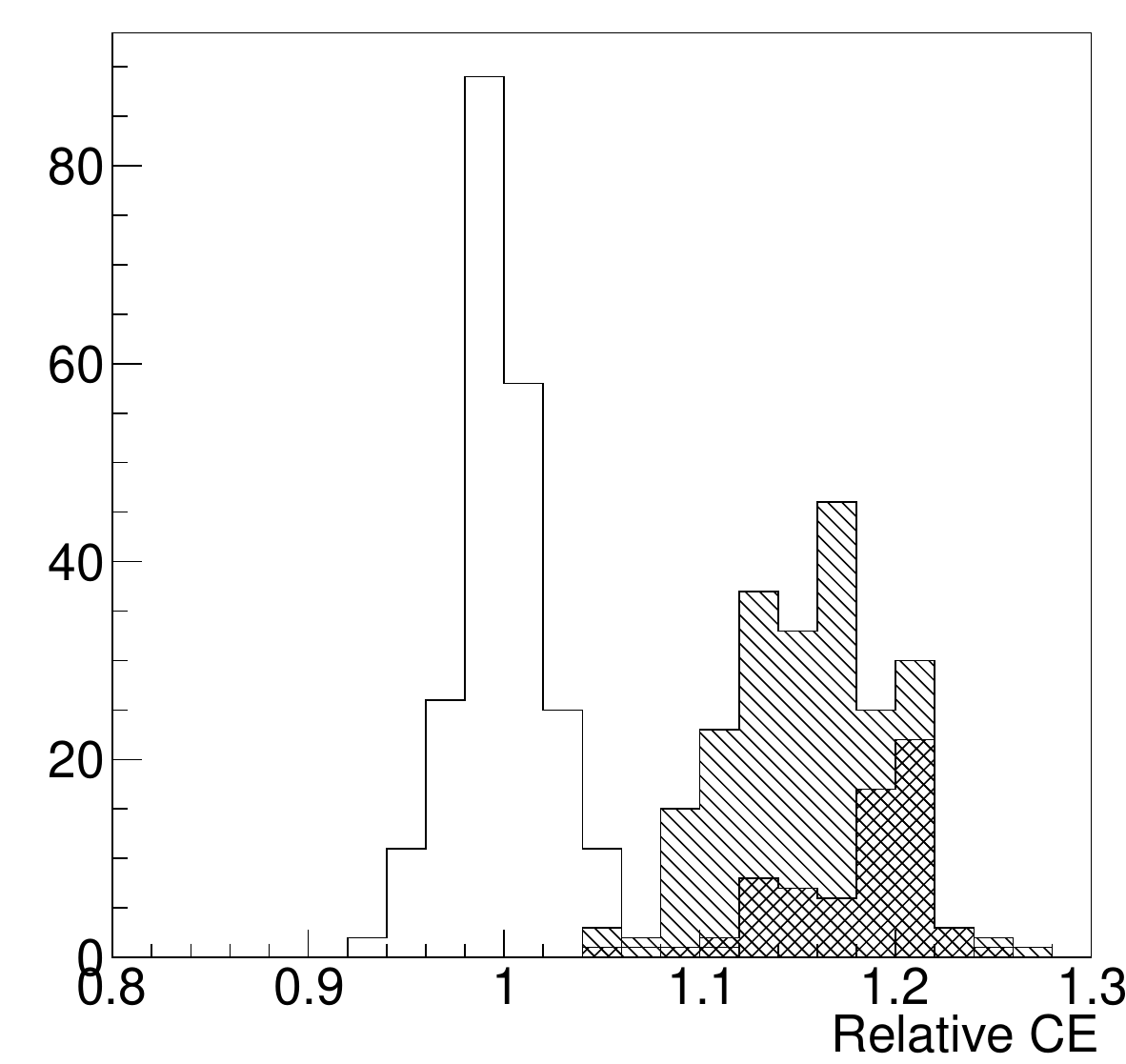}
  \caption{Relative collection efficiency (CE) for the 512 MCP-PMTs installed 
in Belle~II. The CE distribution is separated by PMT type: 
open, hatched, and cross-hatched histograms correspond to conventional-type, 
ALD-type, and life-extended ALD-type PMTs, respectively. All values are 
normalized to the average CE for conventional-type PMTs.}
 \label{fig:MCP-PMT:performanceCETTS}
\end{figure}

The TTS, QE, and CE of all MCP-PMTs were measured at Nagoya University before 
installation in Belle~II. The results are summarized in 
Fig.~\ref{fig:MCP-PMT:gainTTS}~(right),
Fig.~\ref{fig:MCP-PMT:QE}~(right),
and Fig.~\ref{fig:MCP-PMT:performanceCETTS}. 
All PMTs were required to have a TTS less than~50~ps; the average TTS was 34.3~ps. 
In addition, all PMTs were required to have a QE greater than 24\% at $\lambda=360$~nm,
which is approximately the peak wavelength for photons collected by the PMTs.
The average QE at this wavelength was 29.3\%. 
The CE for both ALD-type PMTs and life-extended ALD-type PMTs was typically 
higher than that for conventional-type PMTs, and the timing distributions 
exhibited longer tails.
These features are due to a higher secondary emission rate for ALD-type MCPs 
and improved collection of primary electrons that reflect off the MCP 
surface. The performance characteristics of the PMTs were confirmed 
in bench measurements performed at KEK 
in a 1.5~T magnetic field.

\subsection{PMT module}
\label{subsec:pmt_module}

\begin{figure}[htb]
 \centering
 \includegraphics[height=6cm]{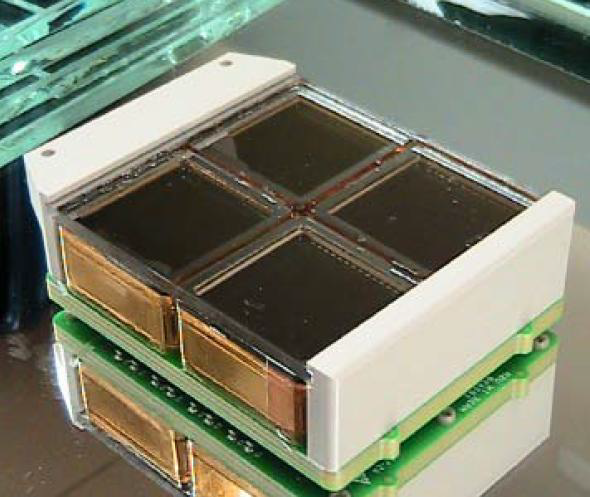}
 \caption{A PMT module consisting of four MCP-PMTs, a wavelength cut filter, 
 and a printed ciruit board.}
 \label{fig:MCP-PMT:PMTmodule}
\end{figure}

Four individual MCP-PMTs were assembled 
into a PMT module as shown in Fig.~\ref{fig:MCP-PMT:PMTmodule}.
A wavelength cut filter (IHU-340 by Isuzu Glass Ltd.~\cite{IsuzuGlass}) 
was mounted to the front face, and a printed circuit board (PCB) was
attached to the back side. The wavelength filter sharply attenuates 
wavelengths below 340~nm, which reduces chromatic dispersion.
The filter was attached using an optically transparent silicone rubber 
(TSE3032 by Momentive Performance Materials~Inc.~\cite{MomentivePerformance}).
The PMT module was mounted to the prism using a ``cookie'' that
was soft-cast from TSE3032 using a mixture ratio of 100:2.
The PCB (referred to as a ``front'' board - see Section~\ref{sec:electronics_overview})
contains pads that make contact with spring-loaded pogo pins mounted on the front-end 
electronics. With this electrical connection, the PCB distributes high voltage to the 
PMT channels and passes PMT signals to the electronics.

%% file: electronics.tex
\section{Readout electronics}
\label{sec:electronics}

\subsection{Overview}
\label{sec:electronics_overview}

\begin{figure}
\centering
\includegraphics[width=13.3cm]{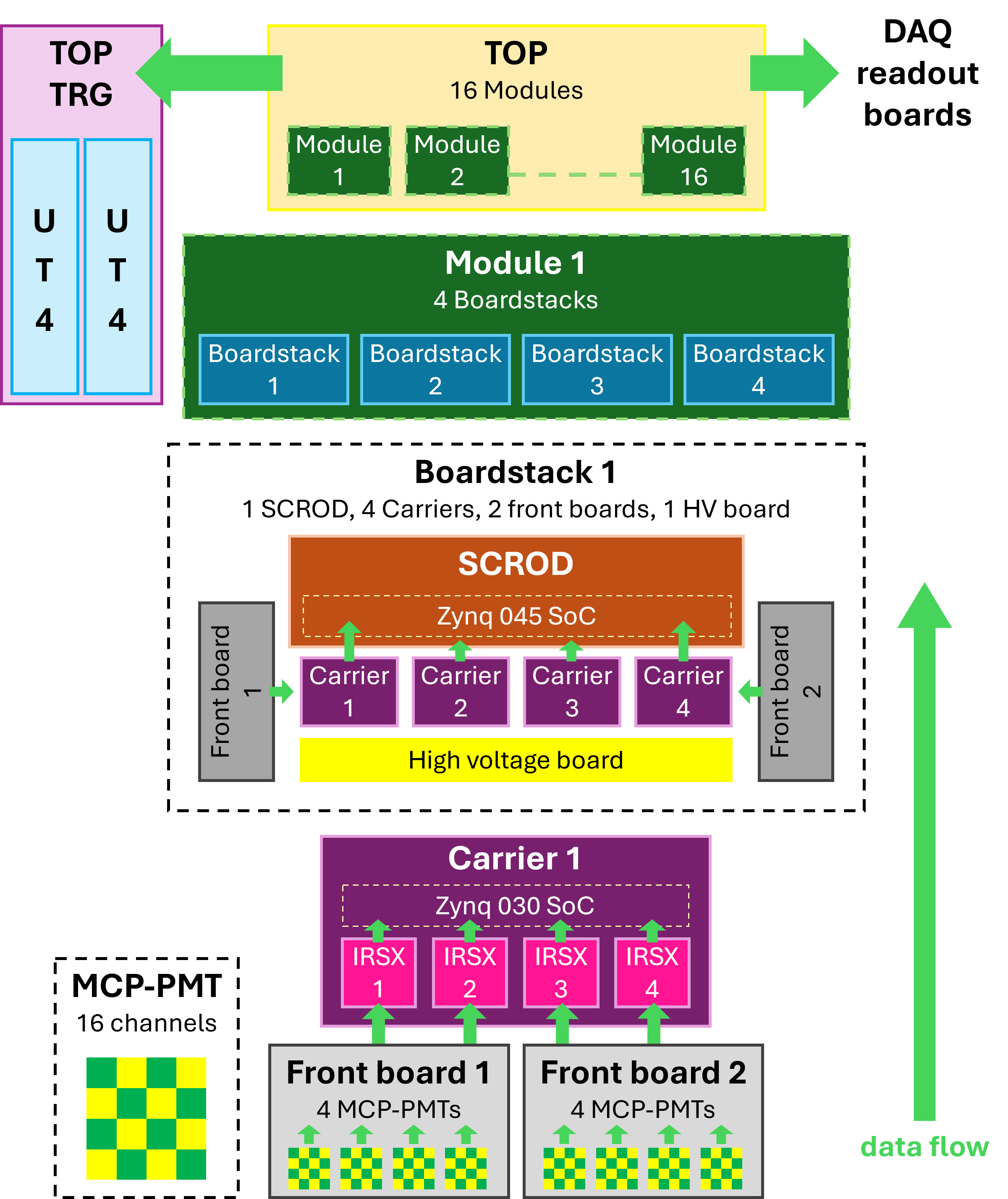}
\vskip0.20in
\caption{Overview of the front-end electronics used to read out a TOP module.
Four PMT modules connect to a front board, and two front boards are read out 
by four carrier boards arranged in a boardstack. There are four boardstacks 
per module reading out all 32 MCP-PMTs (512 channels). }
\label{elec:fig1}
\end{figure}

\begin{figure}
\centering
\includegraphics[width=6.0cm,angle=-90.]{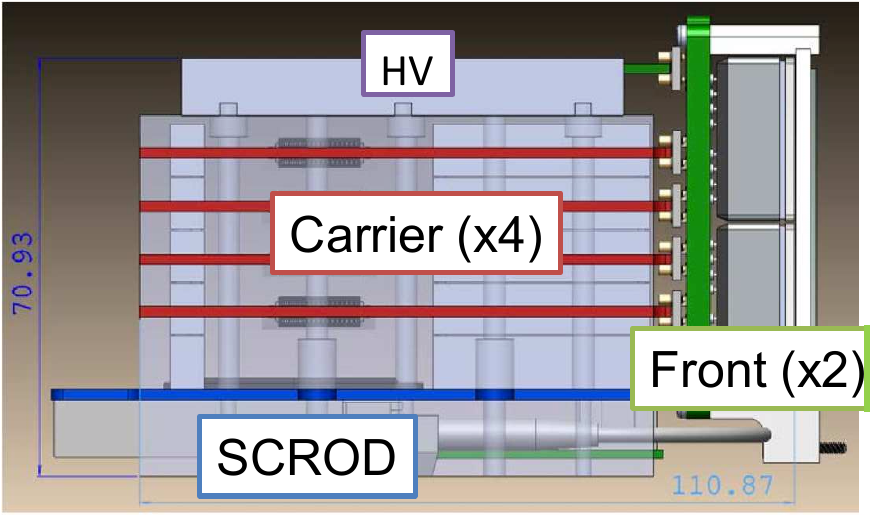}
\vskip0.20in
\caption{Physical layout of a front-end boardstack (see text).
To facilitate assembly and disassembly, the carrier boards and high 
voltage board connect to the front boards via pogo pins and contact pads.}
\label{elec:fig2}
\end{figure}

An overview of the front-end electronics used to read out a TOP module
is shown in Fig.~\ref{elec:fig1}.
The electronics consist of printed circuit boards (PCBs) organized into 
``boardstacks.'' Each TOP module is instrumented with four boardstacks. The PCBs consist of 
four ``carrier'' boards, one high voltage (HV) board, and one ``SCROD'' 
(Serial Control and Readout of Data) controller board; these make electrical connections 
via pogo pins to the PCBs of two PMT modules (see Fig.~\ref{fig:MCP-PMT:PMTmodule}).
The PCBs of the PMT modules are referred to as ``front'' boards.
The physical layout of a boardstack is shown in Fig.~\ref{elec:fig2}.
Each boardstack reads out eight MCP-PMTs for a total of $8\times 16 = 128$ channels.
The carrier boards read out data using a Xilinx Zynq System-on-Chip (SoC), 
which consists of a Field Programmable Gate Array (FPGA) and an ARM 
Processing System (PS) device. These communicate with the other Zynq devices via 
multi-gigabit links on connectors between the boards.  Each SCROD collects data from 
the carrier boards and forwards it to the Belle~II Data Acquisition (DAQ) system and
a TOP interface to the Belle~II trigger logic. Data is transferred via multi-gigabit fiber 
optic links. With this readout, the contribution of the electronics to the single-photon 
timing resolution is approximately 30 ps. Below we give a short summary of boardstack 
operations; for further details, see Ref.~\cite{Kotchetkov:2018qzw}.

\subsection{Boardstack Operation}
\label{sec:electronics_boardstack}

Each carrier board houses four IRSX (Ice Radio Sampler version X) 
ASICs and one Zynq 030 SoC. Linear voltage regulators provide point-of-load 
regulation for the SoC and ASICs. Each SCROD board utilizes a Zynq 045 SoC.

The IRSX ASIC is an eight-channel sampling and digitizing circuit.
In the TOP detector, the sampling rate is 2.714~GSa/s. All eight channels are 
sampled continuously into an analog memory $2^{15}$ samples deep, which corresponds 
to about 12~$\mu$s in duration. Comparators for each channel operating in parallel 
inform the FPGA when the input signal 
exceeds a threshold set by an internal digital-to-analog converter (DAC).
Digitization is carried out in a Wilkinson style, where a time-to-digital circuit 
measures how long it takes a voltage linearly increasing with time to exceed 
a given threshold. This digitization is done in parallel across all channels.
Subsequently, for those channels with a trigger comparater bit set in the FPGA,
the data are read out to the FPGA via high-speed (multi-megabit/s) links.

All carrier, SCROD, and HV boards underwent initial testing at test facilities
set up at Indiana University, the University of Hawaii, the University of Pittsburgh, 
and the University of South Carolina. Final boardstack assembly and testing was 
performed at the University of Hawaii. Boardstacks that were satisfactory were
then shipped to KEK for installation in detector modules.

\subsection{Firmware}

The FPGA's on the carrier and SCROD boards are controlled by firmware written 
in VHDL and Verilog. The firmware on the carrier boards controls the four IRSX ASICs
and keeps track of which channels have valid data. 
When a Belle~II trigger arrives, the firmware commences digitization and readout 
of the associated block of IRSX memory and sends it to the SCROD.
The SCROD firmware consolidates the data from carrier boards,
exchanges slow-control data with these boards, and distributes 
trigger signals from the global Belle~II trigger. 
The data is processed in the PS, properly formatted with headers 
added, and transferred to the Belle~II DAQ system.

\subsection{Trigger}

The TOP detector is also used to create a fast
trigger for the Belle~II detector, taking
advantage of the prompt nature of \cherenkov\ radiation.
This trigger is capable of precise timing resolution,
superior to that of the ECL, which is used to 
initiate the main global trigger of the experiment.

The TOP trigger logic is implemented in two Universal Trigger boards UT4 
(i.e., fourth generation) that are based on a Virtex UltraScale FPGA from Xilinx/AMD. 
Each UT4 board receives data from eight TOP modules via $4\times 8$ optical links. 
The data sent to the trigger boards constitute a separate data stream from
the main data readout, and the trigger data consists of 16-bit ``time-stamps''
encoding photon hit times with a least-count of~2.0~ns. All time-stamps from a 
detector module are input to trigger logic that calculates the beam crossing (collision) 
time using a likelihood-based algorithm. These calculations are performed 
independently for each module, and the results are input to a final level 
of logic that correlates the results from the modules with track
hits from the central tracking detector and makes an overall timing decision. 
This result is subsequently sent to the global trigger decision logic, which 
issues a global trigger for the experiment. The total latency of the TOP 
trigger is about~3.2~$\mu$s.

\subsection{Operational experience}
\label{sec:electronics:SEU}

The readout electronics were installed in 2016 and, after the detector endcaps were closed, 
became inaccessible during the subsequent running period.  During early operation, firmware 
development continued and many issues were resolved.  However, beam-related backgrounds 
were higher than expected, and this required further optimization of the firmware.

After four years of data-taking, there were a handful of component failures:
one SCROD board had a defective DDR memory;
one optical link to a boardstack broke;
one low voltage connection to a boardstack failed;
the JTAG chain in one boardstack ceased to function, 
making it impossible to program the firmware;
one carrier board did not respond to triggers;
three IRSX ASICs could not process events;
and one channel pair became inoperable because of a faulty trigger line. 
In total, 442 out of 8192 channels (5.4\%) were excluded from data-taking. 
However, almost all of these problems were rectified by replacing boardstacks 
during the long Belle~II shutdown in 2022-2023.
During this shutdown, the detector was opened and the electronics could be accessed. 

Another issue that arose during data taking was that of ``single-event upset'' (SEU) events. 
SEU events are caused by beam background particles, especially neutrons, depositing
ionizing radiation within the front-end boards; this disrupts their operation. 
The rate of SEU events increases with beam currents but also depends on the 
settings of accelerator collimators. SEU events affect both the programmable logic (PL) 
of the Zynq SoC devices and the processing system (PS) of the SoC.
While PL errors are usually self-correcting, 
PS errors lead to exceptions such as data aborts that leave the PS in an unrecoverable state.
Since the PS is part of the event processing pipeline, this results in missing data at the interface 
to the DAQ system. To remedy this such that data acquisition could proceed, the affected IRSX ASICs 
were masked, i.e., no longer read out.
Such errors occur at a relatively low rate: a few times per day for the entire TOP detector.

%% file: gas_system.tex
\section{Gas system}
\label{sec:gas}

\begin{figure}[!ht]
\begin{center}
\includegraphics[width=6.2in]{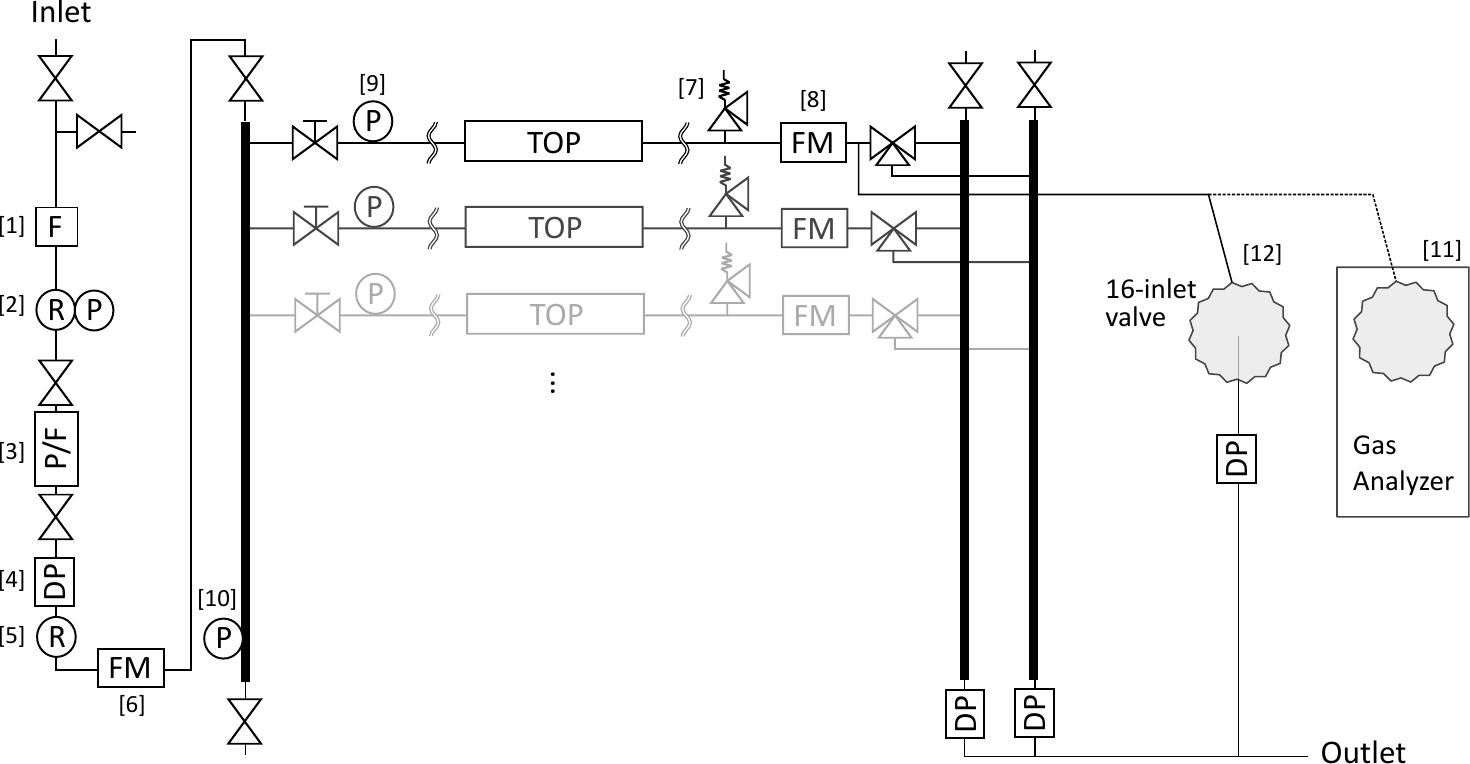}
\end{center}
\caption{Schematic diagram of the gas system. The ID number appearing beside a 
component corresponds to the ID number listed in Table~\ref{gas_system:components}. }
\label{fig:gas_system:design}
\end{figure}

The TOP quartz optics must be maintained in a clean and stable 
environment for the lifetime of the experiment. This environment
must have low moisture and be free of dust and particulates. Such 
contaminants settling on a bar's surface would alter the index of 
refraction at the point of contact, allowing light that would normally 
undergo total internal reflection at that point to refract out. To maintain
the cleanliness of the quartz surfaces, the QBB's are flushed with dry
nitrogen. This gas is obtained from the boil-off of liquid nitrogen
and flows through each QBB at a flow rate of 7--13 liters/hour.
Before entering the QBBs, the gas passes through several filters to 
remove particulates. The output gas is monitored for humidity and contaminants 
by a gas analyzer. The components of the gas system are listed in 
Table~\ref{gas_system:components}, and the layout of the system
is shown in Figure~\ref{fig:gas_system:design}.

\begin{table}[h!]
  \centering 
  \caption{Components of the gas system. The ID number of a component corresponds to the 
    number appearing in Figure~\ref{fig:gas_system:design}.
  \label{gas_system:components}}
  \begin{tabularx}{\textwidth}{rXr}
 \hline
 ID & Component & Number \\
  \hline
 {[1]} & Filter - Fujikin FUFL-915-6.35-2 & 1\\
 {[2]} & Regulator and pressure gauge - Yamato LR-2S-L10-0101-10 & 1\\
 {[3]} & Filter - LiquidGas Fine Purer FPI-2 & 1\\
 {[4]} & Dewpoint Hygrometers - Michell Instruments ED online / TEKHNE TK-100 & 4\\
 {[5]} & Regulator - Yamato IR-1S& 1\\
 {[6]} & Flowmeter - Asbil CMS0020BTTN200100 20L/min & 1\\
 {[7]} & Pressure Relief valves - Circle seals controls 132B-2PP 1kPa & 16\\
 {[8]} & Flowmeter Asbil CMS0002BTTN200100 2L/min & 16\\
 {[9]} & Pressure sensor - Nagano Transmittor KP-15 & 16\\
 {[10]} & Pressure gauge - Dai-ichi-Keiki 0-10kPa & 1\\
 {[11]} & Residual Gas Analyzer – Stanford Research Systems UGA300 & 1\\
 {[12]} & 16-inlet automatic switching valve - GL Sciences EMT4SC16MWE & 1\\
\hline
\end{tabularx}
\end{table}

All gas tubing is stainless steel, with a diameter of 10~mm at the gas manifold and
between the gas supply and the filters. Smaller 6.35~mm-diameter tubing is used between 
the gas rack and a ``tee'' leading to a manual relief valve. Additional pressure-relief 
valves are installed for each QBB to prevent any over-pressure.

To measure possible contamination, we use a Stanford Research Systems UGA300 residual gas analyzer. 
This model is designed to be calibrated against nitrogen gas; thus, pure nitrogen produces 
a null signal, which results in high sensitivity to even small amounts of contamination. 
To sample the gas from each detector module, we use a 16-channel multiplexer. A second multiplexer 
valve is installed to monitor each module's dew point, which is very sensitive to water content. 
The dew point at the inlet (after the filters) is less than the measurement 
limit of $-100^\circ$~C, while the dew point at the outlet is typically in the range
$-40^\circ$~C to $-60^\circ$~C and fluctuates with the dew point in the 
experimental hall. With these two multiplexers, any contamination resulting 
from a leak or outgassing of the epoxies or vacuum grease within a QBB
can be quickly detected. 

%% file: calibration.tex
\section{Calibration}
\label{sec:calibration}

To provide timing information of sufficient precision, the TOP detector 
must be well calibrated. The timing calibration is performed in several 
steps. The first step consists of a time base calibration for 
each of the 8192 electronic channels in order to linearize the time response; it is 
performed by injecting pulses into the IRSX ASICS of the carrier boards
(see Section~\ref{sec:electronics_boardstack}).
The next step is the time alignment of the 512 channels within a detector module.
The channel-by-channel time offsets required are referred to as ``channel $t_0$'s''
and are determined using a laser calibration system. After determining these, all 
detector modules are time-aligned (``module $t_0$'') using collision data. 
These time alignments correct for variations in time delays across the detector. 
Finally, the offset of the bunch-crossing time with respect to the accelerator 
clock is determined, also using collision data (``common $t_0$''). 
The precision attained for each step of the calibration procedure, and the stability of the 
resulting calibration constants, are summarized in Table~\ref{table:calibration}. More details 
of the calibration procedure are provided in Refs.~\cite{Staric:2017lqu,Tamponi:2017dlc}. 

In addition to the above timing calibrations, we also determine ``hot'' and dead channels 
using collision data and mask them. Channel-dependent pulse-height thresholds are set 
to be safely above noise levels. The efficiencies of these thresholds are monitored using 
laser data by fitting the resulting pulse-height distributions and, from these fits, 
estimating the fraction of signal falling below the threshold setting.

\begin{table}
  \caption{Calibration steps for the TOP detector, and the resulting precision.}
  \label{table:calibration}
  \begin{center}
    \begin{tabular}{|c|c|c|c|}
      \hline
      Calibration & Precision & Variation over time & Update frequency \\
                  & (ps)      & (ps)       &                  \\
      \hline
      time base     & 25 & $< 50$ & months \\
      channel $t_0$ & 15 & $< 50$ & months \\
      module $t_0$ & 0.1 & $< 5$ & weeks \\
      common $t_0$ & 
                     10 & 150 & minutes to hours \\
      \hline
    \end{tabular}
  \end{center}
\end{table}

\subsection{Time base calibration}
\label{sec:calibration_tbc}

Each channel of an IRSX ASIC contains an analog sampling array of 128 elements.
The sampling is synchronized using the SuperKEKB accelerator clock, which is 
divided by 24 to produce a 21.2~MHz synchronization clock (``SST-in'') used for the sampling
circuits. The sampling starts at the rising edge of SST-in and proceeds until the next SST-in 
cycle arrives. A delay-locked-loop circuit ensures that all 128 samples are performed within 
a single SST-in period.

The sampling times for a channel are calibrated by injecting a double pulse having a fixed 
time difference between the pulses into the channel input. To ensure all 128 time samples 
are populated, the pulse generator runs asynchronously with the SST-in clock. The sample times 
are determined via a least-squares minimization procedure as described in Ref.~\cite{Staric:2017lqu}. 
Typically, $5\times 10^4$ double pulses per channel are used, which results in
a time-base precision of 25~ps. This calibration is relatively stable over time, 
and new constants are determined only after several months of running.

\subsection{Channel $t_0$ calibration}
\label{sec:calibration_channelt0}

\begin{figure}[ht]
\centering
\includegraphics[width=.95\textwidth]{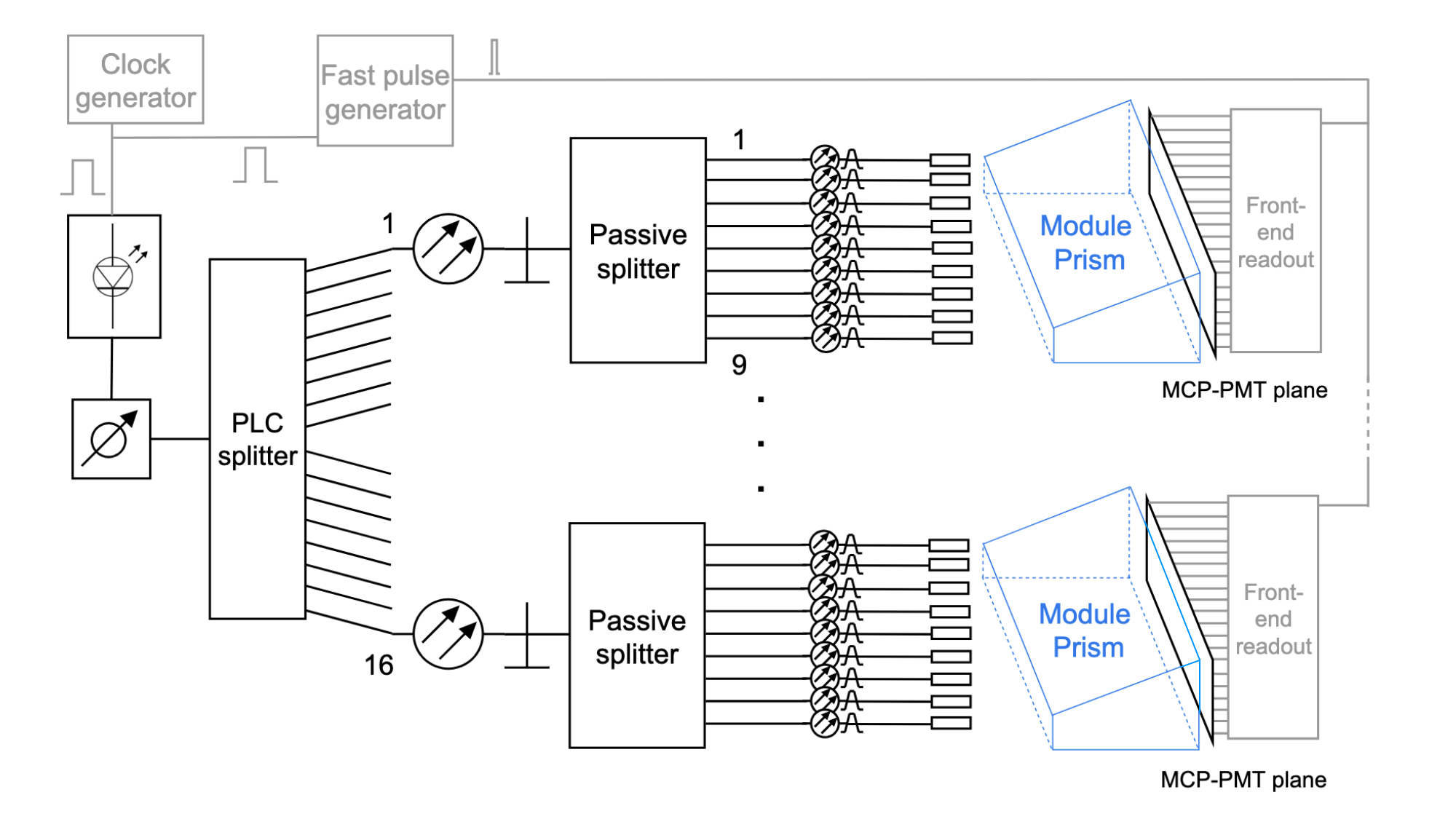}
\caption{Schematic diagram of the laser system used to time-align channels within 
a detector module. 
}
\label{fig:lasersystem}
\end{figure}

To correct for variations in time delays across the channels of a detector module, 
we use a laser calibration system~\cite{Tamponi:2017dlc} built into each QBB as 
described in Section~\ref{sec:barbox_design}. A schematic diagram of the system 
is shown in Figure~\ref{fig:lasersystem}. The system consists of a 405~nm PILAS 
pulsed diode laser~\cite{Pilas_laser}  
followed by a planar-lightwave-circuit (PLC) splitter~\cite{PLC_splitter} 
that divides the laser light among 16 single-mode fibers~\cite{laser_fiber}, 
i.e., one fiber per detector module. 
Each of these primary fibers feeds (via a second splitter) a bundle of nine 
multimode fibers of equal length~\cite{passive_splitter}. These fibers are terminated 
with gradient-index (GRIN) lenses~\cite{fiber_heads} that increase the numerical 
aperture of the system. The GRIN lenses fit into openings in the prism enclosure 
of the QBB such that they are positioned 4.0~cm in front of the tilted 
face of the prism (see Fig.~\ref{fig:mechanics:enclosure}). 
Laser light enters the prism through the tilted face, internally 
reflects off the top surface of the prism, and illuminates the MCP-PMTs.
The reference time for the electronics is provided by a calibration 
pulse injected into the first channel of each boardstack of a detector module.
Both the calibration pulse and the laser are triggered by a common external clock 
that runs asynchronously with the clock provided to the electronics; in this manner 
the laser light signals uniformly populate the digitization window.

\begin{figure}[ht!]
\centering
\includegraphics[width=0.55\textwidth]{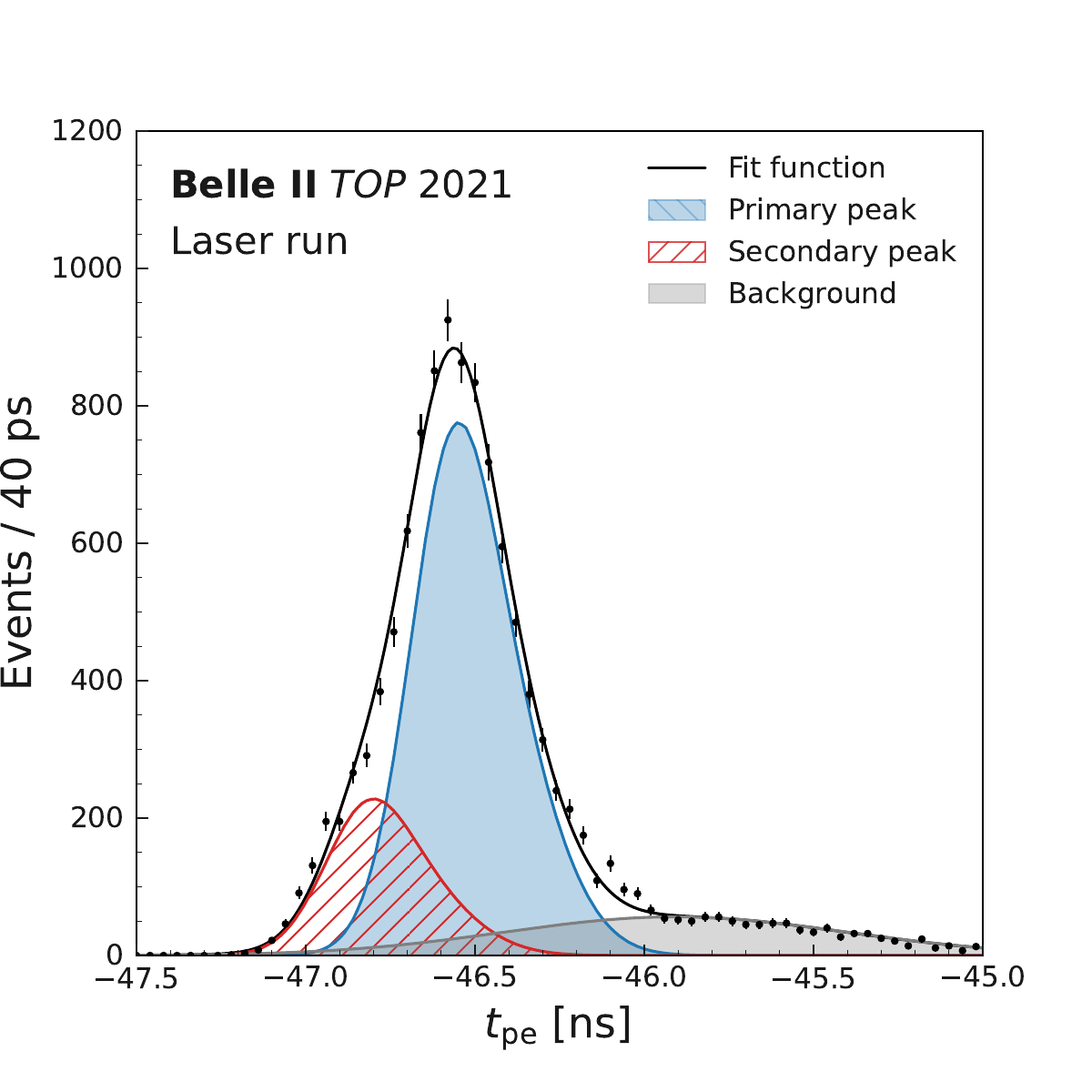}
\caption{Distribution of MCP-PMT hit times from laser light for one channel of 
a detector module. 
The primary peak is due to light internally reflecting off the top surface of the
prism, while the secondary peak is due to light traveling directly to the MCP-PMT
with no reflection. The position of the primary peak is used to determine the 
time offset~($t^{}_0$) of the channel.}
\label{fig:laserlight}
\end{figure}

A typical distribution of hit times from laser light for one channel is shown in 
Figure~\ref{fig:laserlight}. We fit this distribution with the sum of two functions: 
one corresponding to light traveling directly from the fiber to the MCP-PMT, and the 
other corresponding to light reflecting off the upper surface of the prism. As the latter 
signal is much larger, we take this function to define the hit time for the channel. 
Each function consists of a distribution characterizing the transit time spread (TTS) 
of the MCP-PMT convolved with a Gaussian resolution function. The TTS distribution is 
taken from bench measurements performed at Nagoya University before MCP-PMT installation, 
while the parameters of the resolution function are floated in the fit. We also include 
in the fit a broad Gaussian component, to model diffuse light arriving significantly 
later than the main signal. After correcting the fitted hit time for the time of photon 
propagation from the GRIN lens to the MCP-PMT pixel (determined from Monte Carlo simulation), 
we are left with the channel-dependent delay due to transit time and electronics, 
i.e., the desired $t_0$. Determining this $t_0$ for all channels of a detector module, 
and subtracting this off when processing raw data, time-aligns all channels within 
the module.

\subsection{Module $t_0$ calibration}

The time alignment of detector modules relative to each other is performed using 
$e^+e^- \to \mu^+\mu^-$ events, where each muon is required to hit a different module. 
First, the start time of each muon 
is determined by maximizing a muon likelihood (see Section~\ref{sec:reconstruction_likelihood}) 
with respect to a global offset subtracted from the photon hit times. Subsequently, 
the difference between the two hit times is recorded, separately for each pair of detector modules.
After collecting sufficient statistics, the distributions are fitted with a sum of two Gaussian 
functions; this yields, for every module pair, the average time difference. From the overall set 
of time differences, individual detector module $t_0$'s are extracted using a least squares 
minimization procedure. The resulting values are then used as initial values for 
a second, more precise method. In this method, the $t_0$ for a detector module is determined 
by maximizing the sum of muon log likelihoods for all muons traversing the module. The final 
precision on $t_0$ obtained using a sample of $1.0\times 10^5$ muon pairs is better than 10~ps.

\subsection{Common $t_0$ calibration}

Finally, we calibrate the offset between the rising edge of the accelerator clock, which 
synchronizes the front-end electronics, and the bunch-crossing time as determined by the TOP 
(see Section~\ref{sec:reconstruction_likelihood}). This offset depends on the accelerator conditions 
and can change on a daily basis. Thus, this calibration is performed continuously for every 
run. We again use $e^+e^- \to \mu^+\mu^-$ events and maximize the sum of muon log likelihoods 
for a common time offset subtracted from all photon hit times. Using a relatively small 
sample of 1000 $\mu^+\mu^-$ events, we obtain a precision of 10~ps, which is sufficient.

%% file: recon_software.tex
\section{Reconstruction software}
\label{sec:reconstruction}

\subsection{Overview}

\cherenkov\ detectors that record photon hit positions (in contrast to only counting the
number of hits)
have been used in many experiments, and the reconstruction of such data is a well-studied 
field. Charged particles that traverse a medium faster than 
the speed of light in the medium emit photons at an angle $\theta^{}_c$ --\,the \cherenkov\ angle\,-- 
with respect to the particle direction. This angle is given by the formula  $\cos\theta^{}_c = 1/(n\beta)$, 
where $n$ is the index of refraction of the medium and $\beta$ is the particle's velocity divided by 
the speed of light~$c$.
The result is an annulus of photons radiated in the forward direction that can be imaged
onto an array of photodetectors. In \cherenkov\ detectors with an expansion volume, this 
results in a ``ring image'' in which the radius of the ring is proportional to $\beta$,
and the thickness is proportional to the depth of the radiator material.
Such a detector yields a measurement of the velocity $\beta$ that, 
when combined with a measurement of the track's momentum from a tracking detector, 
yields the particle's mass. In this manner the detector distinguishes 
particles that are sufficiently well separated in mass, notably pions and kaons. 

In a TOP detector module, some fraction of radiated photons totally internally reflect 
down the bar and are imaged by the two rows of MCP-PMTs. The nominal ring image is thus
folded upon itself multiple times, and the resulting spatially compressed and time-distorted 
projection precludes calculating $\beta$ from the radius of a simple ring image.
Thus, a new algorithm was developed~\cite{Staric:2008zz,Staric:2011zza} that combines the 
imaging aspect of the detector with its excellent time resolution in an optimal way. 
This algorithm assigns photon hits to a charged particle traversing a quartz bar 
and computes a likelihood value for the particle for each of six particle hypotheses: 
$e$, $\mu$, $\pi$, $K$, $p$, and $d$.

\subsection{Calculation of likelihood}
\label{sec:reconstruction_likelihood}

The likelihood value for a particular charged particle hypothesis is calculated with 
probability density functions (PDFs) calculated analytically using an inverse-ray-tracing 
algorithm~\cite{Staric:2008zz, Staric:2011zza}. For a given hypothesis $h$ 
($h=e$, $\mu$, $\pi$, $K$, $p$, $d$), the likelihood is defined as
\begin{eqnarray}
  \log {\cal L}_h = 
  \sum_{i=1}^{N} \log\Bigl(\frac{N_h S_h(c_i,t_i)+N_B B(c_i,t_i)}{N_h + N_B}\Bigr) + 
\log P_{\mu}(N)\,,
  \label{extlkh}
\end{eqnarray}
where $N_h$ and $S_h(c_i,t_i)$ are the expected photon yield and PDF, respectively, for hypothesis $h$;
$N_B$ and $B(c_i,t_i)$ are the expected yield and PDF, respectively, for background; and $c_i$ and $t_i$ 
are the pixel number and arrival time of photon~$i$. The arrival times are measured relative to the 
bunch-crossing time and thus include the time-of-flight of the particle. The second term in 
Eq.~\ref{extlkh} is the Poisson probability to observe $N$ photons for a mean value $\mu = N_h + N_B$.
The expected signal yield $N_h$ is calculated analytically using the model described in 
Ref.~\cite{Staric:2011zza}, and the expected background yield $N_B$ is estimated 
event-by-event from hits in other modules that have no associated tracks.

\begin{figure}[h]
\hbox{
\hskip-0.06in
\includegraphics[width=8.3cm]{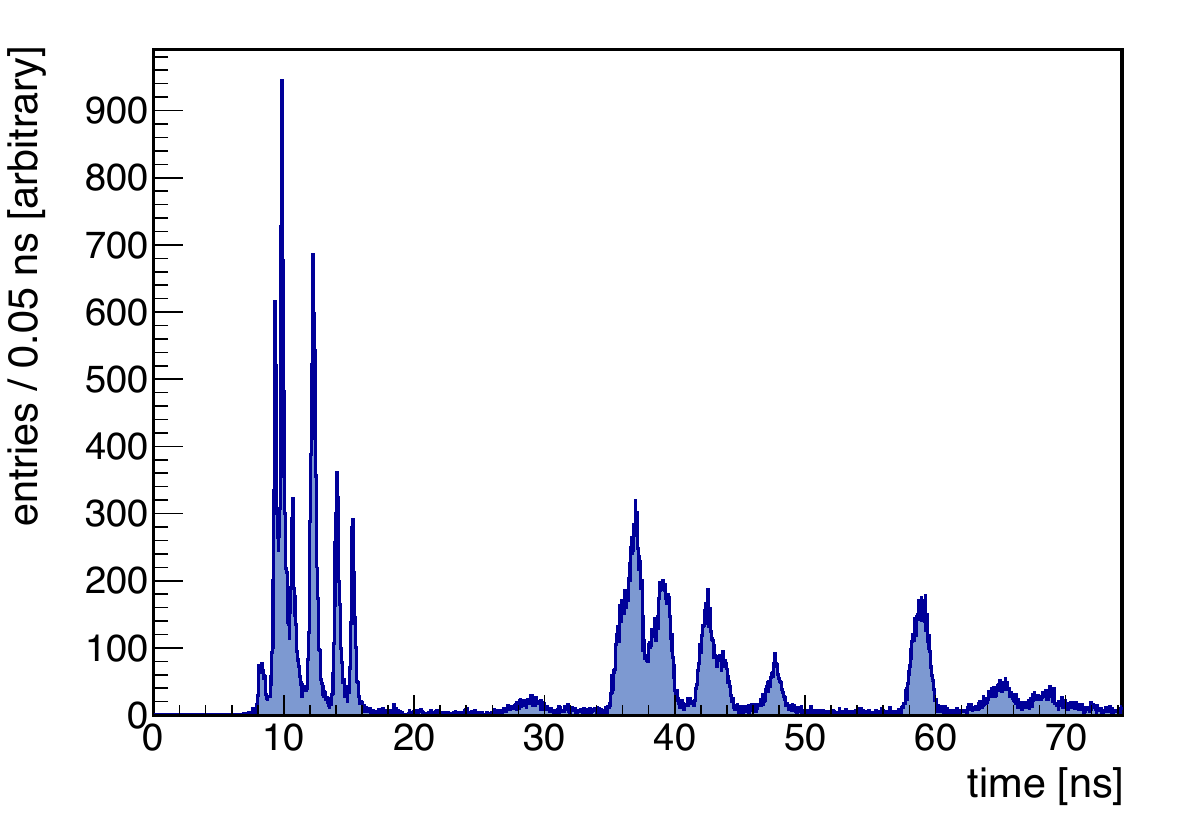}
\hskip-0.06in
\includegraphics[width=8.3cm]{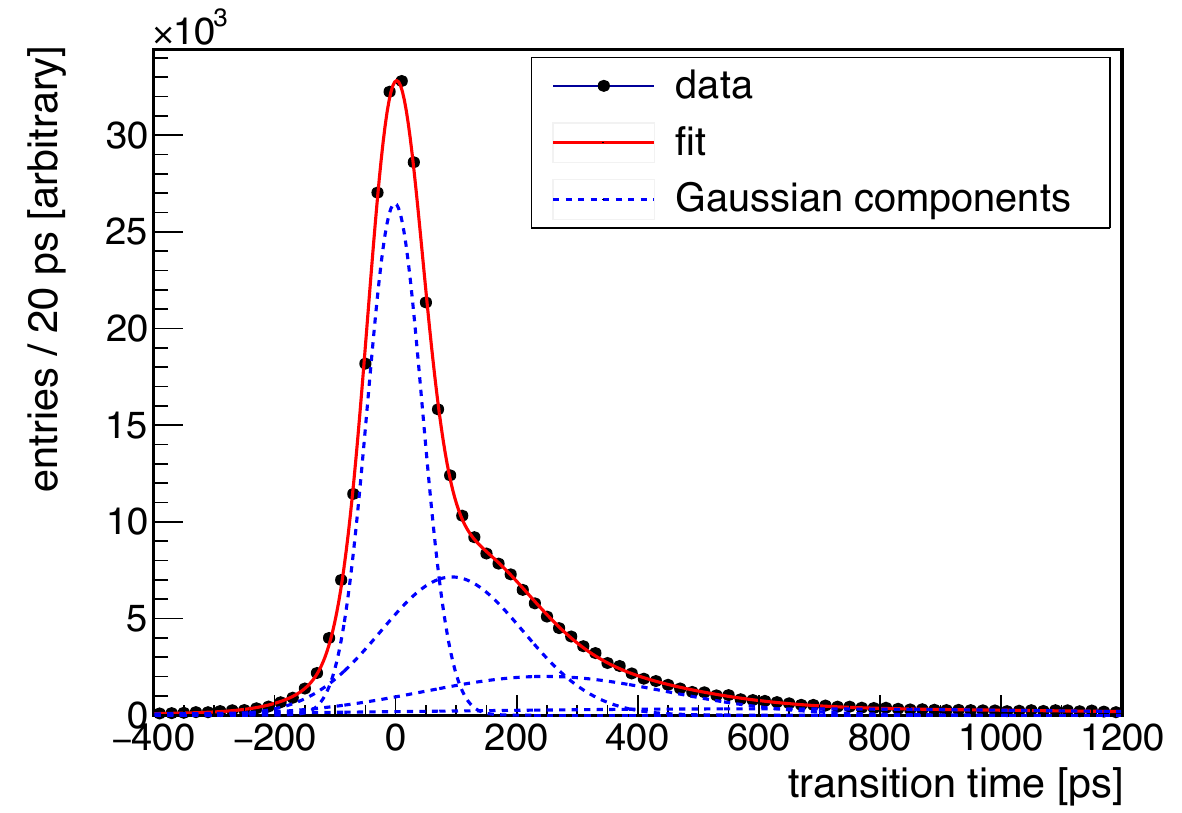}
}
\caption{Typical timing peaks of \cherenkov\ photons for one pixel~(left), and 
the TTS distribution measured on the bench of an ALD-type MCP-PMT. The latter 
is fitted with the sum of four Gaussians. The respective widths ($\sigma$) of 
the Gaussian components, after correcting for the resolution of the electronics,
are 33.5~ps, 114~ps, 199~ps, and 552~ps; the respective fractions are 
46\%, 32\%, 15\%, and 7\%. }
\label{fittedTTS:fig}
\end{figure}

The signal PDF for a given pixel $c$ consists of a number of timing peaks as shown 
in Fig.~\ref{fittedTTS:fig}~(left). The separate peaks correspond to photons that 
underwent different numbers of internal reflections. The number of peaks ($n_c$)
and the fraction of photons in the $k$th peak ($f_{ck}$) are calculated analytically for
different particle hypotheses $h$ using the model described in Ref.~\cite{Staric:2011zza}. 
The shape of each peak is taken to be a Gaussian convolved with the TTS of the MCP-PMT 
(measured on the bench for each MCP-PMT prior to installation) and convolved with 
a second Gaussian corresponding to the electronics resolution. The TTS for each tube is 
modeled with a sum of three or four Gaussian PDF's, depending on the tube type,
as shown in Fig.~\ref{fittedTTS:fig}~(right).
Thus the signal PDF is
\begin{eqnarray}
  S_h(c,t) = \sum_{k=1}^{n_c} f_{ck} 
\Bigl[ \sum_{j=1}^{n_g} f_j\,G(t-t_{ck}-t_j^{\rm TTS};\,
\sigma_{ck} \oplus \sigma_j^{\rm TTS} \oplus \sigma^{\rm el}) \Bigr]\,,
  \label{signalPDF}
\end{eqnarray}
where $t_{ck}$ and $\sigma_{ck}$ are the expected position and width, respectively, 
for photons in the $k$th peak. 
The outermost sum runs over all peaks, and the innermost sum runs over the $n_g$ Gaussian 
PDFs that parameterize the TTS distribution. The parameters $f_j$, $t_j^{\rm TTS}$, 
and $\sigma_j^{\rm TTS}$ are the fraction, position, and width of the $j$th Gaussian. 
The overall width is the sum in quadrature of 
$\sigma_{ck}$, $\sigma_j^{\rm TTS}$, and $\sigma^{\rm el}$, where the last term 
corresponds to the electronic resolution. This is estimated separately for 
each detected photon, as it depends on the amplitude of the electronic signal. 

The background PDF in Eq.~\ref{extlkh} is modeled as uniform for the time window 
$[t_1, t_2]$ in which the photons are measured:
\begin{eqnarray}
  B(c, t) = \frac{\epsilon_c}{\sum_{c'} \epsilon_{c'}} \cdot \frac{1}{t_2 - t_1}\,, 
  \label{bkg_distr}
\end{eqnarray}
where $\epsilon_c$ is the photon detection efficiency (PDE) of pixel $c$, and the sum 
in the denominator runs over all 512 pixels of a detector module. The PDE is a product 
of the quantum and collection efficiencies of the MCP-PMT, and the efficiency due to the 
pulse-height threshold. It is adjusted for the relative decrease of quantum efficiency 
due to tube aging. The quantum and collection efficiencies were measured on the bench
for each MCP-PMT before installation. The PDE's are also included in the computation 
of the expected signal yield $N_h$ in Eq.~\ref{extlkh} and the fractions $f_{ck}$ in 
Eq.~\ref{signalPDF} (due to dependence on pixel number~$c$).

The start signal for photon time measurements is given by the Belle~II first-level trigger. 
However, this trigger is not precise enough to identify the particular bunch crossing that
produced the measured event; hence, the bunch crossing is determined offline using fully 
reconstructed tracks~\cite{Staric:2023svu}.
The method relies on maximizing the sum of log likelihoods for a common time offset subtracted 
from the photon arrival times $t_i$ in Eq.~\ref{extlkh}. The particle identities used for this 
are determined from the specific ionization loss of the track ($dE/dx$) as measured in the 
central drift chamber and the silicon vertex detector. The result is rounded to the nearest 
bunch-crossing time and then used to correct the photon times~$t_i$. The efficiency for
finding the correct bunch-crossing is greater than~99\%.

%% file: operations.tex
\section{Operations}
\label{sec:operations}

\subsection{Firmware programming}

The TOP detector is prepared for data-taking in two steps. First, the 
front-end FPGAs are programmed with dedicated firmware. One USB JTAG 
programmer is fanned-out to all 64 front-end boardstacks in parallel through 
the FTSW system~\cite{FTSW}. The firmware programming is performed via 
Xilinx Zynq programming routines in the OpenOCD~\cite{OpenOCD} software package. 
OpenOCD allows one to change the JTAG clock frequency within a programming script;
we take advantage of this feature to optimize the clock frequencies for individual 
parts of the programming process. For example, all initialisation and handshaking 
is done at a relatively low clock frequency of 200~kHz, while data transfers into 
the Zynq SoC FPGA are performed at~10~MHz.
These settings keep the time needed to program all front-end boards to within 
about four minutes.
After this step, the firmware is configured via Belle2Link~\cite{B2readout},
which is the same architecture as used for data readout. This step was originally 
performed (until November 2021) using 16 COPPER boards~\cite{DAQsystem} (one per 
detector module), but it is now handled by two PCIe40 boards~\cite{Zhou:2020qed}. 
Configuring the firmware in all boardstacks requires about two minutes, with most 
of the time spent on pedestal acquisition and a trigger threshold scan.

\subsection{Data-taking}

There are two configuration settings used for data-taking. In ``GLOBAL'' mode, 
the TOP takes data together with all other detector systems using the common 
Belle II trigger~\cite{DAQsystem}. In ``LOCAL'' mode, the TOP takes data
autonomously. The latter mode is used to take calibration data, which is 
used to determine time base and channel $t^{}_0$ constants as described in 
Sections~\ref{sec:calibration_tbc} and~\ref{sec:calibration_channelt0}. 
In LOCAL mode, a calibration pulse is injected into each ASIC of a detector 
module, or else laser pulses are injected via optical fibers into the prism, 
and triggers are synchronized with these signals.

During GLOBAL mode running, if a boardstack malfunctions due to, e.g., a single-event
upset (SEU) as described in Section~\ref{sec:electronics:SEU}, it can be quickly 
excluded from the readout and recovered by power-cycling and reprogramming, all 
while the rest of the detector continues to take data. It is re-included at the 
next run stop. This leads to a temporary loss of one fourth of the photons in 
the affected module. As the rate of such boardstack malfunctions is only 2-4 
per day for all 64 boardstacks of the TOP detector, and a boardstack is excluded 
only until the next run stop, the impact on detector performance is minimal.

\subsection{Slow control}

The TOP run control, HV control, low voltage (LV) control, chiller
monitoring, and N$^{}_2$ gas monitoring systems operate within the Belle II DAQ software 
framework~\cite{belle2_slc}. These systems are collectively referred to as ``slow control.''
The software used include network communications, a PostgreSQL~\cite{PostgreSQL} 
database, EPICS~\cite{EPICS} archiver appliances, and a Graphical User Interface (GUI).

The communications between various hardware and software components over the network is 
achieved by an architecture called “Network Shared Memory 2” (NSM2), an advanced version 
of the NSM~\cite{NSM} architecture used by the Belle experiment. For each component, 
an NSM2 node reads the status and various parameters from the component, 
converts these to NSM2 format, and distributes them to the Belle II DAQ network. 
The PostgreSQL package is used for managing configuration parameters and logging information. 
All configuration parameters are loaded from a database, and logging information is stored 
and used for troubleshooting.
To store important detector monitoring variables, the EPICS archiver appliance~\cite{Archiver} 
is used. 
Parameters are provided in the format of EPICS Process Variables (PVs). The variables 
provided by NSM2 nodes are converted to EPICS PVs using NSM2CAD routines~\cite{Kim:2020blq}

\begin{figure}
\begin{center}
\includegraphics[width=0.55\textwidth]{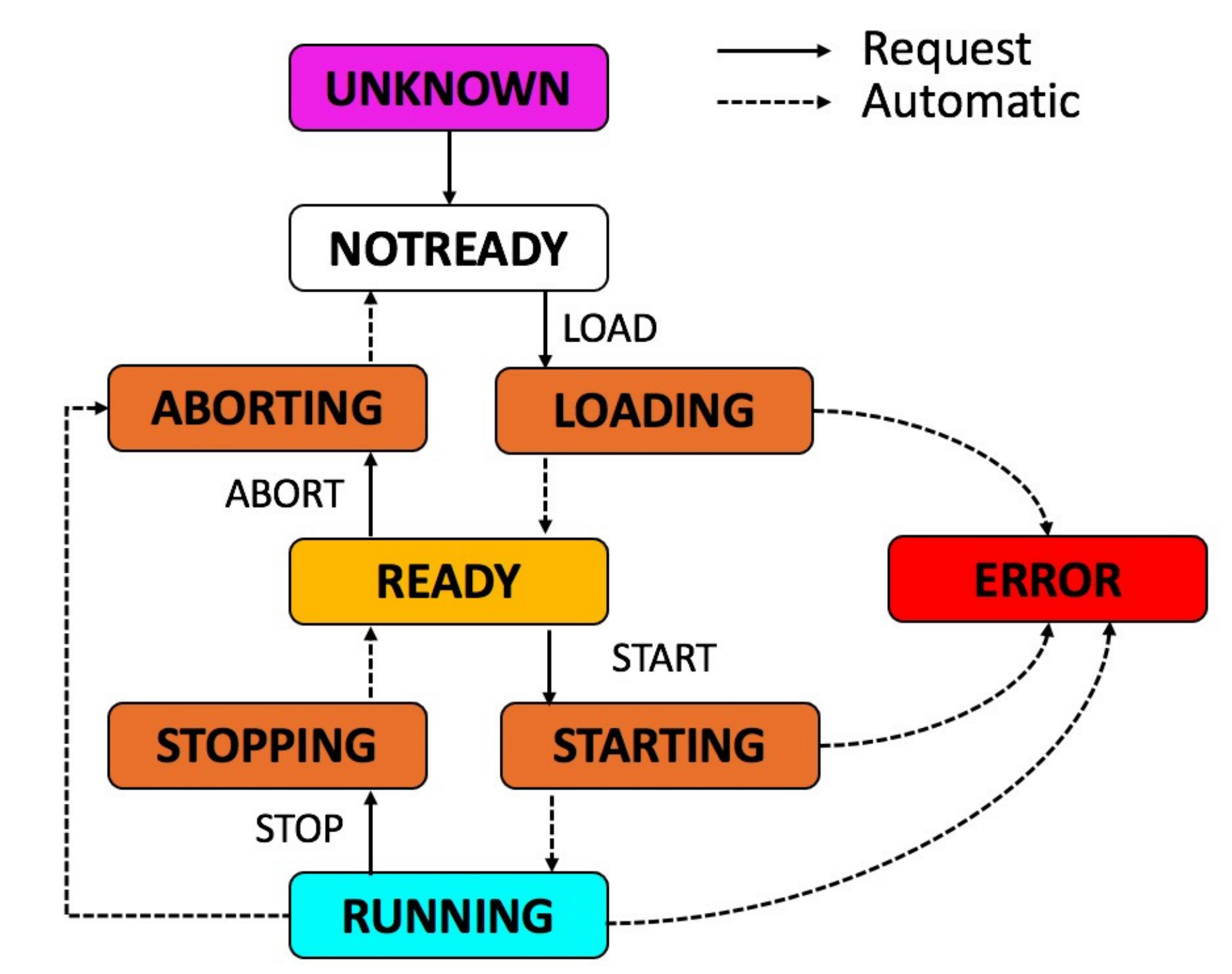}
\vskip0.10in
\caption{The Belle~II run control architecture.}
\label{fig:BelleII_RC}
\end{center}
\end{figure}

The TOP detector follows the Belle~II run control architecture shown in Fig.~\ref{fig:BelleII_RC}. 
The system is initially in the ``NOT\,READY'' state. The ``LOAD'' command loads configuration 
parameters and makes the system ``READY.'' The ``START'' command initiates data taking.
While in the ``RUNNING'' state, one can stop the run quickly if needed. 
If a fault condition arises,
data taking is stopped automatically with an ``ERROR'' state. 
If necessary, one can ``ABORT'' a run and re-load parameters.

\begin{figure}
\begin{center}
\includegraphics[width=0.62\textwidth]{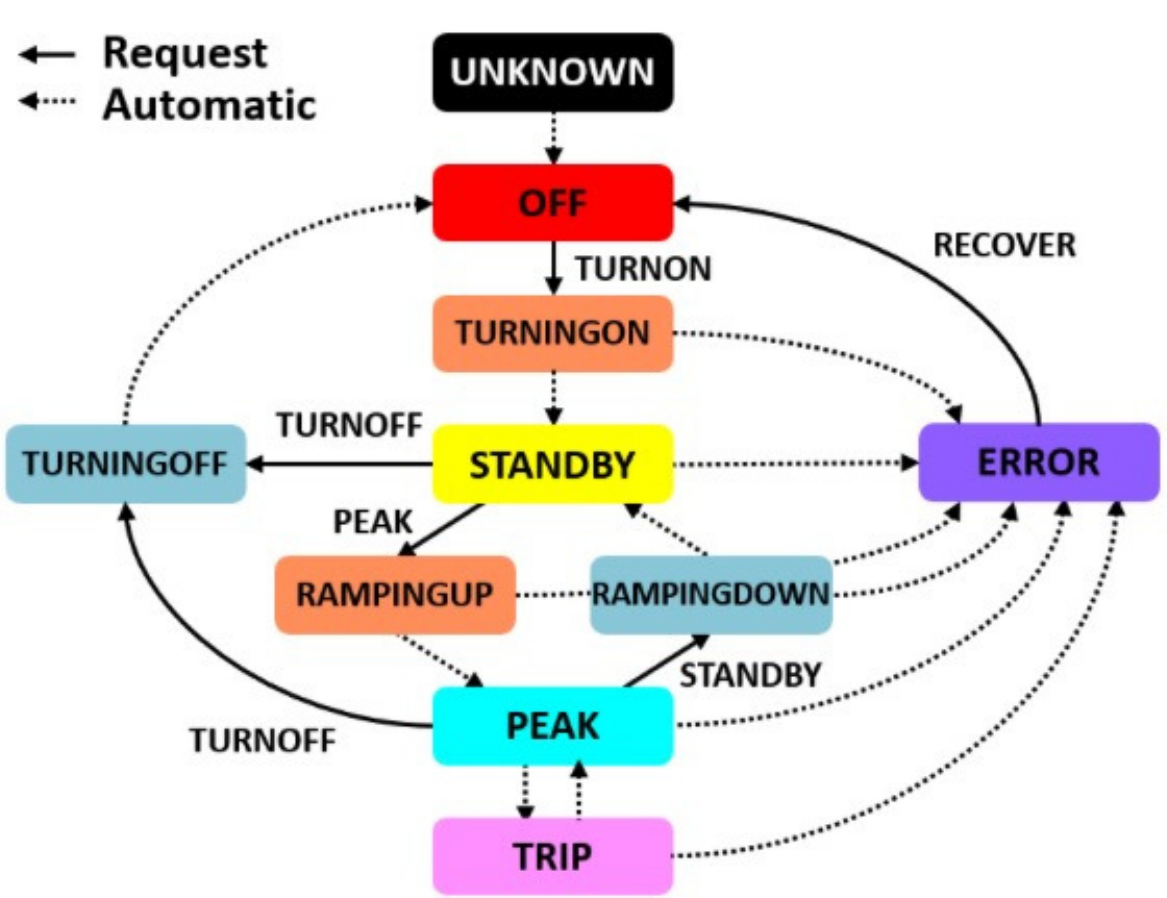}
\vskip0.10in
\caption{The Belle~II high voltage control architecture.
}
\label{fig:BelleII_HVcontrol}
\end{center}
\end{figure}

The TOP HV slow control follows the Belle~II HV control architecture shown in 
Fig.~\ref{fig:BelleII_HVcontrol}. 
Before a run starts, a ``PEAK'' command is issued and the high voltage is raised from 
``STANDBY'' to the nominal values used for data-taking.
These values vary PMT-by-PMT such that each PMT runs at the same nominal gain of $3\times 10^5$.
Trips of individual PMTs are rare and do not change the overall high voltage state.
The history of various parameters of the HV system, as well as those of the LV system, 
chiller, and N$^{}_2$ gas system, are archived with the EPICS~\cite{EPICS} archiver.

\subsection{User interface}

\begin{figure}
\begin{center}
\includegraphics[width=0.99\textwidth]{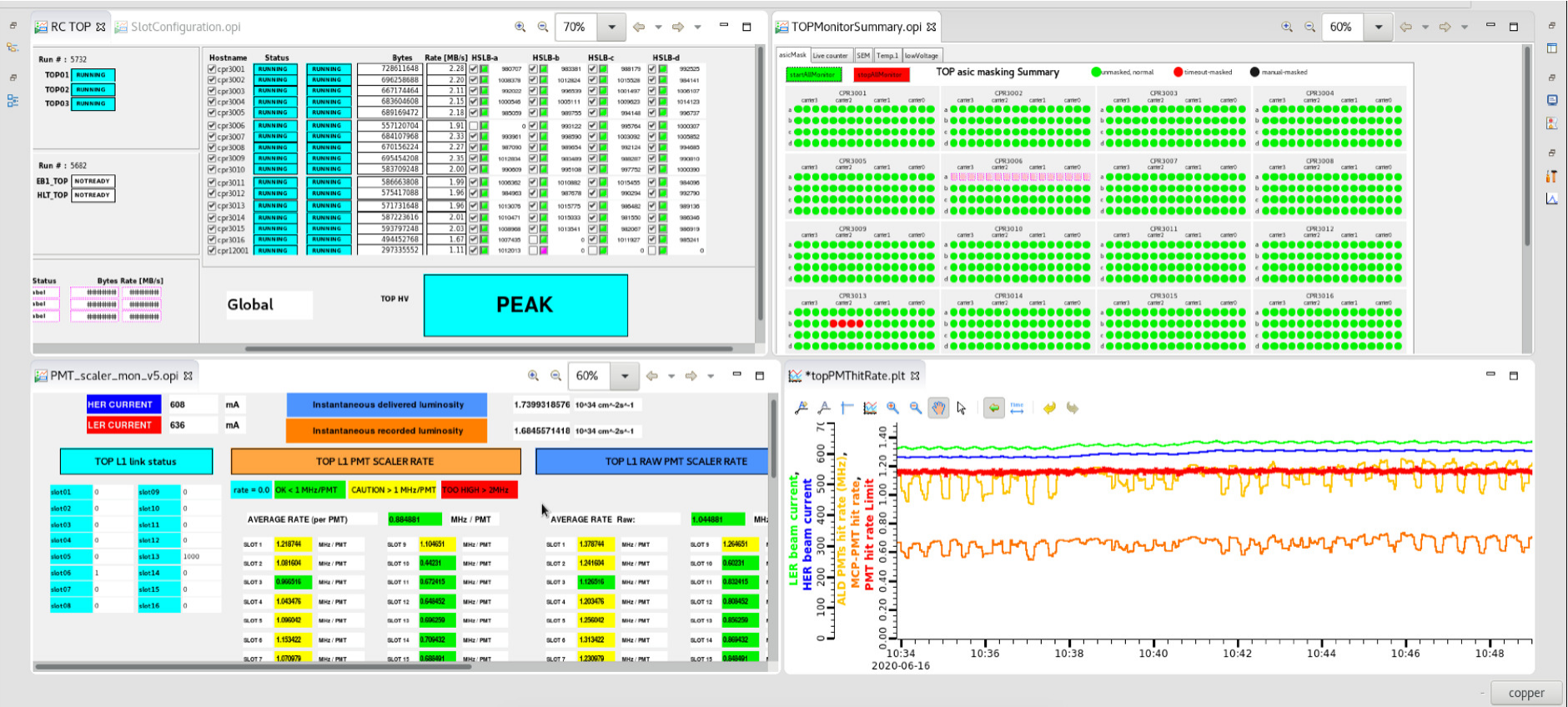}
\vskip0.15in
\caption{Run control and monitoring panels for the TOP. The top-left panel 
is for run control, with the HV status also displayed. The top-right panel 
shows the status of all ASICS in the carrier boards.
The bottom-left panel lists MCP-PMT hit rates, and the bottom-right panel displays
the total MCP-PMT hit rate versus time along with the beam currents. Separate panels 
for HV control, LV control, chiller, and N$^{}_2$ gas monitoring are not shown.}
\label{fig:TOP_panel}
\end{center}
\end{figure}

The main user interface is a CS-Studio~\cite{CSS} GUI displayed in the Belle~II control room 
or accessed remotely via VNC. A typical slow control and monitoring panel is shown in 
Fig.~\ref{fig:TOP_panel}. The panel provides access to run control settings and an overview 
of various parameters read out via the slow control interface. Of particular importance is 
the monitoring of PMT hit rates, which are driven by the level of beam-related backgrounds 
and affect the rate of MCP-PMT aging.
Additional specialized GUIs are used to operate the laser calibration system, 
to include or exclude specific boardstacks from data-taking, and for power cycling 
to clear error states. In addition to these GUIs, a Belle~II-wide Kibana 
system~\cite{Kibana} provides access to error messages and various statistics.

\subsection{Data Quality Monitoring}

\begin{figure}
\begin{center}
\includegraphics[width=0.80\textwidth]{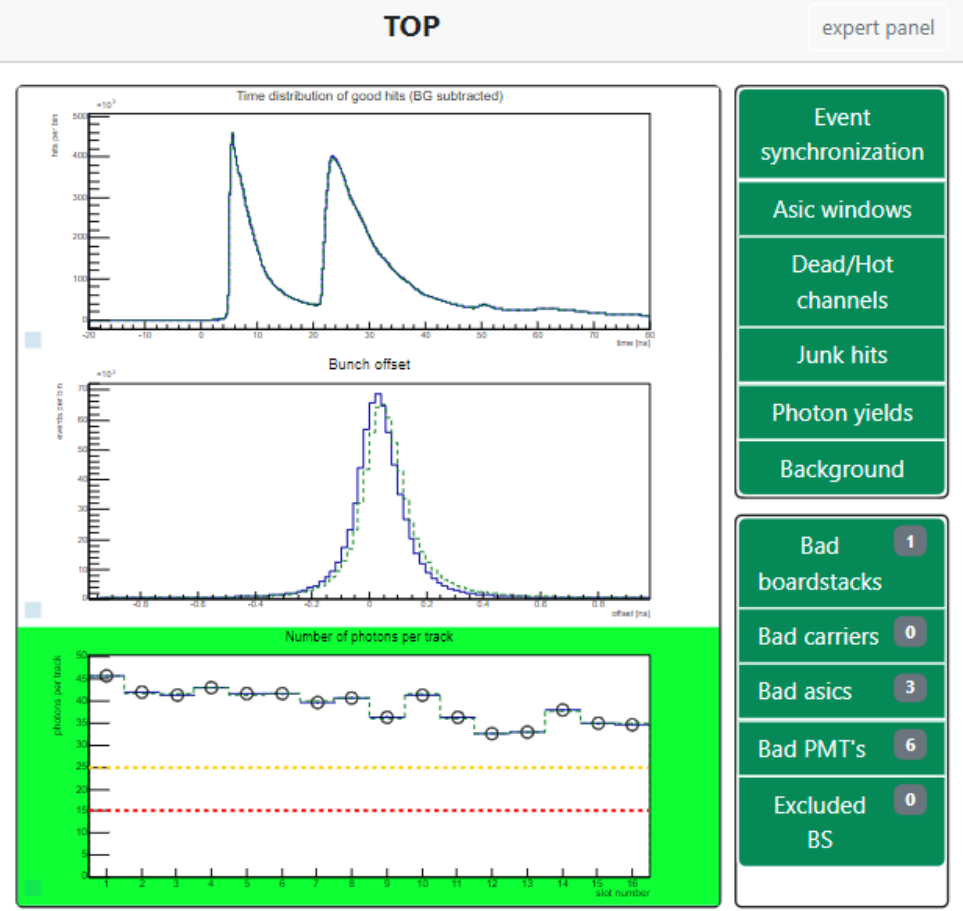}
\vskip0.20in
\caption{DQM summary panel for the TOP.
The top plot shows the time distribution of photons; 
the middle plot shows the time offset with respect to 
the bunch-crossing time; and the bottom plot shows the 
mean number of photons per event for each module. The right 
panel lists the number of inactive detector components.}
\label{fig:DQM_panel}
\end{center}
\end{figure}

Belle~II employs a central data quality monitoring (DQM) system~\cite{DQM}. This
system uses a subset of data to produce histograms in real time during running. The 
histograms are checked by subsystem experts on shift to identify potential detector 
problems. DQM plots for the TOP focus on the arrival times of hits, the number of hits per 
module per event, and the availability of channels for data-taking.
A typical DQM panel is shown in Fig.~\ref{fig:DQM_panel}. 
The distribution of arrival times can show problems with calibration constants, 
faulty boardstack components or PMTs, errors in firmware, or loss of synchronization 
between the clock distributed to boardstacks and the accelerator clock.
Single-event upsets and excessive hit rates can lead to the temporary loss of channels 
until the corresponding boardstack can be reset and reprogrammed. In our experience, 
this leads to an average loss of photon hits across the TOP detector of less than~1\%.

\subsection{Operational experience}

After the TOP modules were installed and the electronics cabled in the Spring of 2016, 
the Belle~II detector endcaps were closed and the TOP became inaccessible until a long shutdown
period of the experiment in 2022-2023. To date, the experiment has recorded almost 600~fb$^{-1}$ 
of data spread over several running periods. These running periods have provided us with valuable 
experience operating the detector.

Overall, the TOP detector has operated robustly, with few issues affecting its performance.
We have not observed any degradation in optical components or epoxies due to radiation damage.
We did observe a drop in quantum efficiency for conventional-type PMTs, which we attribute to 
the integrated charge arising from higher-than-expected beam backgrounds. However, 
during a long shutdown period of the experiment in 2022-2023, almost all conventional-type PMTs 
were replaced with ALD-type PMTs moved from other modules, and the ALD-type PMTs that were moved 
were replaced with new life-extended ALD-type PMTs. We have not observed any drop in quantum
efficiency for life-extended ALD-type PMTs.

Most issues that have arisen concern readout electronics, as discussed in Section~\ref{sec:electronics:SEU}.
With respect to the modules themselves, we observed some delamination of the glue joint between the 
expansion prism and the PEEK frame surrounding it (see Section~\ref{sec:optics:peek_frame} and 
Fig.~\ref{fig:peek-frame-gluing}). This delamination was observed in numerous modules and was not present 
when the prisms were glued to the bars, which took place several weeks after the PEEK frames were 
glued to the prisms. The delamination takes the form of elongated air gaps (bubbles) embedded within 
the epoxy. Fortunately, a large fraction of the glue joint can be inspected and monitored in situ using 
the CCD cameras and LEDs installed in the prism enclosure (see Fig.~\ref{fig:mechanics:enclosure}).
After an initial 2-3 year period of forming, the air gaps appear to have stopped growing, and fewer
new ones have appeared. The total area of the glue joint affected is relatively small -- at the 10\%
level. We have not observed any mechanical consequences of this delamination. We attribute the
delamination to a slight difference in the coefficients of thermal expansion between PEEK and 
Corning 7980 synthetic quartz, and the fact the the quartz surface is very highly polished, 
i.e., there is minimal surface roughness for the epoxy to bind to. 

We have observed a second type of delamination, this one between the glass windows of MCP-PMTs
and the wavelength cut filter of a PMT module. This is caused by a slight rotation of the MCP-PMTs 
in the magnetic field due to residual magnetization of the PMT housing. The phenomenon was observed 
with the CCD cameras and studied offline; the fraction of surface area affected is approximately~7\%. 
We have not observed any degradation in performance of the TOP detector due to these delaminations.  
During the long shutdown of 2022-2023, the new life-extended ALD-type PMTs that were installed 
had a special shim added to the PMT module to prevent such PMT rotation.

To date, we have not observed any leaks in the  gas seals of the modules. To overcome possible failures 
in the seals, we have the ability to increase the nitrogen flow rate by more than a factor of two; however, 
we have not yet needed to make such an adjustment. To assess the gas-tightness of the detector, the
dew point of the outlet gas of each module is monitored. This dew point is relatively stable, 
closely tracking the humidity in the experimental hall. It fluctuates mainly between 
$-40^\circ$~C and $-60^\circ$~C, depending on the season. The variation in dew points among
different modules is small, at the level of $5^\circ$~C.

The alignment procedure for the TOP detector uses $e^+e^-\!\rightarrow\!\mu^+\mu^-$ events and is described in 
Ref.~\cite{Staric:2017lqu}. For each module, small displacements $(\Delta x,\,\Delta y,\,\Delta z)$
from an $(x^{}_0,\,y^{}_0,\,z^{}_0)$ reference position are determined, and small rotation angles
$(\alpha,\,\beta,\,\gamma)$ about the $x,\,y,\,z$ axes from a reference orientation
are determined.  We have checked the stability of the alignment by comparing the results 
for $\Delta x,\,\Delta y,\,\Delta z,\,\alpha,\,\beta,$ and $\gamma$ for three different
alignments performed over a three-year period (May 2019 -- June, 2022). The changes
in displacement are at the level of 1~mm for all 16 modules, and the changes in angles 
are at the level of 1~mrad. Such values are well below the level that would impact 
the detector performance and indicate good mechanical stability.

%% file: performance.tex
\section{Performance}
\label{sec:performance}

\subsection{Particle likelihoods}

Information from the TOP detector is used to determine a 
likelihood ${\cal L}_h^{\rm TOP}$ for a charged-particle hypothesis 
$h = e, \mu, \pi, K, p$, and $d$. For Belle~II data analyses, this likelihood 
is combined with corresponding likelihoods based on information from other 
subdetectors such as the vertex detector, the central tracker, the aerogel 
RICH detector, the electromagnetic calorimeter, and the muon detector system, 
i.e., ${\cal L}_{i} = \prod_{\rm detector} {\cal L}_{i}^{\rm detector}$.
While most of the disciminating power for electrons is due to the
calorimeter, and most of that for muons is due to the muon detector, 
almost all the discriminating power for pions, kaons, and protons 
in the barrel region is due to the TOP.
As much of the physics program of Belle~II focuses on rare and forbidden
decays of $B$ and $D$ mesons and $\tau$ leptons to final states containing 
kaons and pions, it is especially important to distinguish $K^\pm$ tracks 
from $\pi^\pm$ tracks.

We study the performance of the TOP, i.e., its 
ability to discriminate charged kaons from charged pions, using 
$D^{\ast+}\to D^{0}(\to K^{-}\pi^{+})\pi^{+}$ 
decays.\footnote{The inclusion of the charge-conjugate decay mode is implied 
unless noted otherwise.} The low-momentum $\pi^\pm$ originating directly from 
the $D^{*\pm}$, referred to as the ``slow'' pion, is used to identify the
flavor of the $D^{0}$ or $\overline{D}{}^{\,0}$. This in turn identifies 
which of the $D^0$ daughter tracks is the kaon and which is the pion, 
i.e., the charge of the $K$ ($\pi$) must be opposite (match) that of 
the slow pion. To discriminate pions from kaons using TOP information,
we calculate TOP-based likelihood ratios 
\begin{equation}
  {\cal R}^{\rm TOP}_{K} \equiv \frac{{\cal L}^{\rm TOP}_{K}}
{{\cal L}^{\rm TOP}_{K} + {\cal L}^{\rm TOP}_{\pi}}
\hspace{1cm}
  {\cal R}^{\rm TOP}_{\pi} \equiv \frac{{\cal L}^{\rm TOP}_{\pi}}
{{\cal L}^{\rm TOP}_{K} + {\cal L}^{\rm TOP}_{\pi}}\ =\ 1 -   {\cal R}^{\rm TOP}_{K}\,.
  \label{eq:LR}
\end{equation}
A track having ${\cal R}^{\rm TOP}_K>\alpha$, where $\alpha\in [0,1]$, is identified 
as a kaon; otherwise, it is identified as a pion. With this criteria, we define a 
$K$ or $\pi$ identification efficiency as
\begin{equation}
  \varepsilon^{}_{K (\pi)} \ \equiv\ 
\frac{{\rm number\ of\ }K(\pi) {\rm\ tracks\ identified\ as\ } K (\pi)}
     {{\rm number\ of\ }K(\pi) {\rm\ tracks}}\,,
  \label{eq:keff}
\end{equation}
and a $K$ or $\pi$ misidentification rate as
\begin{equation}
 f^{}_{K (\pi)} \ \equiv\ 
 \frac{{\rm number\ of\ }K(\pi) {\rm\ tracks\ misidentified\ as\ } \pi (K)}
      {{\rm number\ of\ }K(\pi) {\rm\ tracks}}\,.
  \label{eq:pifake}
\end{equation}
By definition, $\varepsilon^{}_K + f^{}_K = 1$ and $\varepsilon^{}_\pi + f^{}_\pi = 1$.

\subsection{Data set and Analysis Method}

\begin{figure}[htb]
  \centering
  \includegraphics[width=0.65\linewidth]{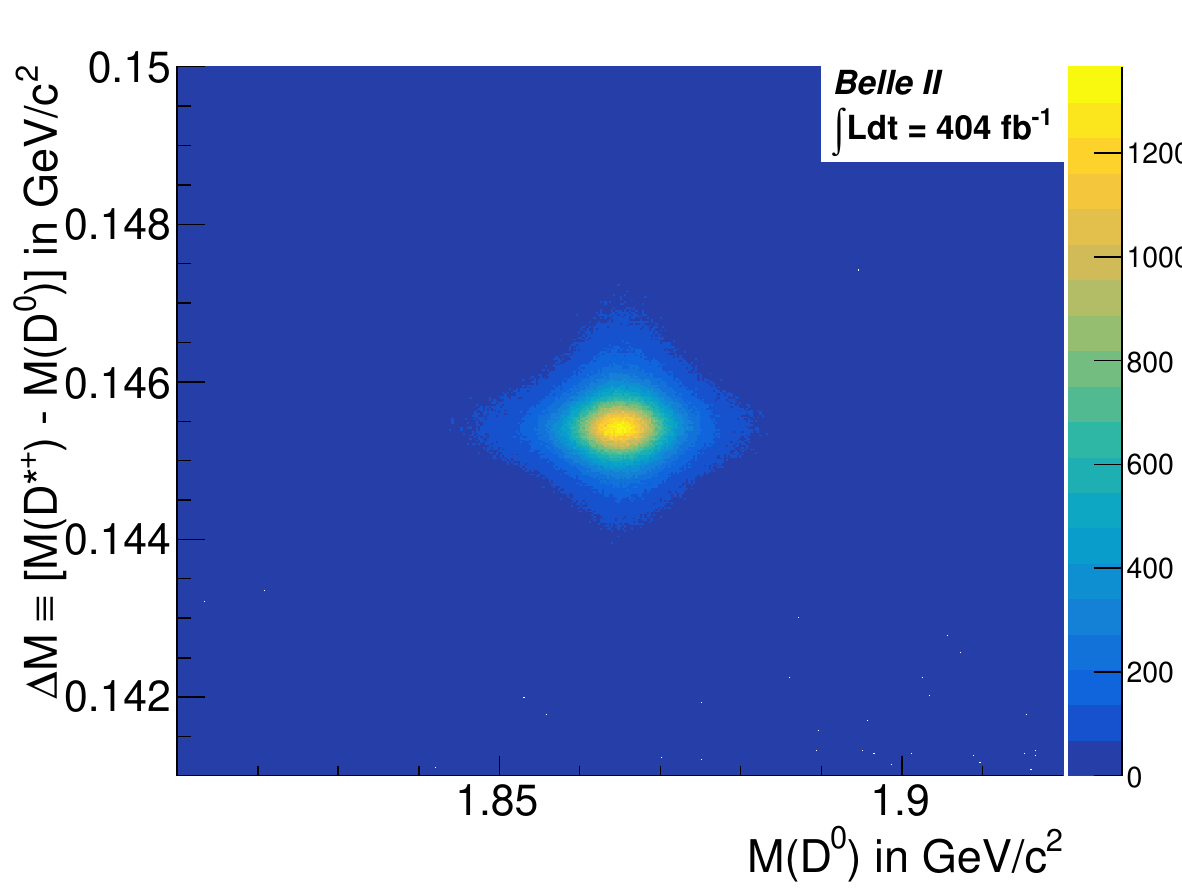}
  \caption{The mass difference $\Delta M = M(D^{\ast +})-M(D^{0})$ versus $M(D^{0})$ 
   for $D^{\ast+}\to D^{0}(\to K^{-}\pi^{+})\pi^{+}$ candidates in data.  }
  \label{fig:dmmd0}
\end{figure}

We have studied the TOP performance using 404~\invfb of $e^+ e^-$ collision data. 
We select tracks that originate near the $e^+e^-$ interaction point
by requiring impact parameters $\delta r < 2$~cm in the $x$-$y$ plane 
(transverse to the positron beam), where $r^2 = x^2 + y^2$, 
and $|\delta z| < 4$ cm along the $z$ axis (parallel to the positron beam). 
To ensure that the tracks are well-measured, we require that they have 
at least 20 hits in the central tracking detector. Two oppositely charged 
tracks are combined to reconstruct a $D^{0}$ meson candidate, and the 
two-body invariant mass $M(D^{0})$ is calculated using kaon and pion mass 
hypotheses for the tracks. Another track is added to reconstruct a $D^{\ast +}$ 
candidate, and the three-body invariant mass 
$M(D^{\ast +})$ is calculated using a pion mass hypothesis for the third track.
We require that the momentum of the $D^{\ast +}$ candidate in the $e^+e^-$ 
center-of-mass frame 
be greater than 2.5~$\gevc$; this value selects $D^{\ast +}$ mesons originating 
from $e^+e^- \to c\bar{c}$ reactions. We define the mass difference
$\Delta M \equiv M(D^{\ast +}) -  M(D^{0})$, which for $D^{\ast +}\to D^0\pi^+$ 
decays peaks at 145~MeV/$c^2$. 
A two-dimensional distribution of $\Delta M$ versus $M(D^{0})$ is shown 
in Fig.~\ref{fig:dmmd0} for $D^{\ast+}\to D^{0}(\to K^{-}\pi^{+})\pi^{+}$ 
candidates; the distribution shows a clear peak centered at 
$M(D^0)\!=\!1865$~MeV/$c^2$ and $\Delta M\!=\!145$~MeV/$c^2$.

\begin{figure}[htb]
\centering
\hbox{
\hskip0.20in
  \includegraphics[width=0.42\linewidth]{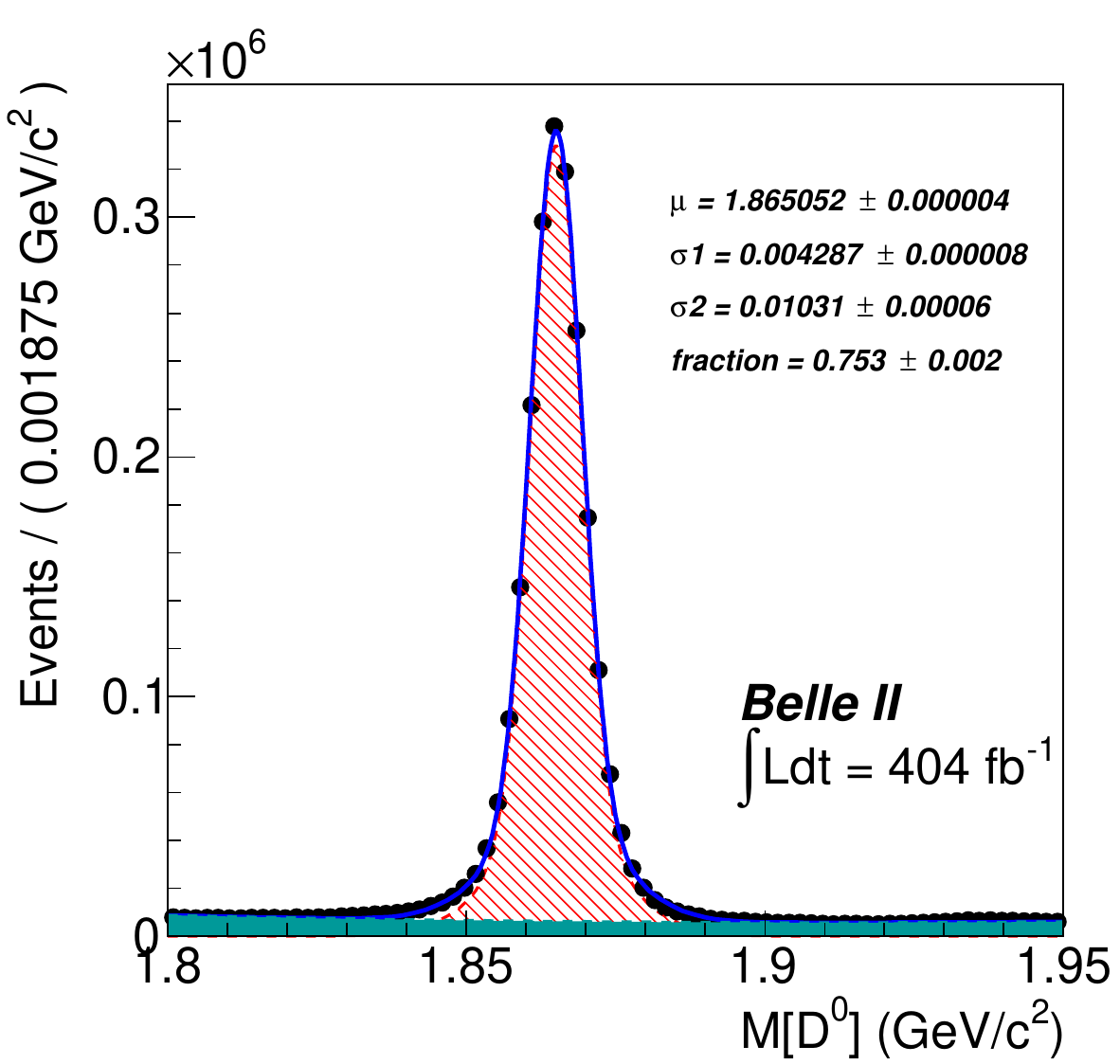}
\hskip0.20in
  \includegraphics[width=0.42\linewidth]{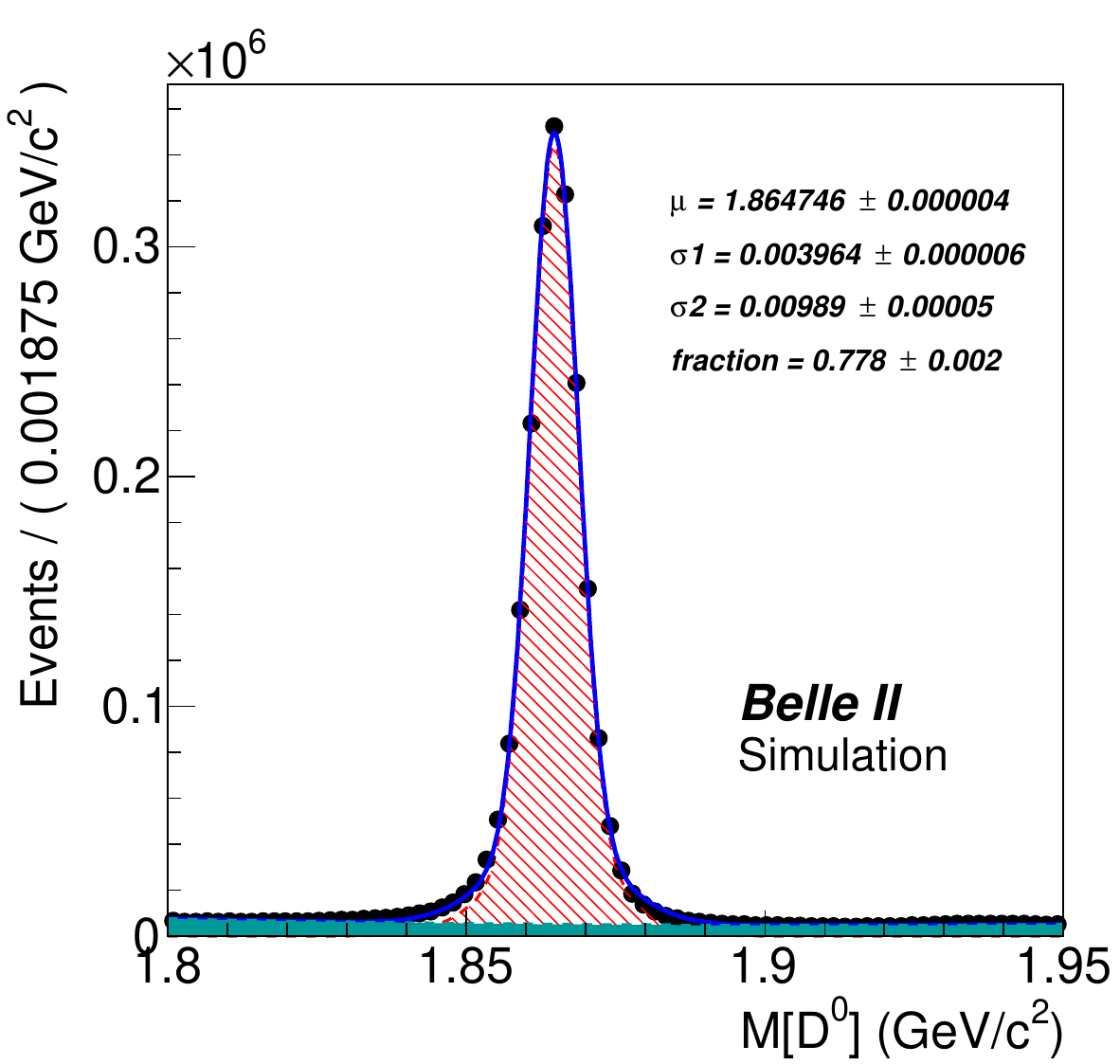}
}
  \caption{$M(D^{0})$ distributions for events satisfying 
  $|\Delta M - 145.4|< 1.5~\mevcc$ for data~(left) and MC simulation~(right). 
   The red hatched area corresponds to the fitted signal, 
   and the blue solid area corresponds to the fitted background.
   The two distributions are in reasonable agreement. 
   For the signal component, the means, standard deviations of the 
   two fitted Gaussians, and fraction for the first Gaussian are listed. }
  \label{fig:md0}
\end{figure}

The $M(D^{0})$ distribution for events satisfying $|\Delta M - 145.4|< 1.5~\mevcc$ 
is shown in Fig.~\ref{fig:md0} for both data and MC-simulated events.
To determine the $D^{\ast+}\to D^{0}(\to K^{-}\pi^{+})\pi^{+}$ signal yield, 
we fit this distribution with the sum of two Gaussian functions 
having a common mean to describe the signal component, and a second-order 
polynomial to describe the background component. We fit event samples
in which the $K$ or $\pi$ track is required to satisfy a kaon identification 
criterion; the fitted signal yield corresponds to either the number of 
correctly identified $K$ tracks in Eq.~(\ref{eq:keff}) or the number of 
incorrectly identified $\pi$ tracks in Eq.~(\ref{eq:pifake}).

For these studies, we also fit the $M(D^{0})$ distributions 
(for events satisfying $|\Delta M - 145.4|< 1.5~\mevcc$) in bins of momentum 
and polar angles. 
The $K^-$ and $\pi^+$ tracks from $D^{\ast+}\to D^{0}(\to K^{-}\pi^{+})\pi^{+}$ 
decays have a broad momentum distribution peaking around 1--2~GeV/$c$ and 
tapering off around 5~GeV/$c$. The polar angle distributions of $K^-$ and 
$\pi^+$ tracks favor the forward region of the detector, i.e., $\cos\theta\simge\!0.7$. 
The distributions of momentum and polar angle for both kaons and pions 
show good agreement between data and MC simulation.

\subsection{Likelihood ratio distribution}

\begin{figure}[!htb]
  \centering
  \includegraphics[width=0.60\linewidth]{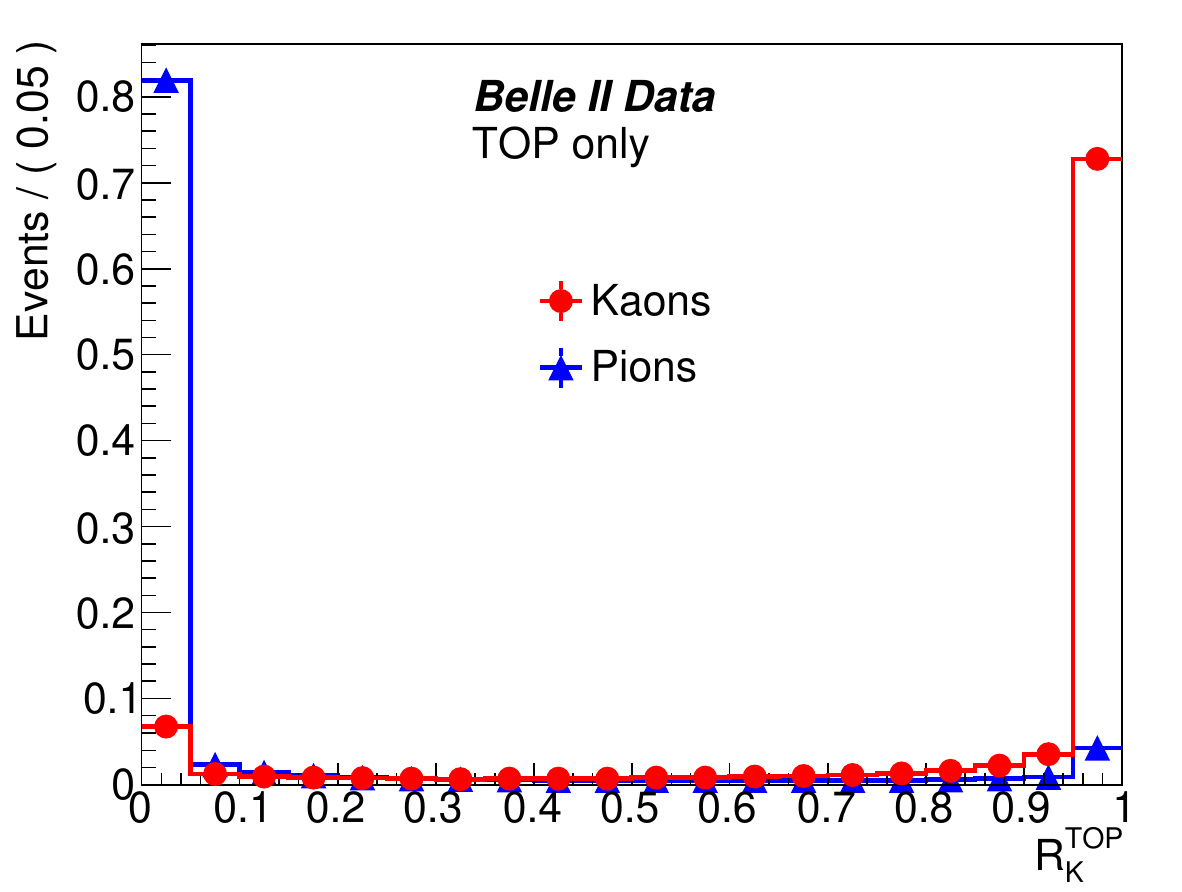}
  \caption{Distribution of the likelihood ratio ${\cal R}^{\rm TOP}_{K}$ for 
    kaon tracks~(circles in red) and pion tracks~(triangles in blue). }
  \label{fig:rkpi}
\end{figure}

The ${\cal R}^{\rm TOP}_{K}$ distribution for kaons and pions is obtained 
using the $_{s}{\cal P}lot$ technique~\cite{PIVK2005356} to remove backgrounds. 
For kaon tracks, the ${\cal R}^{\rm TOP}_{K}$ value should tend towards one, 
whereas for pion tracks ${\cal R}^{\rm TOP}_{K}$ should tend towards zero. 
The ${\cal R}^{\rm TOP}_{K}$ distributions for kaon and pion tracks 
in data are shown in Fig.~\ref{fig:rkpi}; the clear distinction 
seen between kaons and pions is consistent with expectations.

\subsection{Efficiencies and Misidentification rates}

\begin{figure}[!ht]
   \centering
   \includegraphics[width=0.65\linewidth]{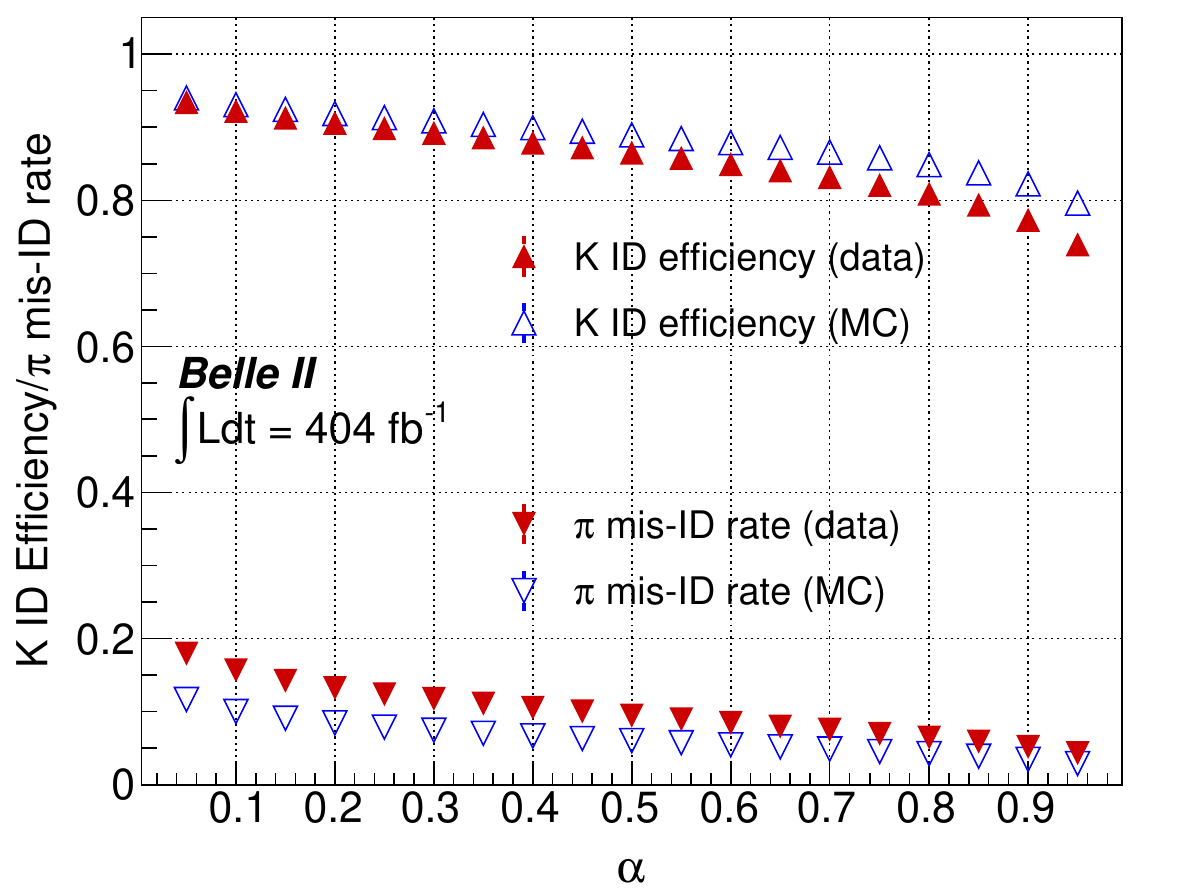}
   \caption{Kaon efficiency $\varepsilon^{}_K$ and pion misidentification rate 
   $f^{}_\pi$ as a function of $\alpha$, for $K$ and $\pi$ tracks satisfying 
   ${\cal R}^{\rm TOP}_{K} > \alpha$. Closed triangles (red) correspond to data, 
   and open triangles (blue) correspond to MC simulation.}
   \label{fig:pid1}
 \end{figure}

\begin{figure}[htb]
\centering
\includegraphics[width=0.65\linewidth]{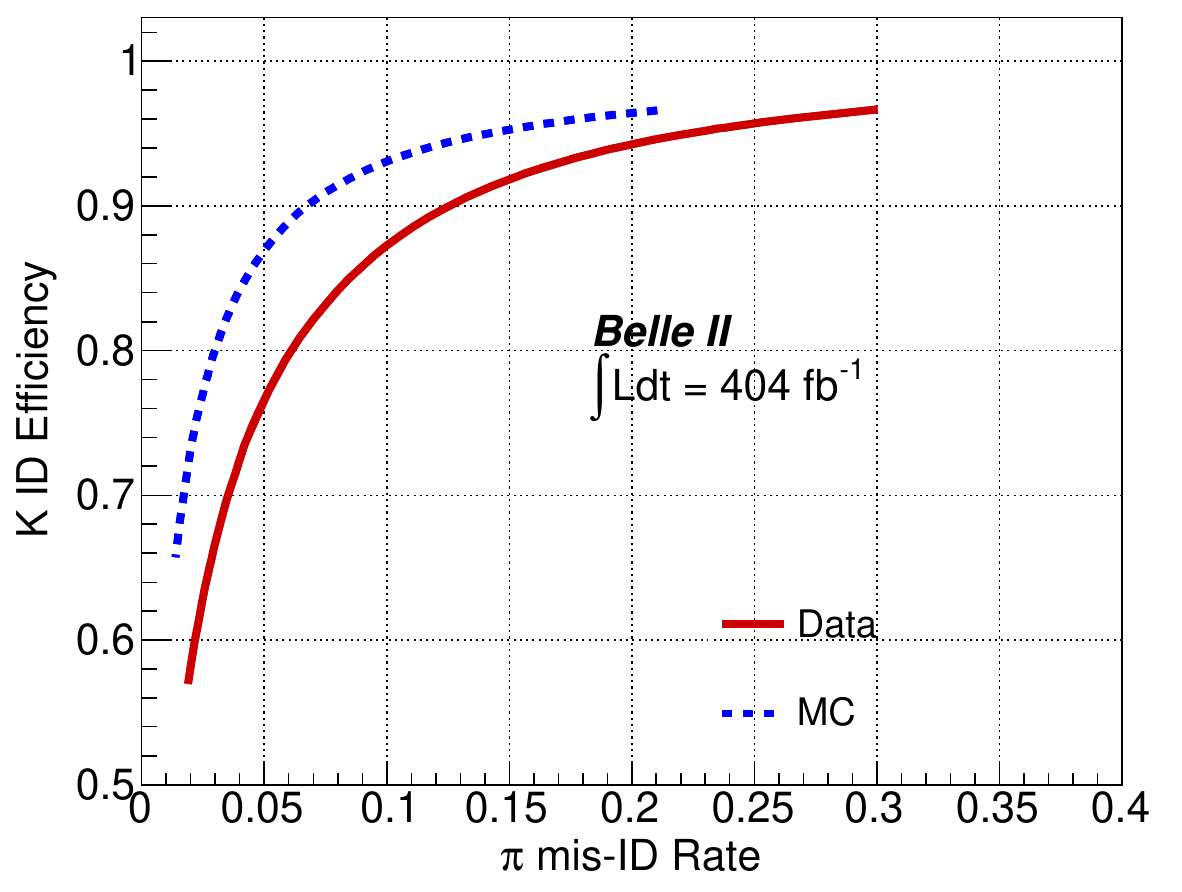}
\caption{Kaon efficiency $\varepsilon^{}_K$ versus pion misidentification rate $f^{}_\pi$.
   The solid contour (red) corresponds to data, and the dashed contour (blue) 
   corresponds to MC simulation. The difference between the two is under study;
   some of the difference is attributed to beam-related backgrounds and small 
   instrumental effects that are not fully simulated.}
\label{fig:roc}
\end{figure}

The kaon and pion identification efficiencies and misidentification rates
are calculated according to Eqs.~(\ref{eq:keff}) and~(\ref{eq:pifake}). 
To assess the TOP performance, we apply the criterion 
${\cal R}^{\rm TOP}_{K} > \alpha$ to kaon and pion tracks and calculate 
$\varepsilon^{}_K$ and $f^{}_\pi$. A plot of $\varepsilon^{}_K$ and $f^{}_\pi$ 
as a function of $\alpha$ is shown in Fig.~\ref{fig:pid1} for both data 
and MC simulation. There is reasonable agreement between the two, 
with the data exhibiting slightly lower efficiency and slightly 
higher misidentification rates.
These differences are currently under study. Some of the difference
is attributed to beam-related backgrounds and small instrumental 
effects that are not fully simulated, e.g., the dependence of
timing resolution on signal pulse-height, and the precision of 
track extrapolation from the Belle~II central tracking chamber.
A plot of $\varepsilon^{}_K$ versus $f^{}_\pi$ is shown in 
Fig.~\ref{fig:roc}; the contours plotted are traced out 
as $\alpha$ decreases from high values to low values. 

\begin{figure}[!ht]
   \centering
\hbox{
   \includegraphics[width=0.50\linewidth]{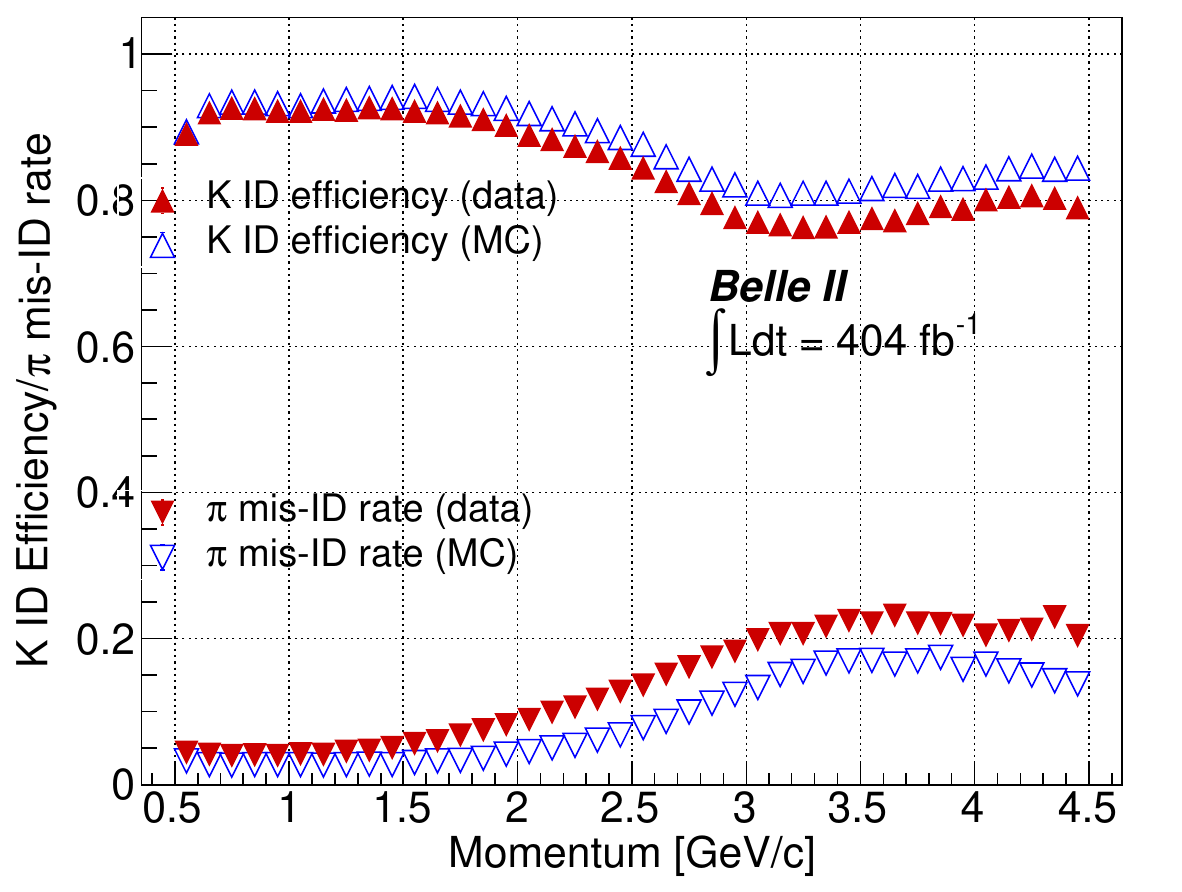}
    \includegraphics[width=0.50\linewidth]{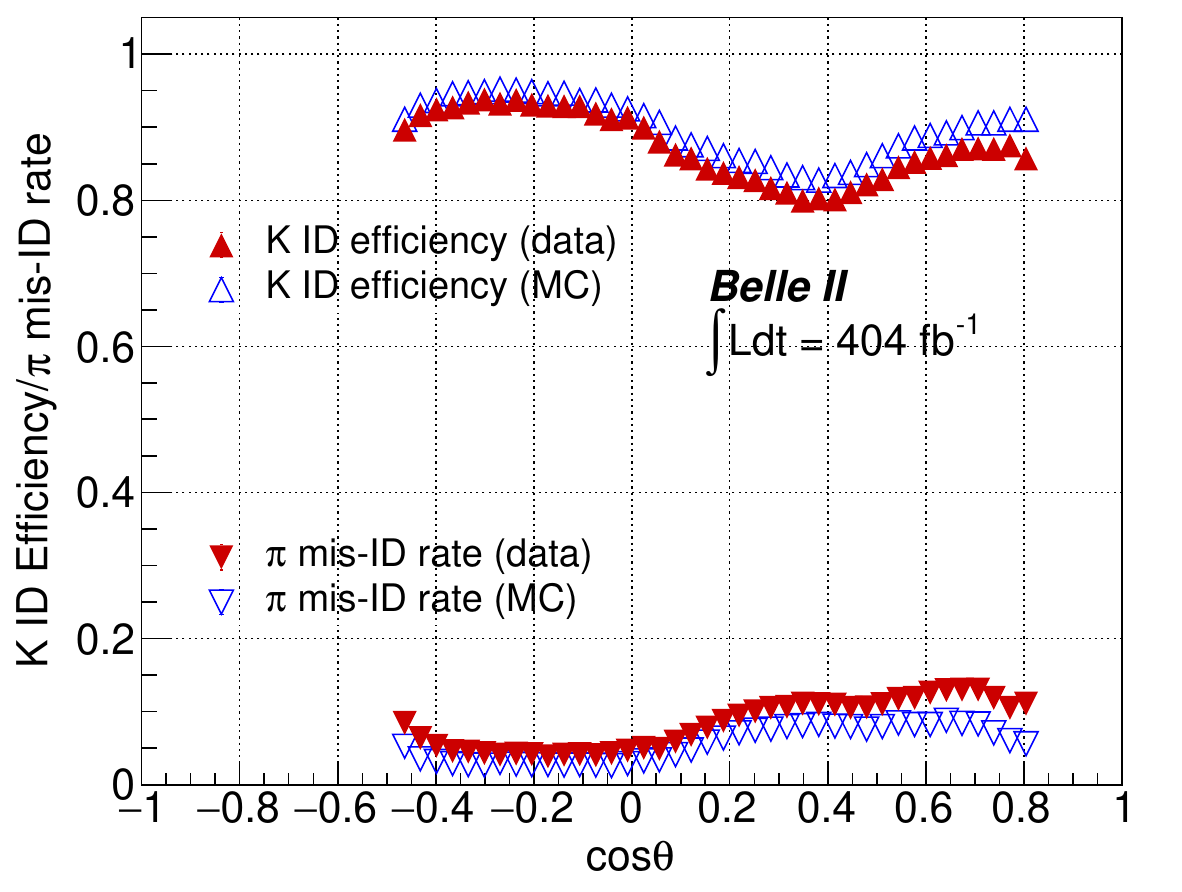}
}
   \caption{Kaon efficiency $\varepsilon^{}_K$ and pion misidentification rate 
   $f^{}_\pi$ as a function of track momentum~(left) and polar angle~(right), 
   for tracks satisfying ${\cal R}^{\rm TOP}_{K} > 0.5$. 
   Closed triangles (red) correspond to data, and open triangles (blue) 
   correspond to MC simulation.}
   \label{fig:pid_R.gt.0.5}
 \end{figure}

We study the performance further by requiring ${\cal R}^{\rm TOP}_{K} > 0.5$
and calculating $\varepsilon^{}_K$ and $f^{}_{\pi}$ as a function of 
track momentum and polar angle with respect to the $+z$ axis. 
The results are shown in Fig.~\ref{fig:pid_R.gt.0.5} for both
data and MC simulation; we observe reasonable agreement between 
the two. There is a drop in $\varepsilon^{}_K$ and rise in $f^{}_{\pi}$
near $\cos\theta\approx 0.35$,
as at this polar angle the number of photons that undergo total internal 
reflection is a minimum. This sensistivity arises due to chromatic dispersion 
(the variation of \cherenkov\ angle with wavelength), which smears the 
likelihood PDFs for kaon and pion hypotheses such that the discriminating 
power between them is reduced for smaller photon yields.
The effect is also visible in the plot of efficiency vs.\ momentum
at $p\approx 3.3$~GeV/$c$ as, for tracks produced in $e^+e^-$
collisions with the beam energies of Belle~II, a track 
momentum near 3.3~GeV/$c$ roughly corresponds 
to a polar angle near $\cos\theta=0.4$.

\begin{figure}[!ht]
   \centering
   \includegraphics[width=0.49\linewidth]{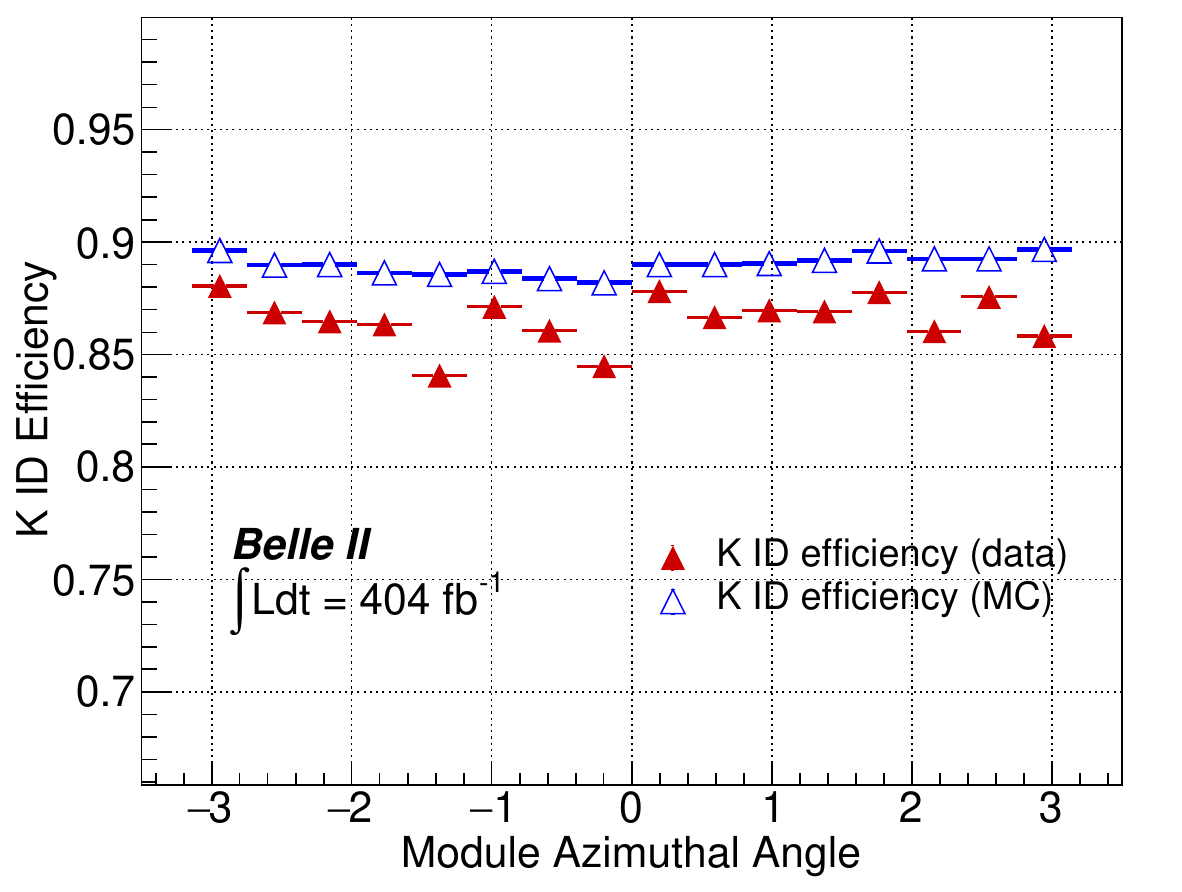}
   \includegraphics[width=0.49\linewidth]{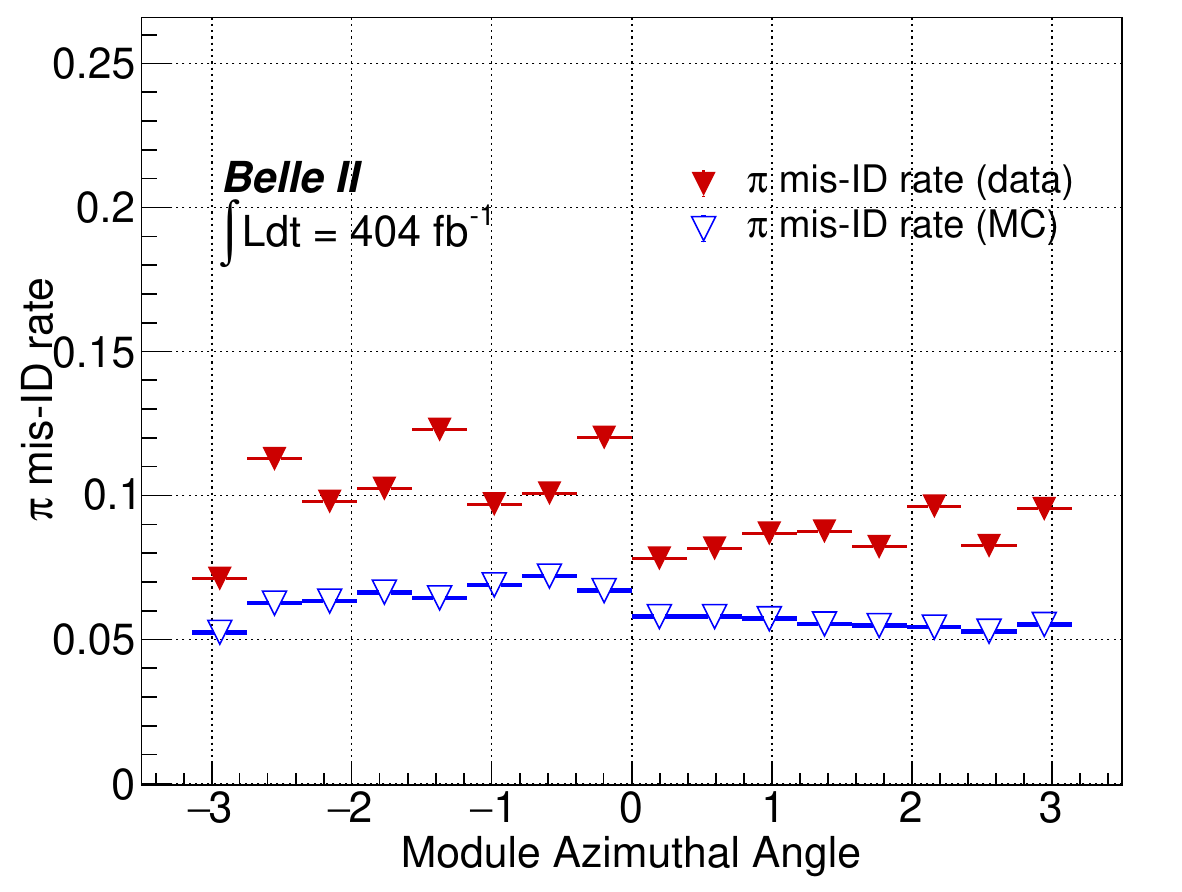}
   \caption{Kaon efficiency $\varepsilon^{}_K$ (left) and pion misidentification 
   rate $f^{}_\pi$ (right) for the 16 detector modules.
   Closed triangles (red) correspond to data, and open triangles (blue) correspond 
   to MC simulation. The modules at $\phi\!=-1.3$ and $\phi\!=-0.2$ show worse performance 
    due to hardware issues that reduced the active channel count by~25\%; the problematic boards
    have since been exchanged. For this data, modules at $\phi\!<\!0$ were instrumented  
   with conventional-type MCP-PMTs, while those at $\phi\!>\!0$ were instrumented 
   with ALD-type MCP-PMTs (see Sections~\ref{sbsctn:pmt_lifetime} and 
   \ref{sbsctn:pmt_massproduction}). }
   \label{fig:pid_phi}
 \end{figure}

Finally, we study the performance of individual detector modules. 
The results for $\varepsilon^{}_K$ and $f^{}_\pi$ are plotted in 
Fig.~\ref{fig:pid_phi}. There is a systematic difference in 
$\varepsilon^{}_K$ and $f^{}_\pi$ between modules at $\phi<0$ and 
those at $\phi>0$ due to different types  of MCP-PMTs used: 
modules at $\phi>0$ were instrumented mainly with conventional-type
MCP-PMTs, while those at $\phi<0$ were instrumented mainly with ALD-type MCP-PMTs
(see Sections~\ref{sbsctn:pmt_lifetime} and \ref{sbsctn:pmt_massproduction}).
During a long shutdown in 2022-2023, almost all conventional-type PMTs were 
replaced with ALD-type PMTs originally at $\phi > 0$, and the region 
$\phi > 0$ was instrumented with new life-extended ALD-type PMTs.
The modules at $\phi\!=-1.3$ and $\phi\!=-0.2$ show worse performance 
due to hardware issues that resulted in turning off one of the boardstacks
in each; this reduced the active channel count by~25\%. The problematic boards
have since been replaced.
The difference observed between data and MC simulation in Fig.~\ref{fig:pid_phi} 
is under study.

%% file: conclusion.tex
\section{Conclusion}
\label{sec:conclusion}

The TOP detector with imaging provides particle identification of charged
hadrons ($\pi^\pm$, $K^\pm$, $p$, $d$) in the barrel region of the Belle~II 
detector and thus plays a crucial role in the physics program of the 
experiment. The detector required four years to construct and another 
year to commission. Since then, it has been fully operational and collecting 
data continuously. The results reported here were obtained from the first 
400~fb$^{-1}$ of data recorded by Belle~II. Almost all Belle~II physics 
measurements to date have used information from the TOP detector 
as an important part of the data analysis. Recently, a 1.5-year shutdown 
period of the experiment and the SuperKEKB collider was completed, during which 
most conventional-type PMTs were replaced with life-extended ALD-type PMTs, 
and several front-end electronics boards with fault conditions were replaced
with new boards.

The detector has thus far performed as designed: 
more than 99\% of channels are fully operational, pion and kaon identification 
efficiencies are greater than 90\%, and kaon and pion misidentification rates 
are approximately 10\% over the momentum ranges of interest. This performance 
has been achieved and maintained despite steadily increasing beam currents and 
luminosity of the SuperKEKB collider, and correspondingly increasing beam backgrounds. 
The operation and maintenance of the TOP detector have been mostly trouble-free. 
The alignment and calibration of the detector have been straightforward. 
The reconstruction algorithm has worked consistently and in accordance 
with expectations based on MC simulation. Sporadic read-out problems 
from a handful of front-end electronics boards have been addressed and 
have had minimal impact on the recorded data. 

Over the next several years, it is planned for SuperKEKB to 
significantly increase its luminosity. Correspondingly, the event rate and 
the level of beam-related backgrounds are expected to increase substantially. 
The TOP detector has been designed to work in such a high-rate environment, 
and we anticipate utilizing its full capabilities under such very-high-luminosity 
running conditions.

%% file: acknowledgements.tex
\section{Acknowledgements}
\label{sec:acknowledgements}

For the construction and operation of the Belle~II TOP detector, we are grateful for 
strong support from 
the Department of Atomic Energy of India, 
SERB grant SRG/2022/001608;
the Istituto Nazionale di Fisica Nucleare, Research Grant BELLE2;
the Japan Society for the Promotion of Science, Grant-in-Aid for Scientific Research Grants
No.~18H05226,
No.~18GS0206,
No.~18071003,
No.~20H05858,
No.~21105005,
No.~22K21347,
No.~23H05433, and
No.~26220706, and
the Ministry of Education, Culture, Sports, Science, and Technology (MEXT) of Japan;  
the Slovenian Research Agency, Grants No.~J1-9124 and No.~P1-0135;
the U.S. National Science Foundation, Research Grant No.~PHY-1913789; 
and the U.S. Department of Energy, Research Awards
No.~DE-SC0012704, 
No.~DE-SC0011784, 
No.~DE-SC0010504, 
No.~DE-SC0021274, 
No.~DE-SC0021616, 
No.~DE-SC0007914, 
No.~DE-AC06-76RLO1830, 
and No.~DE-SC0010073. 

We thank the SuperKEKB team for delivering high-luminosity collisions;
the KEK cryogenics group for the efficient operation of the detector 
solenoid magnet and IBBelle on site;
the KEK Computer Research Center for on-site computing support; the NII for SINET6 network support;
and the raw-data centers hosted by BNL, DESY, GridKa, IN2P3, INFN, 
and the University of Victoria.